\documentclass[a4paper,fleqn,usenatbib]{mnras}

\pdfoutput=1

\usepackage{mathptmx}

\usepackage[T1]{fontenc}
\usepackage{ae,aecompl}

\usepackage{graphicx}	
\usepackage{amsmath}	
\usepackage{amssymb}	
\usepackage{xspace}
\usepackage{nicefrac}
\usepackage[normalem]{ulem}
\usepackage{xcolor}

\newcommand{\oiii}{[\ion{O}{iii}]\xspace}
\newcommand{\ar}[1]{\ion{Ar}{#1}\xspace}
\newcommand{\oii}{[\ion{O}{ii}]\xspace}
\newcommand{\nii}{[\ion{N}{ii}]\xspace}
\newcommand{\sii}{[\ion{S}{ii}]\xspace}

\newcommand{\hii}{\ion{H}{ii}\xspace}
\newcommand{\ariv}{[\ion{Ar}{iv}]\xspace}
\newcommand{\ha}{H$\,\alpha$\xspace}
\newcommand{\hb}{H$\,\beta$\xspace}
\newcommand{\hg}{H$\,\gamma$\xspace}
\newcommand{\haa}{\mathrm{H\,\alpha}\xspace}
\newcommand{\hbb}{\mathrm{H\,\beta}\xspace}
\newcommand{\hgg}{\mathrm{H\,\gamma}\xspace}
\newcommand{\naD}{\ion{Na}{i}\,D\xspace}

\newcommand{\lbol}{$L_{\mathrm{bol}}$}

\newcommand{\broad}{{\sf broad}}
\newcommand{\narrow}{{\sf narrow}}
\newcommand{\narrows}{{\sf narrows}}
\newcommand{\n}[1]{{\sf n#1}}
\newcommand{\bb}{{\sf b}}
\newcommand{\nn}{{\sf n}}
\newcommand{\nb}{{\sf n+b}}
\newcommand{\narrowbroad}{{\sf narrow+broad}}

\newcommand{\integrated}{{\tt integrated}}
\newcommand{\default}{{\tt default}}
\newcommand{\defa}{{\tt def}}
\newcommand{\inte}{{\tt int}}
\newcommand{\proj}{{\tt proj}}

\title[Gauging the effect of Quasar feedback]{Gauging the effect of Supermassive Black Holes feedback on Quasar host galaxies}

\author[B. Dall'Agnol de Oliveira et al.]{B. Dall'Agnol de Oliveira$^{1}$\thanks{E-mail: bruno.ddeo@gmail.com},
        T. Storchi-Bergmann$^{1,2}$, 
        S. B. Kraemer$^{3}$,
        \and
        M. Villar Mart\'in$^{4}$,
        A. Schnorr-M\"uller$^{1}$,
        H. R. Schmitt$^{5}$,
        D. Ruschel-Dutra$^{6}$,
        \and
        D. M. Crenshaw$^{7}$,
        T. C. Fischer$^{8}$
        \\
        $^{1}$Departamento de Astronomia, Universidade Federal do Rio Grande do Sul, IF, CP 15051, 91501-970 Porto Alegre, RS, Brazil\\
        $^{2}$Harvard-Smithsonian Center for Astrophysics, 60 Garden St., Cambridge, MA 02138, USA\\
        $^{3}$Institute for Astrophysics and Computational Sciences, Department of Physics, The Catholic University of America, Washington,\\ DC 20064, USA\\
        $^{4}$Centro de Astrobiolog\'ia (INTA-CSIC), Ctra. de Torrej\'on a Ajalvir, km 4, 28850 Torrej\'on de Ardoz, Madrid, Spain\\
        $^{5}$Naval Research Laboratory, Washington, DC 20375, USA\\
        $^{6}$Departamento de F\'isica, Universidade Federal de Santa Catarina, P.O. Box 476, 88040-900, Florian\'opolis, SC, Brazil \\
        $^{7}$Department of Physics and Astronomy, Georgia State University, Astronomy Offices, 25 Park Place, Suite 600, Atlanta, GA 30303, US\\
        $^{8}$AURA for ESA, Space Telescope Science Institute, 3700 San Martin Drive, Baltimore, MD 21218\\
}

\date{Accepted: 13 April 2021}

\pubyear{2021}

\begin{document}
\label{firstpage}
\pagerange{\pageref{firstpage}--\pageref{lastpage}}
\maketitle

\begin{abstract}

In order to gauge the role that active galactic nuclei (AGN) play in the evolution of galaxies via the effect of kinetic feedback in nearby QSO\,2's ($z\sim0.3$), we observed eight such objects with  bolometric luminosities $L_{bol} \sim 10^{46}\rm{erg\,s^{-1}}$ using Gemini GMOS-IFU's. 
The emission lines were fitted with at least two Gaussian curves, the broadest of which we attributed to gas kinetically disturbed by an outflow. 
We found that the maximum extent of the outflow ranges from $\sim$1 to 8 kpc, being ${\sim}\,0.5\,{\pm}\,0.3$ times the extent of the {\oiii} ionized gas region. Our `{\default}' assumptions for the gas density (obtained from the {\sii} doublet) and outflow velocities resulted in peak mass outflow rates of $\dot{M}_{out}^{\defa}\sim$\,3\,--\,30\,$\rm{M_{\odot}}\,yr^{-1}$ and outflow power of $\dot{E}_{out}^{\defa}\sim\,10^{41}$\,--\,$10^{43}\,\mathrm{erg\,s^{-1}}$. The corresponding kinetic coupling efficiencies are $\varepsilon_f^{\defa}=\dot{E}_{out}^{\defa}/L_{bol}\,\sim7\times10^{-4}$\,--\,0.5\,\%, with the average efficiency being only $0.06$\,\% ($0.01$\,\% median), implying little feedback powers from ionized gas outflows in the host galaxies.
We investigated the effects of varying assumptions and calculations on $\dot{M}_{out}$ and $\dot{E}_{out}$ regarding the ionized gas densities, velocities, masses and inclinations of the outflow relative to the plane of the sky, resulting in average uncertainties of one dex.
In particular, we found that better indicators of the {\oiii} emitting gas density than the default \sii line ratio, such as the {\ariv}$\lambda\lambda$4711,40 line ratio, result in almost an order of magnitude decrease in the $\varepsilon_f$. 

\end{abstract}

\begin{keywords}
galaxies: active -- quasars: emission lines -- ISM: jets and outflows -- quasars: supermassive black holes
\end{keywords}


\section{Introduction}

In Active Galactic Nuclei (AGN), the mass accretion to a nuclear supermassive black hole (SMBH) leads to emission of radiation, accretion disk winds and jets from its vicinity -- the so-called AGN feedback processes \citep{fab12,hec14,har18}. 
In galaxy evolution models \citep[e.g.][]{nel19,sch15,cro16,gon14}, negative feedback from these processes is a required mechanism that helps star formation quenching in the host galaxy, leading to the observed abrupt decrease in the galaxy luminosity function at the high-luminosity end \citep{sil12,naa17}.

The percentage of the AGN bolometric luminosity (\lbol) that couples with the interstellar medium (ISM) of the galaxies -- the AGN coupling efficiency -- seems to have a minimum threshold for it to have a significant impact on the galaxy (e.g. an outflow that can lead to escape of gas from the gravitational potential of the host galaxy). Using galaxy mergers simulations, \citet{dimat05} found a threshold of $\sim$\,5\,\%\,$L_{bol}$ in order to reproduce the relation between the mass of the SMBH (M$_{\rm SMBH}$) and the velocity dispersion $\sigma_*$ of the galaxy bulge \citep{geb00,fer00}. Using empirical estimates of other parameters, \citet{zub18} also got  5\,\%\,$L_{bol}$, but for redshift $z\sim3$ (with 13\,\% for $z\sim0.5$). 
According to the `two-stage' model of \citet{hop10}, this percentage could be lower, with a threshold of  0.5\,\%\,$L_{bol}$.
Note that these efficiencies refer not only to the kinetic power of outflows -- that will be studied in this work -- but to the total energy released.

In the so-called radiative mode feedback, radiation pressure from the AGN accretion disk leads to the formation of winds that couple with the circumnuclear interstellar medium of the host galaxies leading to outflows observed in the ionized gas of the Narrow-Line Region \citep[e.g.][]{sto10,rif13,rif20}.
Most of such studies have been done in the optical using the kinematics measured from the {\oiii}$\lambda$5007 emission line \citep[e.g.][]{cre07,fis13,fis18}.
AGNs can also influence the ISM via the radio-mode (or jet-mode) feedback, where radio jets interact with the gas, which can occur even in non radio-loud quasars \citep[e.g.][]{vil21,jar19}.
Strong outflow powers, in excess of 0.5\%\,\lbol, have been obtained for a few individual sources \citep[e.g.][]{cresci15,che19,cou20}. 
The energy released can also couple strongly with the molecular and neutral phase \citep[e.g.][]{fer15,cic14}. 
Strong outflows have even been seen in the earlier Universe \citep[$z\sim6$,][]{mai12,cic15}, although this seems not to be common  \citep{nov20}. 

Regarding the strength of the outflow power $\dot{E}_{out}$ in larger sample compilations covering $10^{44}\leq L_{bol}\leq10^{48}$\,erg\,s$^{-1}$, \citet{fio17} reported kinetic efficiency 
$\dot{E}_{out}/L_{\rm{bol}}$ of 0.1\,--\,10\,\% for outflows in the ionized phase, and 1\,--\,10\,\% in the molecular phase,
as required by models for a significant impact on the evolution of the host galaxy. Nevertheless, other recent studies \citep[e.g.][]{bar19,fal20,dav20} claim much lower powers, at least for AGN luminosities below $L_{bol}\,
{\sim}\,10^{46}$\,erg\,s$^{-1}$.

A recent study by our group \citep{sto18} of a sample of 9 QSO\,2's with \lbol$\,{\sim}\,10^{46}$\,erg\,s$^{-1}$  at $z\sim 0.3$ investigated the effect of different calculation methods and assumptions in the values of $\dot{E}_{out}$ on the basis of \textit{Hubble Space Telescope} (\textit{HST}) narrow-band images and integrated spectra of the sources. It was concluded that the use of integrated spectra lead to uncertainties in the calculated AGN kinetic power of up to 3 orders of magnitude \citep[see][for a complementary analysis]{rev18}. In order to better constrain the properties used in the calculation of the AGN powers, we have now obtained optical integral field spectroscopy of the above QSO sample. Our goal in the present paper is to revisit and improve the calculations of the AGN powers using this spatially resolved data.

This paper is organized as follows. In Sec.\,\ref{sec:sample} we present the sample, in Sec.\,\ref{sec:obs_red} we describe the observations and data reduction, in Sec.\,\ref{sec:methodology} we describe the analysis methodology, in Sec.\,\ref{sec:results} we present our results, in Sec.\,\ref{sec:discussions} we discuss them and in Sec.\,\ref{sec:conclusions} we present our conclusions.

\section{Sample}\label{sec:sample}

Our sample consists of 7 type 2 QSOs from the 9 objects targeted in \citet{sto18} in a study of \textit{Hubble Space Telescope} (\textit{HST}) narrow-band images combined with Sloan Digital Sky Survey (SDSS) spectra. The 9 type 2 QSOs were drawn from the \citet{rey08} sample, selected for having luminosities $L_{\oiii\lambda5007}$ above $10^{42}$\,erg\,s$^{-1}$, and redshifts in the range 0.1 < z < 0.5: these are luminous QSOs that could still be well resolved by \textit{HST} imaging observations.
To the original sample we added 1 similar object (J120041) from \citet{fis18}.
Incidentally, most of these galaxies have signs of mergers 
(see discussion in  Section 6.1) as seen in the \textit{HST} continuum maps (Fig.\,\ref{fig:hst}), which supports previous findings that a high incidence of mergers is correlated with strong nuclear activity \citep[e.g.][]{tre12}.

Some basic information of the sample objects is presented in Table \ref{tab:sample}: full name identifications (we use short names in the other tables), systemic redshifts ($z$, see Section\,\ref{sec:velocity}), {\oiii}$\lambda$5007 total luminosity (from HST images), angular scales and luminosity distances ($D_L$). 
The angular scale was used to convert spatial distances from arcseconds to kiloparsecs, and $D_L$ to obtain luminosities from fluxes. These two quantities were derived from the redshift, using the cosmological calculator of \citet{ned06} with $\mathrm{H_0=73\,km\,s^{-1}}$, $\mathrm{\Omega_M=0.27}$ and $\mathrm{\Omega_\Lambda=0.7}$, with the redshfits corrected for the CMB dipole model \citep{fix06}.

\begin{table}
	\centering
	\caption{
	\textbf{Sample.}
	(1) galaxy full name (prefix SDSS); 
	(2) redshift;
	(3) observed {\oiii}$\lambda$5007 luminosity (in units of $\mathrm{10^{42}\,erg\,s^{-1}}$);
	(4) spatial scale (in kpc/arcsec);
	(5) luminosity distance (in Mpc).
	}
	\label{tab:sample}
\begin{tabular}{lcccr}
\hline
SDSS name                      & z         & $L_{\oiii}$ & Scale & $D_L$ \\
(1)                            & (2)       & (3)         & (4)   & (5)   \\
\hline
J082313.50+313203.7$^\dagger$  & 0.43320 & 25.0        & 5.46  & 2310  \\
J084135.04+010156.3$^\dagger$  & 0.11045 & 5.37        & 1.95  & 498   \\
J085829.58+441734.7$^\dagger$  & 0.45395 & 11.6        & 5.61  & 2450  \\
J094521.34+173753.3$^\dagger$  & 0.12838 & 6.47        & 2.22  & 583   \\
J123006.79+394319.3$^\dagger$  & 0.40699 & 21.8        & 5.26  & 2149  \\
J135251.21+654113.2$^\dagger$  & 0.20747 & 18.8        & 3.26  & 980   \\
J155019.95+243238.7$^\dagger$  & 0.14294 & 4.83        & 2.42  & 653   \\
J120041.39+314746.2$^\ddagger$ & 0.11586 & 8.33        & 2.04  & 524   \\
\hline
\end{tabular}
\vspace{1ex}
 {\raggedright 
 \textit{HST} data from: $^\dagger$\citet{sto18}, $^\ddagger$\citet{fis18}  \par}
\end{table}

\section{Observations and data reduction}\label{sec:obs_red}
\subsection{Observations}\label{sec:obs}
\begin{figure}
	\includegraphics[width=1\columnwidth]{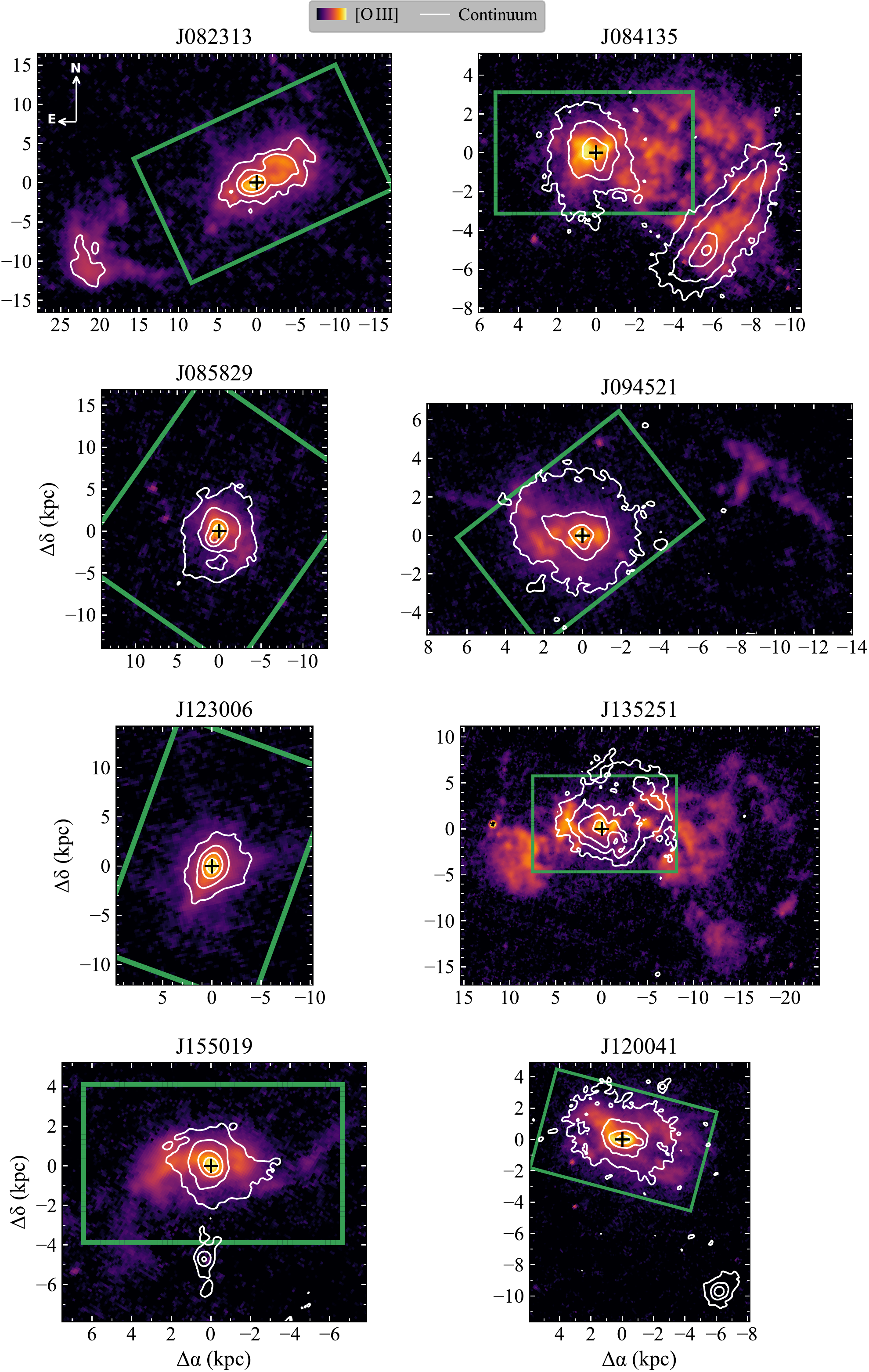}
    \caption{\textit{HST} narrow-band images of our sample. The maps show the \oiii flux distribution of the galaxies, with contours (white) from images of the stellar continuum, both in logarithmic scale. The GMOS-IFU FoVs are highlighted as green rectangles. }
    \label{fig:hst}
\end{figure}

\begin{figure}
	\includegraphics[width=1\columnwidth]{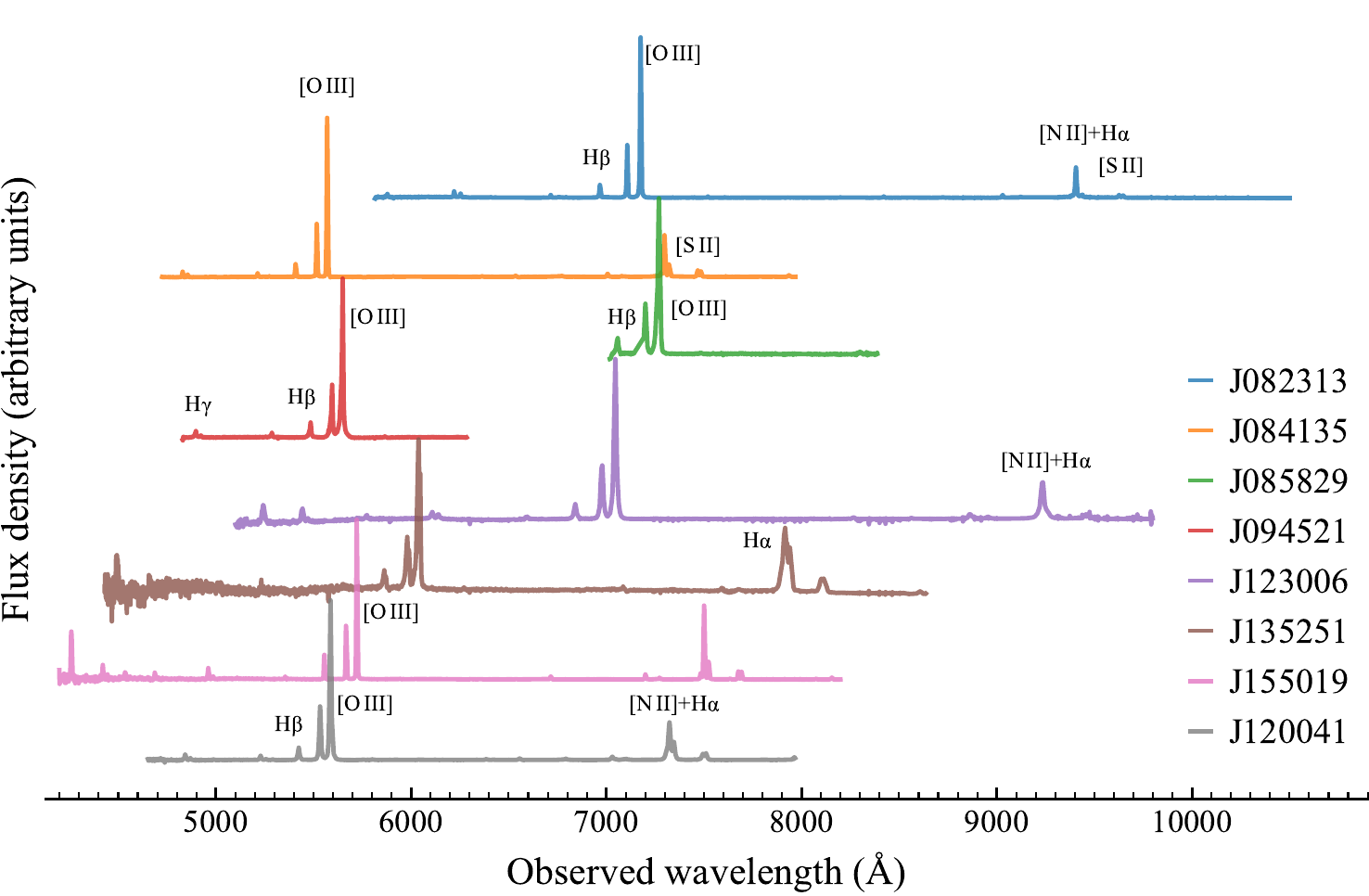}
    \caption{Spectra of the galaxies of our sample, integrated within a seeing-sized radius aperture over the data cube.
    }
    \label{fig:all_specs}
\end{figure}

Observations were undertaken with the Integral Field Unit \citep[IFU,][]{ali02} of the  Multi-Object Spectrographs (GMOS) available at both Gemini telescopes. The details of the observations are displayed in Table \ref{tab:obs}. The GMOS-IFU fields-of-view of each observation are shown as green rectangles over the \textit{HST} \oiii narrow-band images in Fig.\,\ref{fig:hst}, and the spectra of the sample -- integrated over the IFU data cubes -- are displayed in Fig.\,\ref{fig:all_specs}.

The GMOS-IFU instrument offers two configurations, depending on how the fibers are arranged: the two-slit mode covers a field-of-view (FoV) of 5\arcsec$\times$7\arcsec, and is identified as IFU-2 in the table, while the one-slit mode (IFU-R) has a smaller FoV (5\arcsec\,x\,3.5\arcsec) but covers a wider spectral region. For our observations, we opted for the one-slit mode because it enables the observation of more emission lines: from \hb, {\oiii}$\lambda\lambda$4959,5007, to \ha, {\nii}$\lambda\lambda$6548,84 and the pair {\sii}$\lambda\lambda$6718,31. In order to cover these lines and still have an adequate spectral resolution, R400 and B600 gratings were the best choices. When necessary, filters were used to block second-order contamination of the spectra. For each object, several exposures were obtained, aiming to: achieve a signal-to-noise ratio (S/N) above 3 at mid distances between the nucleus and regions at the borders of the IFU FoV; fill the spectral gaps (caused by the gaps between the CCDs) and the spatial holes (caused by dead lenslets/fibers), by means of dithering (offsetting) the observations in the spectral and spatial dimensions, respectively. The total exposure of the combined observations is also shown in the table.

As we found in the Gemini archive similar data for other two nearby QSOs -- J085829 and J094521 -- we have included also these objects in our analysis to enrich our study. These objects are identified in Table \ref{tab:obs} along with the references to the published results.

The archival data does not match all the requirements of our own observations (see Fig.\,\ref{fig:all_specs}). J085829 observations used the two-slit mode (IFU-2), resulting in a much smaller wavelength range, covering only {\hb} and the {\oiii} doublet. In the case of J094521, the use of the B1200 grating, which has a higher resolution, and consequently smaller spectral coverage, also results in a spectrum with only these three emission lines, along with {\hg}.

\begin{table*}
	\centering
	\caption{
	\textbf{Observations and reduction.}
	(1) galaxy short name (full name in Table \ref{tab:sample}); 
	(2) Gemini observation ID; 
    (3) and (4): gratings and filters full names; 
    (5) total exposure time (in seconds); 
    (6) spectral resolution (standard deviation, in \AA);
    (7) FWHM of the PSF (in arcsec), characterising the local seeing;
    (8) multiplicative factor to the flux calibrated data to match the SDSS spectra flux; 
    (9) instrumental resolution (in $\mathrm{km\,s^{-1}}$), corresponding to the rest wavelength range 4861.3\,--\,6730.8{\AA} (calculated from column 6).
    }
	\label{tab:obs}
\begin{tabular}{lcccccccr}
\hline
Name    & Observation ID                 & Grating      & Filter       & Exptime       & Spec. Res. & FWHM$_\mathrm{PSF}$ & $F_{\rm{SDSS}}/F_{\rm{GMOS}}$ & $\sigma_{inst}$ \\
(1)     & (2)                            & (3)          & (4)          & (5)           & (6)        & (7)    & (8)                          & (9)        \\
\hline
J082313 & GN-2018B-Q-207                 & R400\_G5305  & GG455\_G0305 & 6$\times$1040 & 1.00       & 0.68   & 0.42                & 61.9--44.7 \\
J084135 & GS-2018B-Q-110                 & B600\_G5323  & open         & 8$\times$1115 & 0.636      & 0.65   & 0.76              & 39.2--28.3 \\
J085829 & GN-2010B-C-10$^{\dagger\star}$ & R400\_G5305  & i\_G0302     & 2$\times$1800 & 1.06       & 0.59   & 1.14                & 65.2--47.1 \\
J094521 & GS-2010A-Q-8$^{\ddagger\star}$ & B1200\_G5321 & open         & 4$\times$1300 & 0.339      & 0.75   & 0.89              & 20.9--15.1 \\
J123006 & GN-2019A-Q-228                 & R400\_G5305  & GG455\_G0305 & 3$\times$1150 & 1.05       & 0.54   & 0.16                & 65.0--46.9 \\
J135251 & GN-2019A-Q-228                 & R400\_G5305  & open         & 6$\times$900  & 1.00       & 0.55   & 0.61                & 61.9--44.7 \\
J155019 & GN-2018A-Q-206                 & R400\_G5305  & open         & 4$\times$1200 & 1.03       & 0.55   & 0.73                & 63.3--45.7 \\
J120041 & GN-2019A-Q-228                 & B600\_G5307  & open         & 3$\times$1250 & 0.738      & 0.72   & 0.50              & 45.5--32.9 \\
\hline
\end{tabular}
 \vspace{1ex}
 {\raggedright Data from Gemini Archive, with published results in $^\dagger$\citet{liu13a}, and $^\ddagger$  \citet{har14} \par}
 {\raggedright $^{\star}$ Observed in two-slit mode (IFU-2), while the remaining galaxies used one-slit mode (IFU-R) \par}
\end{table*}

\subsection{Data reduction}\label{sec:red}
The data reduction was carried out with the Python package \texttt{GIREDS}\footnote{\url{https://github.com/danielrd6/gireds}}. This software automatizes the process, organizing the fits files and applying all the standard steps -- which include the reduction of both the standard star and the science files -- using the \texttt{IRAF}'s  \citep{tod86} packages provided by Gemini. The steps include (not in order) bias subtraction, cosmic ray removal and sky subtraction, flat fielding, fiber identification and wavelength calibration.

After the relative flux calibration, the software creates a data cube file for each science observation, already corrected for differential atmospheric refraction. Instead of sampling the reduced data using the diameter of the fibers ($\sim$\,0.2\,arcsec), we over-sampled the pixel scale, setting it to 0.1\,arcsec/pixel. Before merging the individual exposures, we removed the telluric absorption contribution from each data cube(see Section\,B1 in the online supplementary material).

At this step, each galaxy had a set of data cubes spatially offset from each other. Using the instrumental offsets (listed in the header), the software merged all observations into a final data cube, containing the calibrated flux and its uncertainties (propagated from the Poission statistics of the uncalibrated data). Furthermore, to obtain an absolute flux calibration -- so that our fluxes matched the SDSS spectra fluxes -- we calculated the flux ratio $F_{\rm{SDSS}}/F_{\rm{GMOS}}$ (listed in  Table\,\ref{tab:obs}) and multiplied our data by this factor. This ratio is the average of the flux density ratio, calculated for 10\,{\AA} windows, where the GMOS-IFU spectrum is the integrated spectra over the corresponding SDSS or BOSS angular aperture. 

The information of the rotation/inversion transformations contained in the World Coordinate System (WCS) were taken from the acquisition images, after scaling them to match the pixel scale of the cubes.
The reference pixel \texttt{CRPIX1,2}, corresponding to the IFU continuum maximum, has the right ascension and declination \texttt{CRVAL1,2} taken from the peak continuum in the \textit{HST} images.

To characterize the local atmospheric seeing, 
using \texttt{DAOPHOT} package from  \texttt{IRAF}, we fitted Moffat functions for the radial profiles of the field stars in the acquisition images. We obtained the full width at half maximum (FWHM) of the instrument point spread functions (PSF) ($\rm{FWHM_{PSF}}$ in Table\,\ref{tab:obs}, for an average Moffat parameter $\beta\sim3.5$). In this work, we also used  $\sigma_\mathrm{PSF}\,{=}\,\mathrm{FWHM_{PSF}}{/}2.355$ for the standard deviation, although the Moffat profiles deviates from a simple Gaussian.

Along with the integral field spectroscopy data, we made use of the available \textit{HST} data from \citet{sto18,fis18}. They consist of narrow-band images centred on {\oiii} and {\ha} emission lines, and in the continuum between these two lines (used to subtract its contribution from the emission line images). We also made use of the Sloan Digital Sky Servey (SDSS) spectra, taken from the Data Release 13 \citep{sdssIV}.

\section{Methodology}\label{sec:methodology}
\subsection{Emission line fitting}\label{sec:fit}
The emission lines profiles can be quite complex, showing a superposition of different components. Therefore, in order to retrieve the ionized gas kinematic and excitation information, we proceeded to model the emission lines with multiple Gaussian curves, which are individually characterized by three parameters: the line-of-sight (LoS) centroid velocity ($v$, also called radial velocity here); velocity dispersion ($\sigma$); and amplitude ($A$). We are considering that each component is measuring the properties of different groups of clouds along the line-of-sight in each pixel. In our fits, we used up to four Gaussians to model each emission line, adopting the same number of components in all pixels of a given object.

Our goal in the present paper is to quantify the AGN feedback via its kinematic coupling with circumnuclear gas of the host galaxy in the ISM. This coupling leads to disturbances in the gas kinematics, the highest velocity ones being observed in the profile wings. We assume that the {\broad} ({\bb}) component is tracing this disturbance and calculate the associated feedback which is usually identified as being due to AGN outflows.
We note that this approach may have limitations (see discussion in Section \ref{sec:broad}), as is probably the case of J094521, that may have a second `outflowing component' (see Section \ref{sec:disc-outflow}).

\begin{figure*}
	\includegraphics[width=1\linewidth]{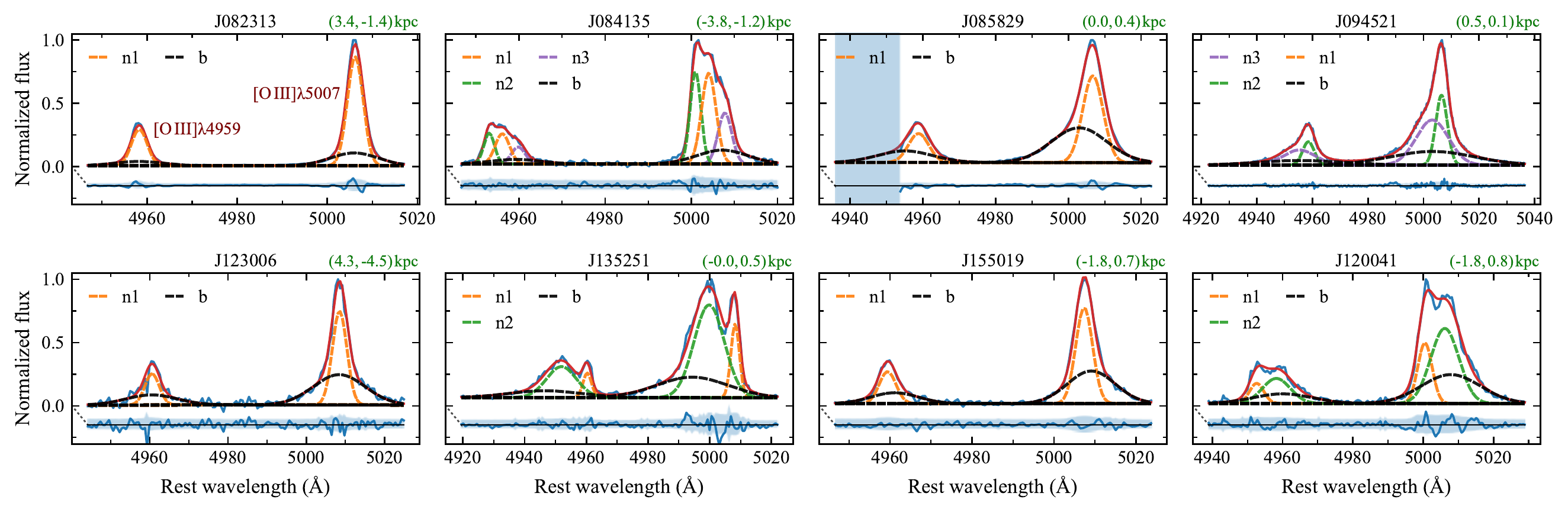}
    \caption{Examples of fitting results for the {\oiii} line profiles for each galaxy of the sample, highlighting the decomposition of the spectral flux density in {\narrow} (\n{1}, \n{2}, \n{3}) and {\broad} (\bb) components. The pixel location (projected distance from the nucleus, in kpc) is shown in top right of each panel. Four objects have more than one {\narrow} component, clearly seem in both {\oiii} profiles. Here, J094521 does not show its {\n{1}} component, that appears in another region of the galaxy, away from the nucleus (see Fig.\,\ref{fig:mapJ094521}, and Fig.\,B9 in the online material). An alternative and larger version of this figure is available in the online material (Fig.\,B5).
    }
    \label{fig:spec-comp}
\end{figure*}

We thus impose that, for every pixel, the {\broad} component should always have the largest $\sigma$. The remaining components are called {\narrows} (\n{1}, \n{2}, ..), which we assume that are originated in clouds that are not kinetically disturbed by an outflow. In our sample, the majority of the QSOs have signs of interactions in the \textit{HST} continuum images (see contours in Fig\,\ref{fig:hst}, and the discussion in Section\,\ref{sec:mergers}).
Hence, the {\narrow} components may be not only modelling or tracing the less disturbed gas in the  the galaxy, but possibly also the gas from the two different galaxies involved in the interaction, with initial non-zero relative velocities.
Possibly because of this, we needed to use more than one narrow component in half of our galaxies, since there are regions where we clearly cannot model their emission lines with only one {\narrow} and one {\broad} component (see Fig.\,\ref{fig:spec-comp}).

We fitted simultaneously the following emission lines, using the same number of {\narrow} and {\broad} components for each line: \hb, {\oiii}$\lambda\lambda$4959,5007, \ha, {\nii}$\lambda\lambda$6548,84 and [S\,II]$\lambda\lambda$6718,31.
For J094521, we also fitted H$\gamma$, used later to obtain  its reddening in Section\,\ref{sec:reddening}.

We defined groups of components that are forced to have the same kinematic properties (centroid velocities $v$ and velocity dispersions $\sigma$) for all emission lines (in units of km\,s$^{-1}$):
$\sigma_{\hbb,{\bb}} = ... = \sigma_{\sii,{\bb}}$;
and $v_{\hbb,{\bb}} = ... = v_{\sii,{\bb}}$,
and equivalently for each {\narrow} component. 
The velocity dispersion $\sigma_x$ refers to the observed value ($\sigma_{x,obs}$) corrected by the broadening caused by the instrumental dispersion ($\sigma_{inst}$): $\sigma_{\haa}^2 = \sigma_{obs,\haa}^2 - \sigma_{inst,\haa}^2$. 
Note that $\sigma_{inst}$ (in km\,s$^{-1}$) is different for each emission line rest wavelength (see Table\,\ref{tab:obs}). This correction is based on the one presented by \cite{gal19}.

To decrease the number of fitting parameters, we fixed the ratios between doublet lines according to their expected ratios \citep{ost06}: 
{\oiii}$\lambda\lambda5007/4959=3$ and {\nii}$\lambda\lambda6584/6548=3$.
We also added the following flux constraints, to avoid non-physical solutions: $\haa/\hbb > 2.74$ (intrinsic recombination value for $T_e=2{\times}10^4$\,K and $n_e=10^2\mathrm{\,cm^{-3}}$, also according to \citet{ost06}, and $0.436<{\sii}\lambda\lambda6718/6731<1.496$ \citep[asymptotic values for high and low densities from][]{pro14}.
These bounds/constraints were applied to each component.

The fitting procedure was performed using the publicly available software \texttt{IFSCUBE} \citep{ifscube}. This program uses \texttt{SCIPY} \citep{scipy} implementations of non-linear minimization and allows the addition of constraints and bounds to the parameters.
For each spectrum, the chi-square of the difference between the observed and the modelled flux densities is minimized, where the flux uncertainties and degrees of freedom are considered.
Initially, a guess to the parameters of an initial pixel is given. Then, the program fits the rest of the pixels, updating the initial guess based on the successful fits of the neighbouring spectra.

\subsubsection{Constraining the number of components}
To obtain the number of the components needed to fit the emission lines, first, we fitted only the {\oiii}$\lambda\lambda$4959,5007 emission lines, due to their high S/N. We varied the initial parameter guesses, trying to keep the spatial continuity in the resulting maps of the parameter.

Next,  we ran the fitting procedure for all emission lines simultaneously.
However, depending on the region, some components may be very weak. Therefore, we started the new fit in a high S/N pixel, that displayed all components, using the resulting parameters (kinetic part) of the previous {\oiii} fit 
as the new initial guesses.

\begin{figure}
    \centering
	\includegraphics[width=0.95\columnwidth]{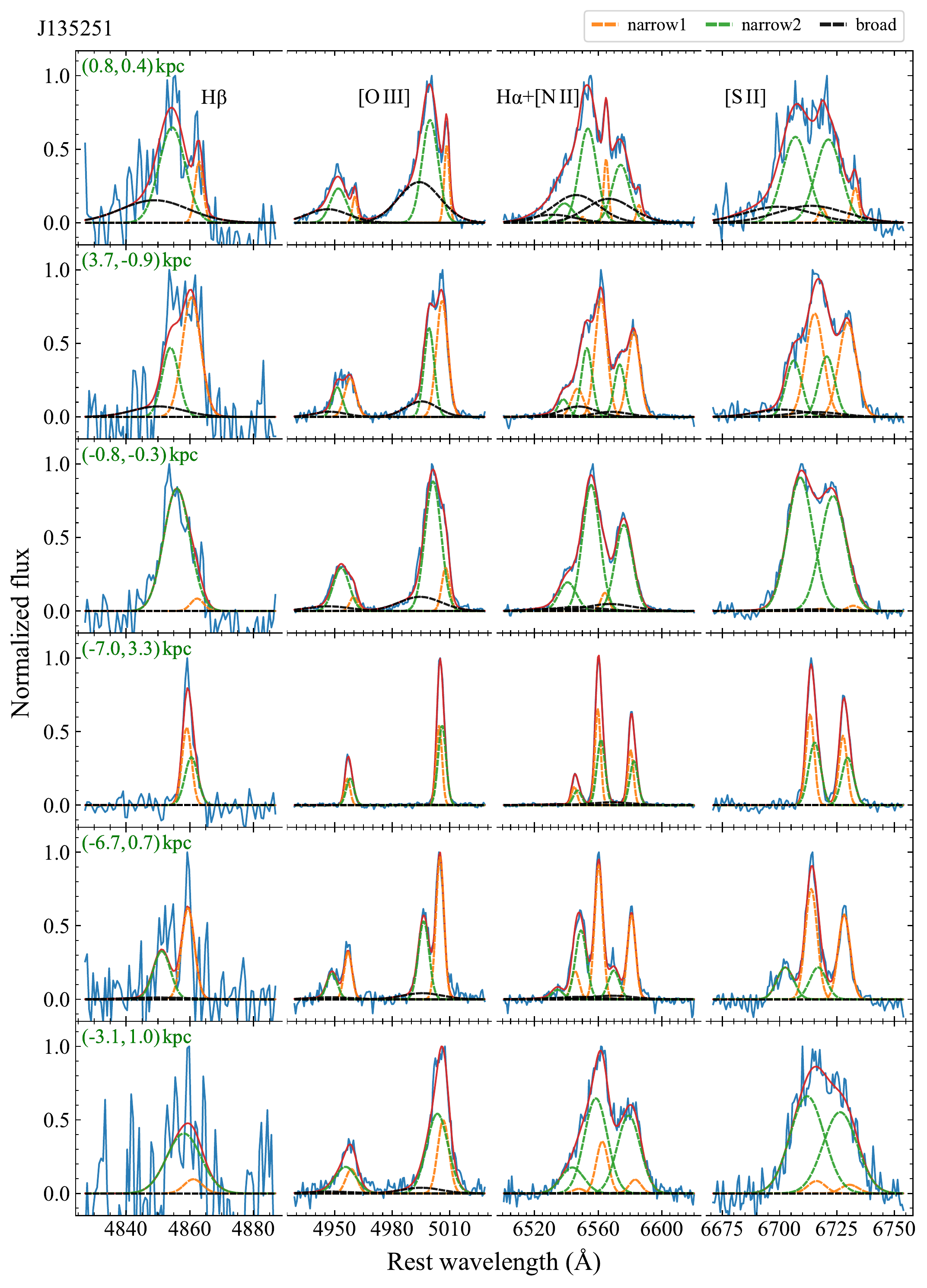}
     \caption{Fitting results for a few spectra of J135251. Each row of the figure corresponds to a different position in the galaxy, identified by $(\Delta \alpha, \Delta \delta)$, the distance from the nucleus (in green at the upper left corner of each row). The columns are zoom in the emission lines: \hb, \oiii, (\ha+\nii) and \sii.}
    \label{fig:specJ135251}
\end{figure}

In Fig.\,\ref{fig:spec-comp}, we present results of the fitting process for representative {\oiii} emission-line profiles of each QSO, with the Gaussian profiles superimposed on the original spectra. Along with the {\broad}, four QSOs needed more than one {\narrow} component in the models: J084135, J094521, J135251 and J120041. 
Examples of fits of profiles from other spaxels are displayed for J135251 in Fig.\,\ref{fig:specJ135251}, while the corresponding figures of the remaining QSOs are presented in the online supplementary material (Figs.\,B6--B13), where different rows correspond to line profiles from different spaxels over the FoV, with the columns displaying a zoom-in on the emission line profiles of \hb, \oiii, \nii, \ha and \sii.
We chose to show pixels with different characteristics, highlighting the fact that the number of components needed in the models may vary over the FoV. 
For example, J094521 (Fig.\,B9) has a low-$\sigma$ component ({\narrow}1) that is only visible away in pixels outside the nuclear region (mainly at $\sim$\,3\,kpc to the East, as it can also be seen in the maps of Fig.\,\ref{fig:mapJ094521}). We can also observe that the multiple components appear not only in {\oiii}, but also in the other emission lines (e.g. first row of Fig.\,\ref{fig:specJ135251}, for J135251). This reinforces the decision of fixing the kinematics among components of different emission lines.

During the profile fitting process, we identified that the nuclear spectra of J082313 were best fitted using an additional (weak) Broad Line Region component to the H$\alpha$ emission line profiles (see Section\,B2 in the online supplementary material).

\subsection{Gas excitation and physical properties}\label{sec:nebular}

From the emission-line ratios, we have checked the gas excitation over the whole emitting regions via the BPT diagram \citep{bal81} that is presented in the online supplementary material (Section\,B3). This diagram shows a range of {\nii}/{\ha} ratios but high {\oiii}/{\hb} ratios everywhere putting all points in the Seyfert excitation region.

\subsubsection{Density}\label{sec:density}
\begin{figure}
	\includegraphics[width=0.95\columnwidth]{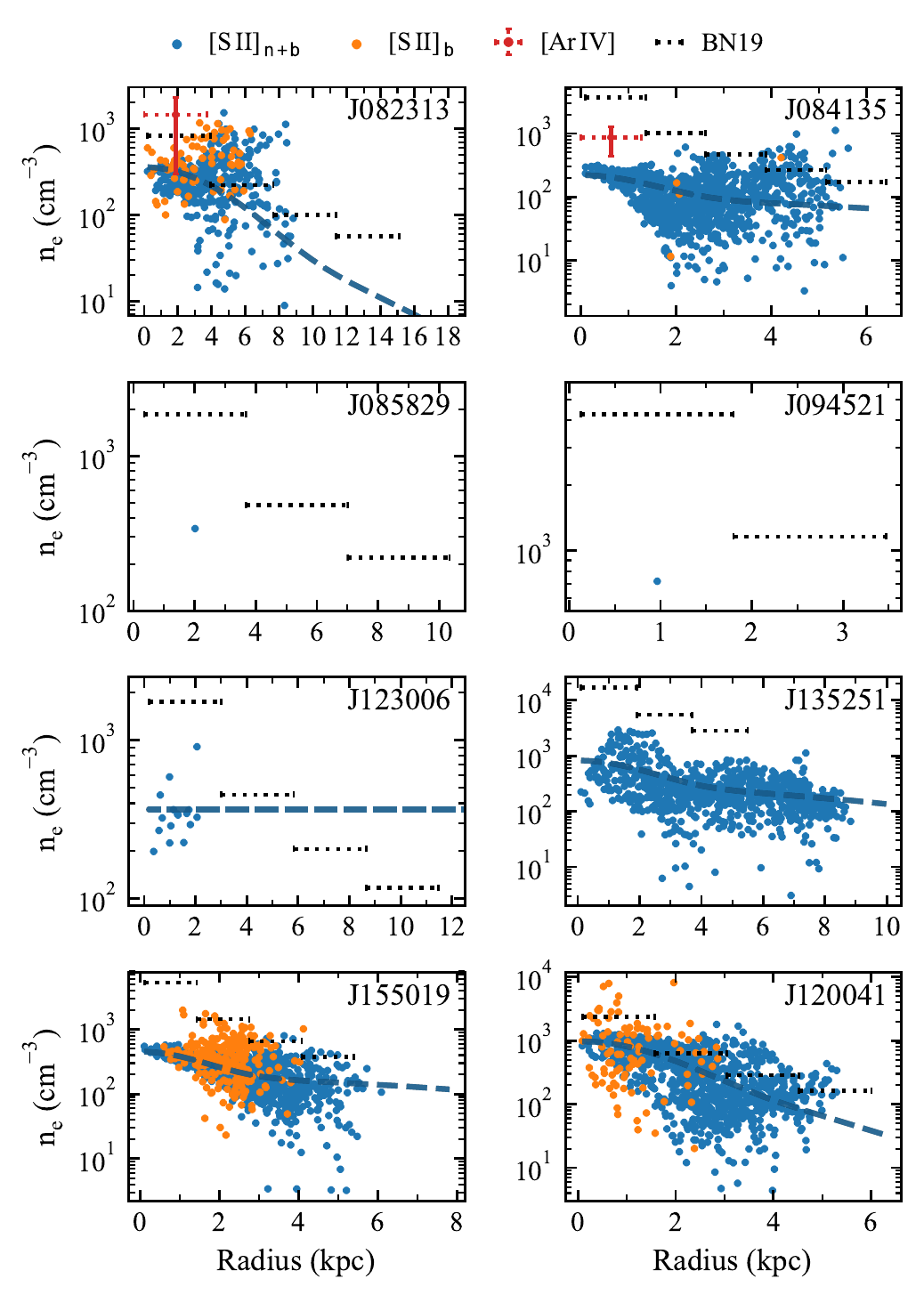}
    \caption{Comparison of gas density values obtained from different indicators. Orange circles were obtained using only the {\broad} component of the {\sii} lines, while blue circles were obtained from the integrated profiles (\nb).  In red, the values obtained from the {\ariv} lines in seeing-size aperture spectra. Also shown are values from the ionization parameter method \citep[BN19]{bar19}, for successively increasing radii (black dotted bars). 
    The dashed blue lines are the fits made to $n_{e,\sii\nb}$ (see Section\,\ref{sec:snr}).
    J085829 and J094521 have $n_{e,\sii\nb}$ measured from their SDSS spectra.}
    \label{fig:ne}
\end{figure}

In order to obtain a measurement of the electron density ($n_e$), we have first used the ratio between the pair of lines {\sii}$\lambda\lambda$6718,31. Assuming a typical NLR electron temperature in AGNs of $T_\mathrm{e}=10^4K$, we obtained $n_e$ following \citet{pro14}. The two objects  -- J085829 and J094521 -- that didn't have {\sii} coverage in the IFS spectra, had $n_e$ measured from the  the SDSS spectra. The resulting $n_e$ values as a function of radial distance from the nucleus  are shown in Fig.\,\ref{fig:ne}. Whenever possible, we performed two calculations for $n_e$: using only the {\broad} components ($n_{e,\sii\bb}$, in orange), and using the integrated profile comprising all components ($n_{e,\sii\nb}$, in blue). Both values display significant uncertainties due to the closeness of the lines and their multiple components (e.g. Fig.\,\ref{fig:specJ135251}), with the {\broad} one being more affected due to its lower S/N. Since the resulting $n_{e,\sii\nb}$ and $n_{e,\sii\bb}$ values are consistent within the errors, we were not able to verify if $n_{e,\sii\bb}>n_{e,\sii\nb}$, as found by other authors \citep[e.g.][]{vil14}. Therefore, due to its lower variance and higher spacial coverage, we decided to use $n_{e,\sii\nb}$ in the calculations with the method we have called `{\default}' along this paper. In the following sections we define the other assumptions and parameters of this {\default} method (see Table\,\ref{tab:met}). Figs.\,\ref{fig:mapJ135251} and \ref{fig:mapJ082313}--\ref{fig:mapJ120041} show the spatial distribution of $n_{e,\sii\nb}$ for all targets.

\citet{pro14} updated the expression for another electron density tracer, the {\ariv}$\lambda\lambda$4711,40 emission-line ratio, which is best suited to calculate the gas density associated with the {\oiii} emitting clouds due to the more similar ionization potential than that of the {\sii} lines (see discussion in Section \ref{sec:disc-met-ne}).
Using a PSF-size aperture spectra, we could measure these lines for two objects (J082313 and J084135) which show in the spectra {\ariv} lines with S/N > 3. We fitted these lines with a Montecarlo method (see Section\,B4 in the online material), obtaining higher $n_{e,\ariv}$ values than those obtained from the {\sii} lines (red errorbars in Fig.\,\ref{fig:ne}).

Finally, we have also investigated the method of calculation of the gas density described in \citet[BN19]{bar19}, which is based on the ionization parameter $U$.
We calculated $U$ using their expression, that depends on the line ratios of $\oiii{/}\hbb$ and $\haa{/}\nii$ (obtained from SDSS, in the case of J085829 and J094521).
The equation depends also on the AGN bolometric luminosity (\lbol), and the extent of the region  associated with the outflow $R_{out}$, with a $n_{e,\rm{BN19}}\propto L_{bol}\,U^{-1}\,R_{out}^{-2}$ dependence. Fig.\,\ref{fig:ne} displays the values calculated for different radii, separated by a  $\rm{FWHM_{PSF}}$ width (black dotted bars). 
{\lbol} (Table\,\ref{tab:out_prop}) were calculated in \citet{sto18}, from the \cite{tru15} relation between and {\lbol} and the reddening-corrected \citep{lam09} {\oiii}$\lambda$5007 luminosity (Table\,\ref{tab:sample} displays the values before the reddening correction). For J120041 (that is not in \citet{sto18} sample), we applied the same method. In Section\,\ref{sec:disc-met-ne}, we discuss the effect that different density tracers have in the outflow properties.

\subsubsection{Reddening}\label{sec:reddening}
To measure the extinction caused by foreground dust, we compared the observed ratio of hydrogen Balmer emission lines with its intrinsic recombination value for a given electron temperature and density, assuming the same reddening value for all the components. The color excess $E(B-V)$ was obtained using the expression from \citet{rev18}, where $R_\haa$ and $R_\hbb$ were obtained from the reddening law of \citet{sav79}. Assuming that the extinction is the same to all components,  the observed ratio $(\haa/\hbb)_{obs}$ was computed using the sum of the fluxes of all components of each emission line, while the intrinsic value was set to $(\haa/\hbb)_{int}=3.1$ \citep{hal83}. To obtain the corrected absolute flux from a given emission line, we used $  F(\lambda)_{int} = F(\lambda)_{obs} \cdot10^{0.4 \cdot R_\lambda \cdot E(B-V)} $ \citep{sea79}, with $F(\lambda)_{obs}$ and $F(\lambda)_{int}$ being the observed and the corrected fluxes, respectively, of a given emission line, with $R_\lambda$ also obtained from \citet{sav79}. For J094521, we used instead the ratio $\hgg/\hbb$, because its spectra do not cover the \ha region (see Fig.\,\ref{fig:all_specs}), using an intrinsic value $(\hgg/\hbb)_{int}=0.469$ \citep[case B, $\mathrm{T_e=10^4\,K}$]{ost06} in this case. The spacial distributions are shown in Figs.\,\ref{fig:mapJ135251}, and \ref{fig:mapJ082313}--\ref{fig:mapJ120041}.
For J085829, we used a single value of $E(B-V)=0.27$\,mag, obtained from the SDSS spectrum. 

\subsection{Ionized gas mass}\label{sec:mass}
The mass of the ionized gas ($M_{\hii}$) can be obtained from the luminosity of {\hii} emission lines. Ignoring the small contribution from ions other than \hii and \ion{He}{ii} to the total gas mass, we can use \citep{sto18}:
\begin{equation}\label{eq:mii}
 M_{\hii} =  \frac{(1.4\,m_p)\,L(\hbb)}{n_e\,\alpha_{\hbb}^{eff}\,(h\,\nu_{\hbb})},
\end{equation}
where $m_p$ is the proton mass, $L(\hbb)$ is the \hb luminosity (calculated from its flux and $D_L$ from Table\,\ref{tab:sample}), $h\,\nu_{\hbb}$ is the energy of the transition, $\alpha_{\hbb}^{eff}=3.02\,{\times}\,10^{-14}\,\rm{cm^{3}\,s^{-1}}$ is the effective recombination coefficient of {\hb} (for $n_e\,{=}\,10^2$cm$^{-3}$, and $T_e\,{=}\,10^4$\,K).
The factor 1.4 is to consider the contribution of He to the total mass. \ha has a higher S/N over most of the FoV, covering a larger spatial extent. Therefore, instead of directly using the \hb luminosity, we used the \ha luminosity: $L_{\hbb} = L_{\haa}/3.1$ (based on the intrinsic ratio assumed in the reddening correction). However, we used \hb in J085829 and J094521, since their spectra do not cover the \ha line (see Fig.\,\ref{fig:all_specs}).

\subsection{Outflow velocity}\label{sec:velocity}
We made the hypothesis that the {\broad} component is tracing the outflow. In order to weight in particular the effect of the highest velocity gas that contributes to the wings of the emission line profile, we used the following parametrization for the outflow velocity ($v_{out}$):
\begin{equation}
    v_{95} = \left| v_{\broad} \right| + 2\,\sigma_{\broad},
\end{equation}
where $v_{\broad}$ is the centroid velocity (velocity shift relative to the galaxy systemic velocity) of the {\broad} component, and $\sigma_{\broad}$ its velocity dispersion.
This definition is similar to $V_{max}$ from  \citet{rup13,fio17}.
In this equation, we are assigning values of velocity at the extreme ends of the {\broad} component as representative of the outflow velocity. In Section \ref{sec:disc-methods}, we test other $v_{out}$ definitions. 

The galaxy systemic velocity was adopted to correspond to the mean value of one of the {\narrow} components, inside a seeing-size radius: {\narrow{1}} for most galaxies and {\narrow{2}} for J094521 and J120041. The criterion used for this assumption was to consider as systemic, the component that best represents the velocity field of the host galaxy, that we assumed to be the component with lowest mean velocity dispersion inside this region, as broader components may contain, for example, contribution from outflows or gas from interacting companion galaxies. In this way we avoid adopting as systemic velocity  those from components that are not present in the galaxy nucleus (e.g. {\n{1}} in J094251, as shown in Fig.\,\ref{fig:mapJ094521}). Table\,\ref{tab:sample} displays the corresponding systemic redshifts.

The fitting process imposed that the {\broad} components of all emission lines from a given spectrum have the same kinematics, hence we could use any emission line to calculate $v_{out}$. We chose {\oiii}$\lambda$5007 because its higher S/N ratio over the FoV (in relation to the other lines), allows its measurement up to larger distances from the nucleus.

\subsection{Mass outflow rate and power}\label{sec:outflow}
In order to characterize the feedback, we obtained the ionized gas mass outflow rate ($\dot{M}_{out}(r)$) and the outflow power ($\dot{E}_{out}(r)$) as a function of radial distance from the nucleus. These quantities were calculated as rates crossing rings at increasing radii $r$ from the nucleus. Following \citet{shi19}, we define:

\begin{equation}\label{eq:mdot}
    \dot{M}_{out}(r) = \frac{M_{\hii}(r)\,v_{out}(r)}{\delta r},
\end{equation}

\begin{equation}\label{eq:edot}
    \dot{E}_{out}(r) = \frac{1}{2} \dot{M}_{out}(r) \,v_{out}^2(r),
\end{equation}
where $\delta r$ is the width of an annular aperture, at a given radius $r$ with origin at the nucleus. The position of the nucleus was adopted to correspond to the peak of the continuum emission, while $\delta r=0.5\,\sigma_{\rm{PSF}}$ (where $\sigma_\mathrm{PSF}=\mathrm{FWHM_{PSF}}{/}2.355$). 
$M_{\hii}(r)$ is the integration of $M_{\hii}(x,y)$ over all pixels inside each $\delta r$ annulus, while the $v_{out}(r)$ is the average velocity value within the annulus (see Fig.\,\ref{fig:vout}). 

In the {\default} method, $M_{\hii}$ (Eq.\,\ref{eq:mdot}) was calculated using only the {\broad} component, adopted to correspond to the outflowing gas. However, we tested the effect of using all components ({\broad}+{\narrow}) in Section\,\ref{sec:disc-met-mass}. 
The above equations imply that we are observing the $v_{out}$ component that radially crosses each annulus, but we are actually measuring only the line-of-sight component of $v_{out}$ and the sky-plane component of $\delta r$. In Section\,\ref{sec:disc-met-vout}, we discuss how the projection effects in $v_{out}$ and $\delta r$ affect the calculations.

\subsection{Signal-to-noise ratio}\label{sec:snr}
For each component, we only used pixels with S/N > 3, where the signal (S) is the peak flux emission and the value of the noise (N) has been adopted as the standard deviation of the continuum flux close to each emission line. The result is that some data is discarded, limiting the spatial extent of the components measurements. However, in this way we can be more confident on the properties derived from the remaining data. Another issue is that each emission line has a different S/N over the FoV and consequently covers a different extent. For example, there are regions in which {\oiii}$\lambda$5007 is strong, but the gas density could not be determined. Therefore, to compute $\dot{M}_{out}$ and $\dot{E}_{out}$ in these regions, we extrapolated the values of $n_e$, $E(B-V)$, F$_\haa$ and F$_\hbb$. This was done by modeling the radial profiles of these quantities via the fit of two 1D Gaussian curves over the corresponding radial profiles and replacing the missing values by the ones extrapolated using this model. As an example, Fig\,\ref{fig:ne} shows the resulting fits for $n_{e,\sii}$ (for J123006, we used the average).

\section{Results}\label{sec:results}

\subsection{Maps}\label{sec:maps}
\begin{figure}
	\includegraphics[width=1\columnwidth]{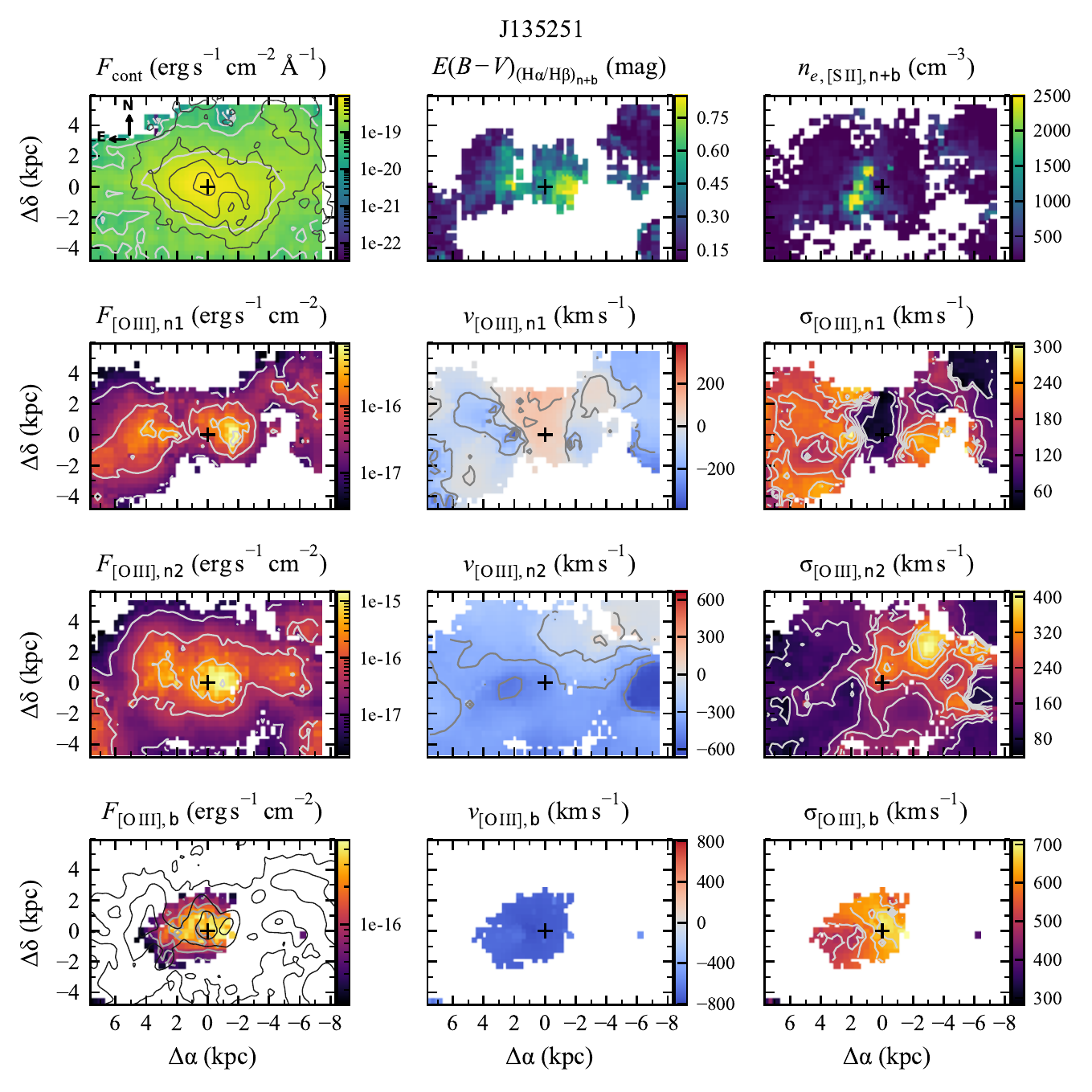}
\caption{Maps of the measured properties of J135251.
In the first row, from left to right, we show: the continuum flux ($F_{\rm{cont}}$), the gas reddening ($E(B\,{-}\,V)$), and the electron density ($n_e$) maps.
The remaining rows display maps of the parameters of each {\oiii}$\lambda$5007 component: flux ($F_{\oiii}$, left); radial velocity ($v_{\oiii}$, centre); and the velocity dispersion ($\sigma_{\oiii}$, right column).
 The last row refers to the {\broad} component, while the remaining middle rows refer to the {\narrow} components (\n{1} and \n{2}, in this case).
We overplotted the \textit{HST} contours (in black, starting in 3\,$\rm{\sigma_{sky}^{HST}}$) of the continuum (top left) and \oiii (bottom left) images.
North is up and East is left, with the right ascension and declination distances given relative to the continuum peak (black cross).
The systemic velocity were calculated from $v_{\oiii,\n{1}}$.
} 
    \label{fig:mapJ135251}
\end{figure}

We present the result of the fitting process, with maps of the fitted parameters displayed in Fig.\,\ref{fig:mapJ135251} for J135251, and Figs.\,\ref{fig:mapJ082313}--\ref{fig:mapJ120041} for the remaining QSOs (Larger size versions are available in the online material, Figs.\,B14--B21). The continuum flux densities were measured in 400{\AA} wide spectral windows centred at $\lambda$5300{\AA}. We chose to show only the maps for {\oiii}$\lambda$5007 because they have the highest S/N over the whole FoV, and the kinematics of the other emission lines are the same -- as imposed by the fitting procedure.

\subsection{Outflow radius}\label{sec:radius}

\begin{figure}
	\includegraphics[width=0.95\linewidth]{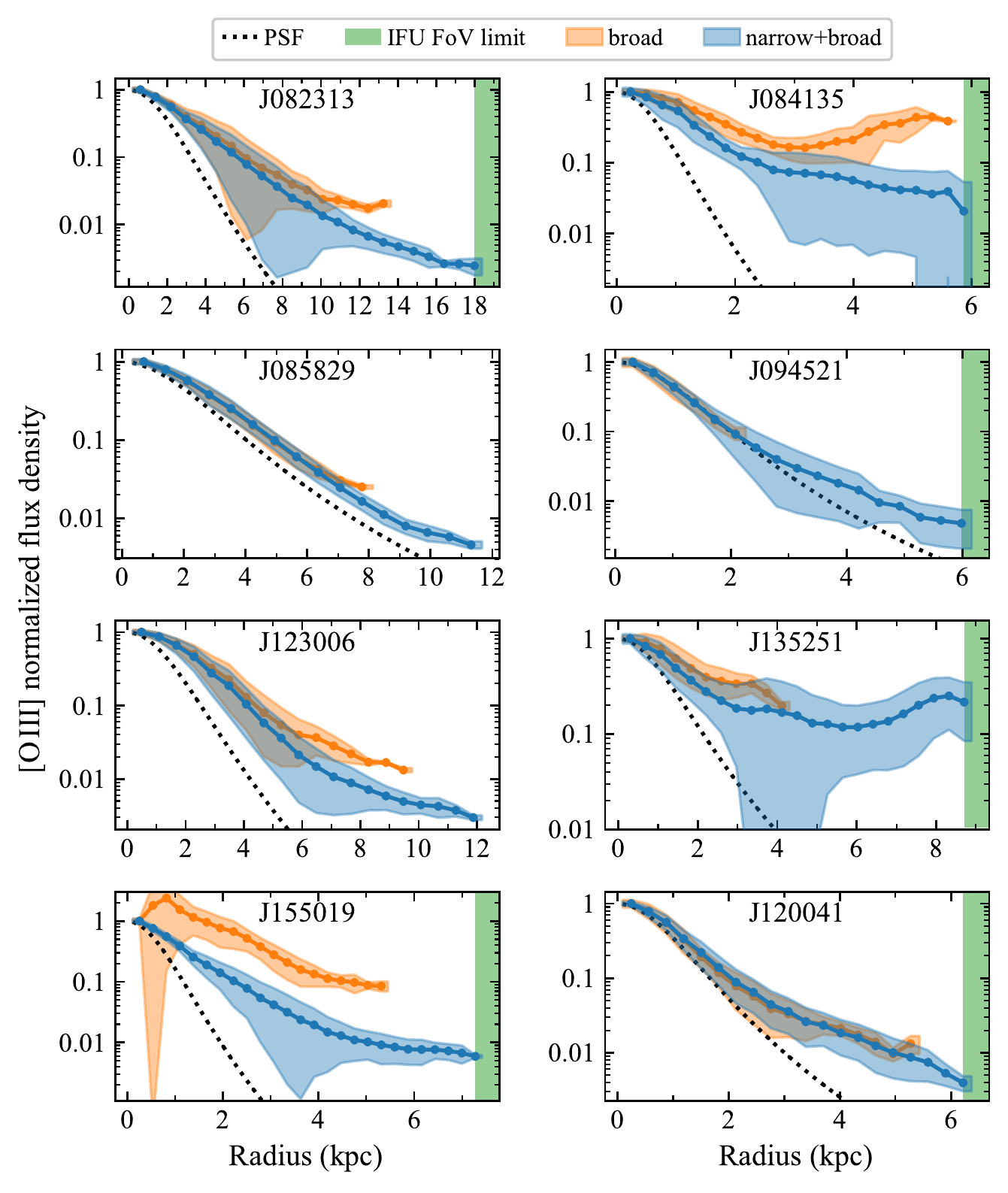}
    \caption{Radial profiles of the normalized {\oiii} flux density, both for the {\broad} and {\narrowbroad} components.
    The blue line is the mean value of $F_{\oiii,\nb}$, inside annuli $\rm{\sigma_{PSF}}{/}2$ wide, while the orange line is calculated from $F_{\oiii,\bb}$. 
    The shaded regions are the $F_{\oiii}$ standard deviation of the azymuthal mean value. 
    The $R_{\bb}$ and $R_{\nb}$ extents in Table\,\ref{tab:out_prop} correspond to the radial profiles limits.
    The green vertical regions mark the limit of the IFU FoV, and the black dotted line is the PSF radial profile.}
    \label{fig:radius_out}
\end{figure}

In order to investigate the extent of the outflowing component, in Fig.\,\ref{fig:radius_out} we present the flux density radial profiles of {\oiii}, both for the {\broad} and {\narrowbroad} components. The radial profiles correspond to the mean azimuthal value of $F_{\oiii}$ inside annuli with $\rm{\sigma_{PSF}}{/}2$ width, with the shaded regions showing the standard deviation of the mean. 
Note that the variation is not the $F_{\oiii}$ uncertainty, but mostly reflects its variation over the pixels inside each annulus: there are fewer pixels close to the nucleus and at the outer parts of FoV, resulting in a smaller $F_{\oiii}$ standard deviation. Isolated pixels were discarded, avoiding spurious data, that could lead to an overestimation of the outflow radii.

We measured the radial extent of each component (Table\,\ref{tab:out_prop}), defining the radius $R$ as the maximum radial distance reached, including only pixels with $F_{\oiii}$ S/N > 3 (see profiles in Fig\,\ref{fig:radius_out}). Both $R_{\bb}$ and $R_{\nb}$ uncertainties were set equal to $\rm{\sigma_{PSF}}$. Note that some galaxies have these measurements limited by the GMOS-IFU FoV (green region in the figure, and marked by ($\dagger$) in Table\,\ref{tab:out_prop}).

In order to compare the radial extent obtained above with those obtained using the narrow-band images of \citet{sto18}, we re-measured the {\oiii} flux distribution extents in the \textit{HST} images ($R_{\oiii}$), defining them as the maximum distance from the nucleus where the corresponding fluxes could be measured, limited by the contours at  $3\,\rm{\sigma_{sky}^{HST}}$, where $\rm{\sigma_{sky}^{HST}}$ is the flux standard deviation in a region of sky. These are shown as the outermost black contours in the lower left panel of Figs.\,\ref{fig:mapJ135251} and \ref{fig:mapJ082313}--\ref{fig:mapJ120041}, although it is not completely visible in some galaxies (due to the small IFU FoV, as shown in Fig.\,\ref{fig:hst}). The $R_{\oiii}$ uncertainties are equal to the difference in the extent values measured at the thresholds corresponding to 2 and 4\,$\rm{\sigma_{sky}^{HST}}$.
The above radial sizes are also shown in Table\,\ref{tab:out_prop}.

\subsection{Radial profiles of $\dot{M}_{out}$ and $\dot{E}_{out}$}\label{sec:radial}
\begin{figure*}
	\includegraphics[width=.9\linewidth]{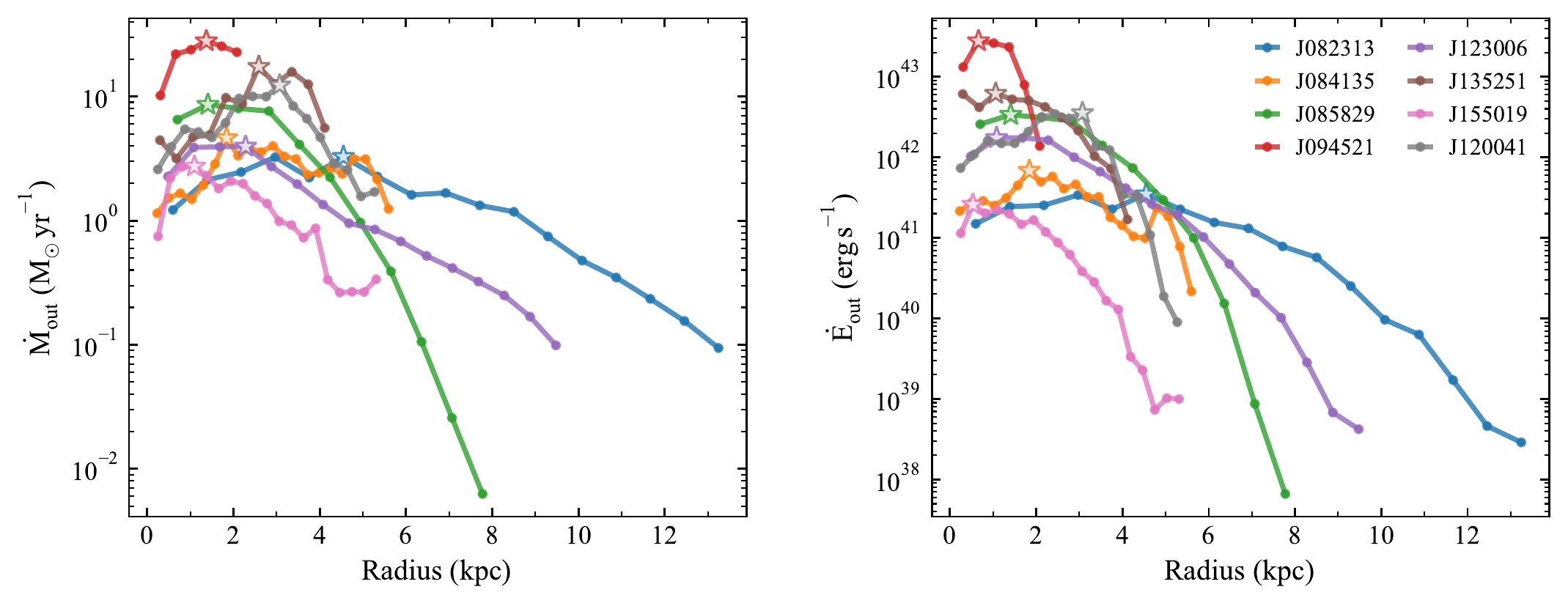}
    \caption{Radial profiles of $\dot{M}_{out}(r)$ (left) and $\dot{E}_{out}(r)$ (right) for the QSOs of our sample.
    The peak values are highlighted by stars, and correspond to $\dot{M}_{out}^{\defa}$ and $\dot{E}_{out}^{\defa}$ listed in Table\,\ref{tab:out_prop}, and  presented in Fig.\,\ref{fig:literature} as filled blue stars.
    }
\label{fig:rates_radial}
\end{figure*}

Using Eqs.\,\ref{eq:mdot} and \ref{eq:edot}, we obtained radial profiles of $\dot{M}_{out}(r)$ and $\dot{E}_{out}(r)$ for the galaxies in our sample. Fig.\,\ref{fig:rates_radial} shows the rates of these quantities crossing $\delta r=\sigma_{\rm{PSF}}{/}2$ wide rings, at different radii $r$. As described in Section\,\ref{sec:outflow}, the outflow velocity $v_{out}(r)$ is the mean value calculated from $v_{out}(x,y)$ within the annuli, and $M_\hii(r)$, the sum of $M_\hii(x,y)$ along the annuli.
 
More specifically, the figure shows the profiles resulting from our {\default} assumptions, namely: ionized gas mass $M_{\hii}^{\bb}$, calculated using only the outflow contribution (the {\broad} component); electron density $n_{e,{\sii_{\nb}}}$, calculated from the {\sii} ratio, using the full profile ({\narrowbroad}, see Section\,\ref{sec:density}); outflow velocity $v_{95}$ (see Section\,\ref{sec:velocity}).
Note, however, that the {\default} assumptions are not necessarily the best choice. For example, the use of {\sii} lines ratios as electron density estimators have been questioned by other authors \citep[e.g.][]{dav20}. In order to take these aspects into account, we test the effects of different assumptions in Section\,\ref{sec:disc-methods}.

We present `characteristic' $\dot{M}_{out}$ and $\dot{E}_{out}$ values -- defined as the maximum values along the corresponding radial profiles (stars in Fig.\,\ref{fig:rates_radial}) -- following what has been also adopted by other authors \cite[e.g.][]{shi19,fal20} in Table\,\ref{tab:out_prop}. In this table, we list the values for the {\default} assumptions, together with the minimum and maximum values obtained for these properties according to the tests discussed in Section\,\ref{sec:disc-methods}.

\section{Discussion}\label{sec:discussions}
\subsection{Maps}\label{sec:mergers}
\subsubsection{Multiple components and interactions}\label{sec:discuss
ion:mergers}

Diverse and complex scenarios are seen in the maps of the flux distributions and kinematics of the ionized gas from our galaxies (Fig.\,\ref{fig:mapJ135251} and \,\ref{fig:mapJ082313}--\ref{fig:mapJ120041}).
Besides the {\broad} component, present in all QSOs, 4 of them needed more than one {\narrow} Gaussian in order to model its emission-line profiles (see Section\,\ref{sec:fit}): J084135, J094521, J135251 and J120041.
This could be a consequence of the high incidence of mergers in our objects, given that most show signs of interaction in their \textit{HST} images (Fig.\,\ref{fig:hst}). 
The superposition of emission from gas with origin in different galaxies -- that have a non-zero relative velocity -- and their perturbation from the merger, can lead to more than one {\narrow} component. 

In J084135 and J135251, besides the highly disturbed ionized gas (mapped by \oiii), we can see the remains of another galaxy in the continuum maps of (Fig.\,\ref{fig:hst}), probably the result of a major merger.

J155019 and J120041 have continuum blobs to the South and Southwest of the nucleus, respectively, possibly signaling companions (assuming the same redshift) -- that along with the disturbed {\oiii}  (including tidal features in J155019) -- suggests a minor merger. 
In the case of J085829, the only signature of interaction seems to be a second continuum peak, ($\sim$1.1\,kpc to the southeast of the nucleus), suggesting that the interaction 
is close to settling.

Fig.\,\ref{fig:hst} shows that the {\oiii} emission of J094521 extends beyond the body of its host galaxy. Using additional \textit{HST} observations of this QSO, \citet[][Fig.\,A10]{vil21} pointed out the presence of a huge tidal tail, confirming the existence of an interaction. 

J082313 has {\oiii} and continuum emission to the south-east detached from the main body of the galaxy. The continuum flux distribution could indicate an interacting companion, but the broadband filter of our HST observations \citep[see][]{sto18} is contaminated by the \oiii lines -- the same being true for J085829 -- and therefore we are not sure about the actual contribution from stars.
In addition, the J082313 GMOS-IFS continuum -- regardless of its weak S/N -- seems to be not totally correlated with the \textit{HST} one (black contours in the upper left panel of Fig.\,\ref{fig:mapJ082313}), weakening the merger hypothesis. 

Therefore, only J123006 and J082313 (and possibly J085829) do not have signs of interactions, and the three of them needed only one {\narrow} component for the fit, supporting that the second (an third) {narrow} components are associated with interactions.

We have found that each component can be quite complex and requires a detailed analysis. But it is not the purpose of the present work to dissect the results of each component -- we save that for a future work. As an example, the \narrow{1} component of J094521 (Fig\,\ref{fig:mapJ094521}) has a very low velocity dispersion, and is visible only outside the nucleus. 
Interestingly, high spatial resolution VLA and e-MERLIN radio maps of this QSO, presented by \citet{jar19}, seem to correlate with the flux distributions of our {\narrow{1}} component. The interpretation that these authors propose is that the radio jet is being deflected while it pushes a cloud of gas (although they didn't decompose the {\oiii} profile in multiple components). The connection between our {\narrow{1}} component and the local radio emission reinforces our results and the need of multiple Gaussian components, which can originate not only in outflows and mergers, as discussed above, but also in interactions with a radio jet.

\begin{table*}
	\centering
	\caption{
	\textbf{Outflow properties.} 
	Radial sizes ($R$) in units of kpc, mass outflow rates ($\dot{E}_{out}$) in units of $\mathrm{M_{\odot}\,yr^{-1}}$, outflow powers ($\dot{E}_{out}$) in units of $\mathrm{10^{42}\,erg\,s^{-1}}$ and kinetic coupling efficiency ($\varepsilon_f=\dot{E}_{out}/L_{bol}$) in \%. 
	(2)--(3) Radius of the {\broad} and {\narrowbroad} components, from the GMOS-IFS (see Fig.\,\ref{fig:radius_out});
	(4) {\oiii} radius, from the \textit{HST} images (see Section\,\ref{sec:radius});
	(5) bolometric luminosity (in units of $\mathrm{10^{45}\,erg\,s^{-1}}$); 
	(6)--(8) Peak mass outflow rate, power, and kinetic coupling efficiency, using the {\default} assumptions; 
	(9)--(11) Corresponding range of results from the tests with summary in Table\,\ref{tab:met}, and data in Tables\,\ref{tab:methods1} and \,\ref{tab:methods2}.
	}
	\label{tab:out_prop}
\begin{tabular}{ccccccccccccc}
\hline
& \multicolumn{2}{c}{GMOS-IFS} & & \textit{HST} & & \multicolumn{3}{c}{\default} & & \multicolumn{3}{c}{{tests range}} \\ \cline{2-3}\cline{5-5}\cline{7-9}\cline{11-13}\\[-1\medskipamount]
Name & $R_{\bb}$ & $R_{\nb}$ & & $R_{\oiii}$ & log(\lbol) & $\dot{M}_{out}^{\defa}$ & $\dot{E}_{out}^{\defa}$ & $\varepsilon_{f}^{\defa}$ & & $\dot{M}_{out}$ & $\dot{E}_{out}$ & $\varepsilon_{f}$ \\
(1) & (2) & (3) & & (4) & (5) & (6) & (7) & (8) & & (9) & (10) & (11) \\
\hline
J082313	&	13.3$\pm$1.6	&	18$\pm$1.6$^*$	&	&	26.3$\pm$1.4	&	46.7	&	3.3	&	0.35	&	0.0007	&	&	[0.22,\,8.6]	&	[0.023,\,0.91]	&	[0.00005,\,0.002]	\\
J084135	&	5.6$\pm$0.5	&	5.9$\pm$0.5$^*$	&	&	9.3$\pm$0.8	&	45.9	&	4.6	&	0.68	&	0.009	&	&	[0.27,\,24]	&	[0.049,\,3.7]	&	[0.0006,\,0.05]	\\
J085829	&	7.8$\pm$1.4$^\ddagger$	&	11.3$\pm$1.4	&	&	5.4$\pm$0.8	&	46.3	&	8.6	&	3.4	&	0.02	&	&	[1.7,\,16]	&	[0.65,\,13]	&	[0.003,\,0.07]	\\
J094521	&	<2.1$\pm$0.7$^\dagger$	&	6$\pm$0.7$^*$	&	&	12.7$\pm$0.8	&	45.8	&	28	&	28	&	0.4	&	&	[7.3,\,360]	&	[4.7,\,460]	&	[0.07,\,7]	\\
J123006	&	9.5$\pm$1.2	&	11.9$\pm$1.2	&	&	6.2$\pm$1.2	&	46.6	&	4	&	1.7	&	0.004	&	&	[1,\,7.6]	&	[0.25,\,3.3]	&	[0.0008,\,0.01]	\\
J135251	&	4.1$\pm$0.8	&	8.7$\pm$0.8$^*$	&	&	19.3$\pm$1.8	&	46.6	&	17	&	6.2	&	0.01	&	&	[0.53,\,71]	&	[0.27,\,52]	&	[0.002,\,0.3]	\\
J155019	&	5.3$\pm$0.6	&	7.3$\pm$0.6$^*$	&	&	6.5$\pm$1.7	&	45.7	&	2.7	&	0.26	&	0.005	&	&	[0.49,\,12]	&	[0.041,\,1.9]	&	[0.0001,\,0.005]	\\
J120041	&	5.3$\pm$0.6$^\ddagger$	&	6.2$\pm$0.6$^*$	&	&	5.4$\pm$0.5	&	45.8	&	12	&	3.6	&	0.05	&	&	[3.1,\,31]	&	[0.52,\,10]	&	[0.001,\,0.02]	\\\hline
\end{tabular}
\vspace{1ex}
 {\raggedright Measured from an $^\dagger$ unresolved, or $^\dagger$ barely resolved, {\broad} component (see Fig.\,\ref{fig:radius_out}) \par}
 {\raggedright $^*$ Radial size limited by the GMOS-IFU Field-of-View \par}
 \end{table*}
 
\subsubsection{The {\broad} component}\label{sec:broad}

As previously pointed out, the {\broad} component is assumed to be tracing the kinematic coupling between the AGN and the host galaxy, which we are generally referring to as `outflows'. Here, we analyse its characteristics.

The velocity dispersion maps of the {\broad} component ($\sigma_{\oiii\bb}$) show values reaching a maximum of $\sim$\,900\,km\,s$^{-1}$ for J094521, while in J084135 it reaches only 300\,km\,s$^{-1}$. The remaining galaxies achieve $\sigma_{\oiii\bb}$ peak values between 400\,km\,s$^{-1}$ and 700\,km\,s$^{-1}$. The different degrees of disturbance for each object, as revealed by these values, indicates that the feedback is not equally strong for different objects and/or may vary due to varying outflow axis orientation relative to the LoS in an ambient affected by extinction \citep{bae16}. 

Part of the sample has radial velocity maps ($v_{\oiii\bb}$) with a predominance of negative values (J085829, J094521 and J135251). This is expected in a scenario where the outflow axis makes a relatively large angle with the plane of the sky, since the radiation coming from the receding part of the outflow should be more affected by dust extinction \citep{bae16}. In particular, this component in J085829, J094521, and J120041 is barely -- or not (in the second case) -- spatially resolved (compare the spatial profile of $F_{\oiii\bb}$ with that of the PSF in Fig.\,\ref{fig:radius_out}). Along with the negative velocity values of J085829 and J094521, this is a sign that the outflow axes of these two QSOs are more aligned with our LoS (high inclination $i$ relative to the plane of the sky) than the other targuets.

On the other hand, J123006 has only positive velocity values, ranging between $\sim$\,10\,km\,s$^{-1}$ in the most perturbed region, and 100\,km\,s$^{-1}$ in the least perturbed one. Therefore, the region with characteristics of an outflow (higher $\sigma_{\oiii\bb}$) has velocities close to zero. But this may be due to uncertainties in the calculation of its systemic velocity.

Other QSOs -- J082313, J084135, J155019 and J120041 --  have a mix of positive and negative velocities, with |$v_{\oiii\bb}$| maxima between $\sim$\,80\,--\,200\,km\,s$^{-1}$. Radial velocity values centred on zero -- specially if the velocity field has a small velocity range -- can be the result of an outflow axis more aligned with the plane of the sky (small inclination $i$).
We note that our outflow definition includes not only the AGN winds, but also gas that is only moderately disturbed by it without changing significantly its original bulk motion.
The scenario is analogous to the `disturbed rotation' discussed by \citet{fis18}, where it was used a FWHM$_{\oiii\bb}$\,>\,250\,km\,s$^{-1}$ threshold to identify disturbed gas in interacting systems \citep[]{bel13,ram17}. In almost all spaxels of our sample, this threshold is achieved.

Another possible source of disturbance in the gas kinematics is the interaction with a radio jet. Although our sample comprises essentially radio-quiet objects, \citet{vil21} has recently presented VLA radio maps of J094521 and J084135, showing a correlation between the \oiii extended emission and the VLA radio maps, revealing that in these galaxies the radio emission probably plays a role in the gas kinematics.

Let's also consider the possibility that our hypothesis is not correct and the {\broad} component -- at least in a percentage of the pixels fitted -- could be tracing gas that is not in outflow (e.g. gas perturbed by mergers). We indeed observe regions inside the galaxies that have lower $\sigma_{\oiii\bb}$. For example, J155019 (see Fig.\,\ref{fig:mapJ155019}) has a more perturbed region with $\sigma_{\oiii\bb}$ close to 500\,km\,s$^{-1}$, while having values $\sim$\,100\,km\,s$^{-1}$ perpendicular to what seems to be its outflow axis. 
These less perturbed regions usually appear in spectra with lower S/N of $F_{\oiii\bb}$ (which makes the emission line decomposition harder). A consequence of the weaker intensity, is that associated flux introduces a small effect in the calculations of $\dot{M}_{out}$ and $\dot{E}_{out}$, although affecting the extent measurements.

\subsection{Outflow radius}\label{sec:disc-radius}

Here we analyse the extent of the gas that couples with the outflows, separating it from an usually more extended one, that may include gas with ordered motion due to orbits in the galaxy potential, along with other kinematic deviations that can be associated with galaxy interactions.

We define the outflow radius as the extent of the region showing the {\broad} component $R_{out}=R_{\bb}$. Table\,\ref{tab:out_prop} displays their values, that reach up to $\sim$\,2\,--\,13\,kpc, with an average of $6.6\pm3.5$\,kpc. The largest extent is observed in J082313, and the smallest in J094521 that, as we suggested above, is not spatially resolved and probably has its outflow more directed towards the observer. 

On the other hand, we associate the radial size measured from the \textit{HST} {\oiii} images ($R_{\oiii}$) with the Extended Narrow Line Region \citep[ENLR,][]{ung87}, since it includes the contribution from all ionized gas, comprising all kinematic components.
The same region is comprised by  the extent of the full profile ($R_{\nb}$) measured in the IFS data, given that it contains not only the outflowing component, but also the remaining {\narrow} ones.
The $R_{\oiii}$ values range between $\sim$\,5\,--\,26\,kpc, while $R_{\nb}$ varies between $\sim$\,6\,--\,18\,kpc. 
However, the IFU FoV limits the extent of $R_{\nb}$ in 6 QSOs of the sample (see Figs.\,\ref{fig:hst}). Therefore, in order to later compare the radial sizes of the outflow and the ENLR, we further define $R_{\rm{ELNR}}=\max(R_{\oiii},R_{\nb})$, which results in values with an average of $13\,{\pm}\,7$\,kpc.

Some previous studies on QSO\,2's extended gas emission report lower $R_{out}$ values than ours. One reason for this discrepancy may be the different definitions for the outflow radii. While we identify it as corresponding to the maximum extent over which we observe the broadest component with S/N ratio in the flux $F_{\oiii,\bb}\ge3$, \citet{fis18}, for example, considered the presence of an outflow only where there were components with peak velocities $|v_{\oiii}|>300$\,km\,$s^{-1}$, while \cite{kar16} used a threshold for the presence of an outflow corresponding to $\sigma_{\oiii}>\sigma_{*}$, where $\sigma_{*}$ is the stellar velocity dispersion. These authors found values for the extent of the outflow below $\sim$\,2\,kpc.

Another issue is the beam-smearing due to the PSF of the observations \citep{hus16}, that result in oversized extents when not corrected for it. To evaluate this effect, we first measured the FWHM/2 of the  $F_{\oiii,\bb}$ radial profiles (FWHM$_{out}$/2) in Fig.\,\ref{fig:radius_out}. We then corrected it for the beam smearing using the approximation $\rm{FWHM}_{out,0}=\sqrt{\rm{FWHM}_{out}^2 - \rm{FWHM_{PSF}}^2}$ \citep[e.g.][]{tad18}, that is valid for PSF and $F_{\oiii,\bb}$ distributions with 2D-Gaussian spacial profiles (which is not exactly true in both cases, but is an approximation).
Finally, we corrected $R_{out}$ adopting for it the same relation  between the corrected and measured FWHM: $R_{out,0} = R_{out} \left( \frac{\rm{FWHM}_{out,0}}{\rm{FWHM}_{out}} \right)$. We did not attempt to propagate errors, since this method is highly uncertain. The new corrected values for the outflow radii $R_{out,0}$ range between $\sim$\,1\,--\,8\,kpc, with an average of $4\pm2$\,kpc, with the value of J094521 presented as an upper limit, as it is not resolved. We applied the same percentage correction to $R_{\nb}$, from which we have the new $R_{\rm{ELNR},0}=\max(R_{\oiii},R_{\nb,0})$. The corrected radial sizes are displayed in Table\,\ref{tab:psf_radius}, along with the remaining corrected extents cited in this Section.

We can also compare our values with those obtained for similar objects by \citet{vil16}, who adopted as the radius of the outflow half the FWHM of the flux distribution of the outflowing gas, obtaining values lower than 1\,--\,2\,kpc after correcting for beam-smearing. In comparison, our FWHM$_{out,0}$/2 measurements returned values between $\sim$\,0.5\,--2.5\,\,kpc, 4\,$\pm$\,2 times lower than $R_{out,0}$, on average. The FWHM$_{out,0}$/2 extent can be viewed as a measure of the radial size of the bulk content of the outflow, while our $R_{out,0}$ measurement includes also the contribution of lower luminosity gas farther out. 
$R_{out,0}$ is thus expected to be larger than FWHM$_{out,0}$/2. Although we have tried to correct its value by the beam-smearing effect of the PSF, the method we have applied does not effectively remove its contribution to the wings of the spatial profile, which could result in an overestimation of $R_{out,0}$ in the less resolved cases (e.g. J085829; see Fig.\,\ref{fig:radius_out}).

\begin{table}
\centering
	\caption{
	\textbf{PSF broadening correction.}
	Sizes in units of kpc, where the subscript `0' identifies the quantities for which the PFS correction was applied.
	(2) FWHM of the radial profile of the outflowing component;
	(3) Outflow extent;
	(4) ENLR extent;
	(5) Fraction of the radial extent affected by the outflow.
	}
    \label{tab:psf_radius}
\begin{tabular}{lcccr}
\hline
Name & FWHM$_{out,0}$ & $R_{out,0}$ & $R_{ENLR,0}$  & $R_{out,0}/R_{ENLR,0}$ \\
(1)  & (2) & (3) & (4) & (5)\\
\hline
J082313 & 2.3  & 7 & 26 & 0.3 \\
J084135 & 2.3  & 5 & 9  & 0.6 \\
J085829 & 2.6  & 5 & 7  & 0.7 \\
J094521 & 1.2 & <\,1 & 13 & 0.1 \\
J123006 & 3.6  & 8 & 9  & 0.9 \\
J135251 & 3.2  & 4 & 19 & 0.2 \\
J155019 & 4.8  & 5 & 7  & 0.7 \\
J120041 & 0.94 & 3 & 5  & 0.6 \\
\hline
\end{tabular}
\end{table}

Comparing the corrected outflow and ENLR extents, we obtained  $R_{out,0}/R_{\rm{ENLR},0}$ values ranging between $\sim$\,0.1 for J094521, and 0.9 for J12006, with an average of $\sim0.5\pm0.3$. In comparison, \citet{fis18} found a $\sim$\,0.2 value for this ratio, with J120041, in particular, having $R_{out}=1.07$\,kpc, 3 times lower than our value. The smaller values of $R_{out,0}$ relative to $R_{\rm{ENLR},0}$ indicate that not all ionised gas is disturbed by the outflows. Therefore, it is essential to isolate the {\broad} component, instead of simply using the extent of the total ionized gas as the outflow radius for obtaining the outflow properties.

\subsection{Radial profiles of $\dot{M}_{out}$ and $\dot{E}_{out}$}
To explore the outflow strength, in Fig.\,\ref{fig:rates_radial} we plot the $\dot{M}_{out}(r)$ (left) and $\dot{E}_{out}(r)$ (right) radial profiles -- with their maxima highlighted (stars) -- calculated using the {\default} assumptions.

The values of $\dot{M}_{out}$ within 1\,kpc (radial size not corrected for  beam-smearing) from the nucleus range from $\sim\,$0.1 to 10\,$\mathrm{M_{\odot}\,yr^{-1}}$, while the maximum values reach 3 to 30\,$\mathrm{M_{\odot}\,yr^{-1}}$ at radii of 1 to 5\,kpc.
The outflow powers range from $10^{40}$ to $10^{43}\,\mathrm{erg\,s^{-1}}$ within the inner kiloparsec, with maxima between $\sim10^{41}$ and $10^{43}\,\mathrm{erg\,s^{-1}}$.

\begin{figure}
    \centering
	\includegraphics[width=0.8\columnwidth]{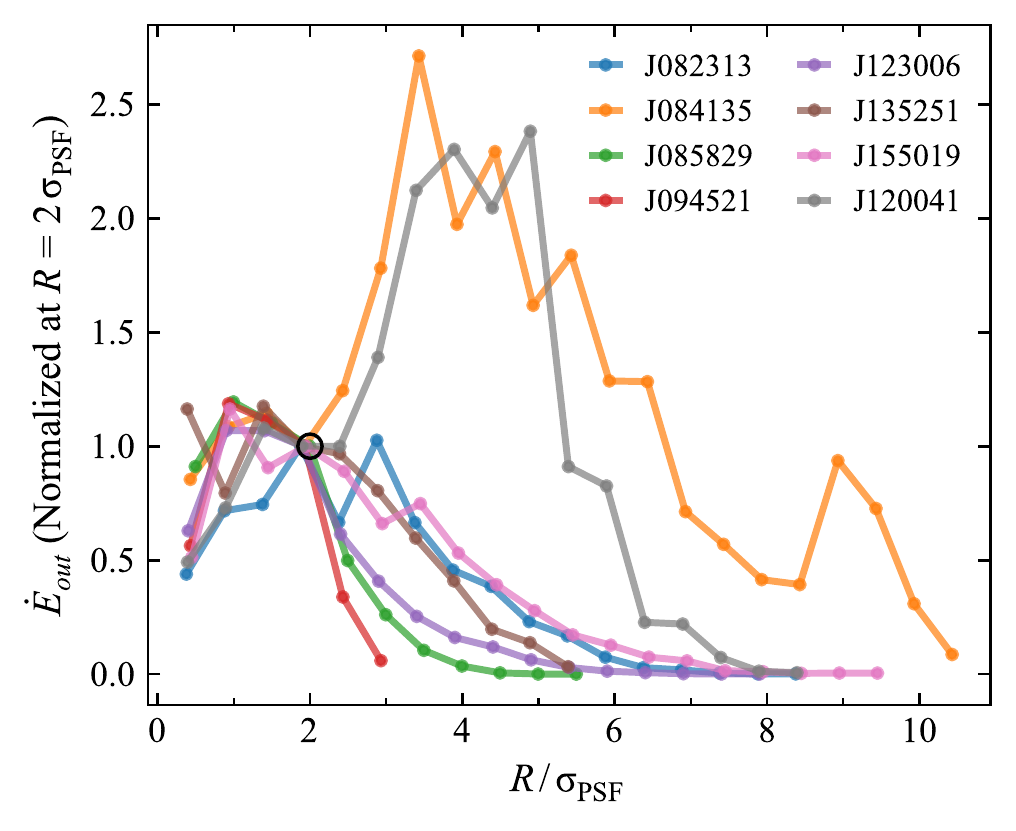}
    \caption{Same as the right panel of Fig.\,\ref{fig:rates_radial}, but for the outflow power $\dot{E}_{out}$ plotted as a function of the radius normalized by the PSF standard deviation: $R/\sigma_{\rm{PSF}}$. The power is plotted in linear scale, and is normalized at $R=2\,\sigma_{\rm{PSF}}$ (black circle).}
    \label{fig:power_psf}
\end{figure}

Since the PSF may affect the radial profiles of $\dot{M}_{out}(r)$ and $\dot{E}_{out}(r)$, we also plotted $\dot{E}_{out}(r)$ as function of the radius normalized by the $\rm{\sigma_{PSF}}$ in  Fig.\,\ref{fig:power_psf}. For a better comparison between the profiles, we also normalized the $\dot{E}_{out}(r)$ values at $R=2\sigma_{\rm{PSF}}$.
A comparison between these radial profiles shows that J084135 and J12004 seem to form a different group, with  $\dot{M}_{out}(r)$ and $\dot{E}_{out}(r)$ peaking outside the nuclear region.
Coincidentally J084135 is a clear major merger and J12004 has signs of a minor merger, and the velocities may be partly affected by the interactions.
A similar trend -- peak in the outflow away from the nucleus -- has been observed by other authors in local Seyferts, although for much smaller scales  \citep[e.g.][]{cre15,rev18b,shi19}.

The other QSOs show a more continous decay in  the $\dot{E}_{out}(r)$ radial profiles, with J085829 and J094521 being the most compact, which could be caused by their unresolved {\broad} component (Fig.\,\ref{fig:radius_out}), probably due to the fact that their outflow is more aligned with our LoS, as suggested by the dominance of negative values in the $v_{\oiii,\bb}$ maps (Figs.\,\ref{fig:mapJ085829} and \ref{fig:mapJ094521}). In this scenario, the outflow axis of the two sources with $\dot{E}_{out}(r)$ peaking at the largest distances, J084135, J120041  and J082313, would be more perpendicular to the LoS. This hypothesis is reinforced by the mix of negative and positive values of $v_{\oiii,\bb}$ (Figs.\,\ref{fig:mapJ084135} and \ref{fig:mapJ120041}) over the FoV, and the extended {\oiii} emission. The remaining QSOs show $\dot{E}_{out}(r)$ radial profiles somewhat more extended than the two most compact ones.

\subsection{Mass outflow rates and powers}\label{sec:disc-outflow}
\begin{figure*}
	\includegraphics[width=0.95\linewidth]{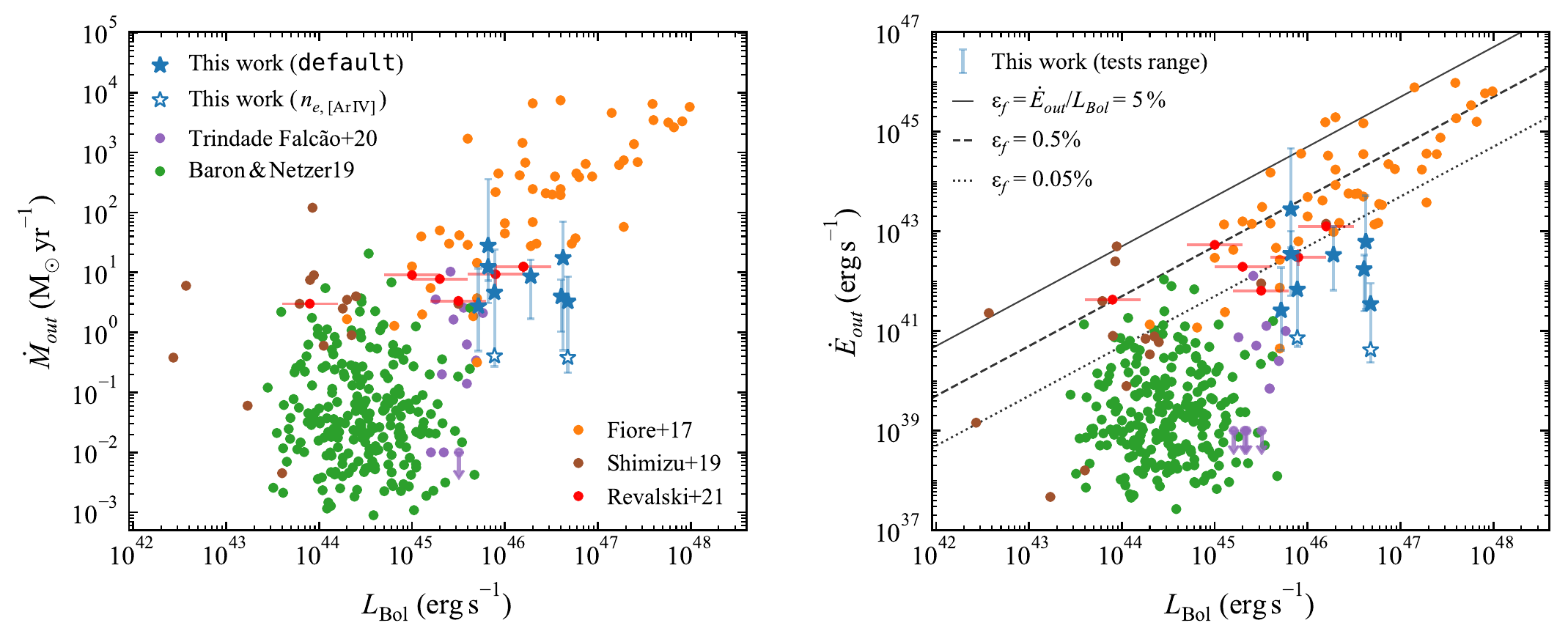}
    \caption{Mass outflow rate ($\dot{M}_{out}$, left) and outflow power ($\dot{E}_{out}$, right) as a function of the AGN bolometric luminosity \citep[based on the similar figure of][]{shi19}. Our data, for the {\default} assumptions, are presented as filled blue stars and the blue bars correspond to variations due to the tests listed in Fig.\,\ref{fig:comp_method}. For J082313 and J084135, we add blue stars corresponding to powers calculated using the gas density obtained from the {\ariv} line ratio. We added the following data from the literature: orange circles, for the ionized outflows from \citet{fio17}; brown circles, for the data collected by \citet{shi19}; purple circles, for \citet{fal20}; red errorbars, for \citet{rev21}; 
    and green circles, for \citet{bar19}.
The dotted, dashed and solid black lines correspond to kinetic coupling efficiencies $\varepsilon_f$ of 0.05\%, 0.5\% and 5\%, respectively.
    }
    \label{fig:literature}
\end{figure*}

In order to compare our data with previous results from the literature, in Fig.\,\ref{fig:literature}, we plot the maximum values of the $\dot{M}_{out}(r)$ and $\dot{E}_{out}(r)$  profiles as a function of \lbol, inspired by Figure\,19 of \citet{shi19}, and \citet{fio17}. Note that when referring to the peak values, we drop the `$(r)$' $\rightarrow$ for $\dot{M}_{out}$ and $\dot{E}_{out}$. We list these values in Table\,\ref{tab:out_prop}, together with the kinetic coupling efficiencies $\varepsilon_f=\dot{E}_{out}/L_{bol}$. We also list in this table the range of values obtained for these properties that result from tests we describe in the following Sections.

The blue stars in Fig.\,\ref{fig:literature} correspond to the radial maxima obtained from the {\default} assumptions: $\dot{M}_{out}^{\defa}$ and $\dot{E}_{out}^{\defa}$ (also shown as stars in the radial profile in Fig.\ref{fig:rates_radial}). On average, the values are $\overline{\dot{M}_{out}^{\defa}}=10\pm8.8\,\mathrm{M_{\odot}\,yr^{-1}}$ and $\overline{\dot{E}_{out}^{\defa}}=(5.5\pm9.3){\times}10^{42}\,\mathrm{erg\,s^{-1}}$, with the corresponding kinetic coupling efficiencies $\varepsilon_f^{\defa}$ ranging between 0.0007\,--\,0.4\,\%. The blue bars refer to the range in the results obtained from assumptions distinct from the {\default} ones (see Fig.\,\ref{fig:comp_method}), where we highlight with blue open stars, the values obtained when using the electron density traced by the {\ariv} emission-line ratio, what could be done only for two QSOs (see discussion in Section\,\ref{sec:density}). With the exception of J094521, for which $\dot{M}_{out}^{\defa}$ and $\dot{E}_{out}^{\defa}$ fall on top of the $\varepsilon_f=0.5\%$ threshold, the others QSOs consistently fall below the \citet{fio17} average powers for the same luminosities.

We note that the {\narrow{3}} component of J094521 shows only negative $v_{\oiii,\nn{3}}$ values, in the circumnuclear region, as well as high ${\sigma}_{\oiii,\nn{3}}$ > 300\,km\,s$^{-1}$ in the nucleus and its Northwest (see Fig.\,\ref{fig:mapJ094521}), in the direction of the {\oiii} blob seen in Fig.\,\ref{fig:hst}. 
J135251 also shows a similar disturbed region in the {\narrow{2}}  (Fig.\,\ref{fig:mapJ135251}). These components may also be associated with outflows. 
Adding their contribution, the $\dot{M}_{out}^{\defa}$ and $\dot{E}_{out}^{\defa}$ values would increase by factors of $\sim$\,2\,--\,3. However, in order to perform an homogeneous analysis for all objects, we keep using only the {\broad} component in the {\default} method.

We note that the assumptions made to obtain $\dot{M}_{out}$ and $\dot{E}_{out}$ in the different works included in Fig.\,\ref{fig:literature} are not the same. 
\citet{fio17} used homogeneous assumptions to redo the calculations from retrieved data of previous publications, using a constant low gas density of $n_e=200$\,cm$^{-3}$, and an outflow velocity similar to ours $v_{95}$. 
\cite{bar19} calculated the electron density from the ionization parameter ($n_{e,\rm{BN19}}$, see Section\,\ref{sec:density}), with the $R_{out}$ corrsponding to the dust sublimation radius, whose temperature came from spectral energy distribution (SEDs) fitting. 
\cite{fal20} used photoionization models to calculate the ionized gas mass and density,  and deprojected their LoS velocity used in $v_{out}$, and  -- similarly to our work -- used the peak values of the $\dot{M}_{out}(r)$ and $\dot{E}_{out}(r)$ radial profiles. These two latter works also obtained weaker outflow properties, compared to \citet{fio17}. 
On the other hand, \citet{rev21} -- also using photoionization models --  found higher values, showing a similar trend compared to that of \citet{fio17}.
We also added in Fig.\,\ref{fig:literature} the low-luminosity AGNs data, compiled from the literature by \cite{shi19}, corresponding to a variety of methods of calculation and resulting in large scattering of $\dot{M}_{out}$ and $\dot{E}_{out}$ values.

\subsection{Influence of different calculation methods and assumptions on $\dot{M}_{out}$ and $\dot{E}_{out}$}\label{sec:disc-methods}

\begin{table}
	\centering
	\caption{
	\textbf{Parameters of the tests}.
	Each line shows the parameters used in each test, with the {\default} ones listed in the first line.
	To be used together with Fig.\,\ref{fig:comp_method}; it shows the parameters used to obtain the $\dot{M}_{out}$ and $\dot{E}_{out}$ values listed in Tables\,\ref{tab:out_prop}, \ref{tab:methods1} and \ref{tab:methods2}.
	}
	\label{tab:met}
	
\begin{tabular}{lccccr}
\hline
 Test                & & Mass                    & Density                 & $v_{out}$              & $\delta r$
 \\[-1.3\medskipamount] \\ \cline{1-2}\cline{3-6}  \\[-1.3\medskipamount] 
 {\default}       & & $ M_{\hii_{\,\,\bb}} $      & $n_{e,\sii_{\nb}}$       & $v_{95}$               & $ \rm{\sigma_{PSF}}/2 $
 \\[-1.3\medskipamount] \\  \cline{1-1}\cline{3-6}  \\[-1.3\medskipamount] 
{\tt integ}.    & & $\sum M_{\hii_{\,\,\bb}}  $ & $\bar{n}_{e,\sii_{\nb}}$ & $\bar{v}_{95}$         & $ R_{\bb} $                          \\
{\nb}            & & $ M_{\hii_{\,\,\nb}} $      & -                       & -                      & -                                    \\
$v_{68}$         & & -                       & -                       & $v_{68}$               & -                                        \\
{\proj}             & & -                       & -                       & $v_{95,0}(i)$ & $ \rm{\sigma_{PSF}}/2{/} \rm{cos}(i) $ \\
$n_{e,\rm{BN19}}$ & & -                       & $n_{e,\rm{BN19}}$        & -                      & -                                         \\
$n_{e,\ariv}$    & & -                       & $n_{e,\ariv}$           & -                      & -                                         \\
\hline
\end{tabular}
\end{table}

As already mentioned, the {\default} method does not necessarily represent the best estimate for the feedback properties. It provides, however, an homogeneous method to be applied to all QSOs: ionized gas mass $M_{\hii,\,\bb}$, calculated using only the outflowing component ({\broad}); spatially resolved electron density $n_{e,{\sii_{\nb}}}$, calculated from the {\sii} line ratio ({\narrowbroad}); and $v_{95}$ as the outflow velocity. The radial profiles correspond to the mass outflow rates and powers crossing annuli with $\delta r = \rm{\sigma_{PSF}}{/}2$ width, and therefore, does not assume that these rates are radially constant. From the peak of these curves, we obtained $\dot{M}_{out}^{\defa}$ and $\dot{E}_{out}^{\defa}$.

\begin{figure}
	\includegraphics[width=0.95\linewidth]{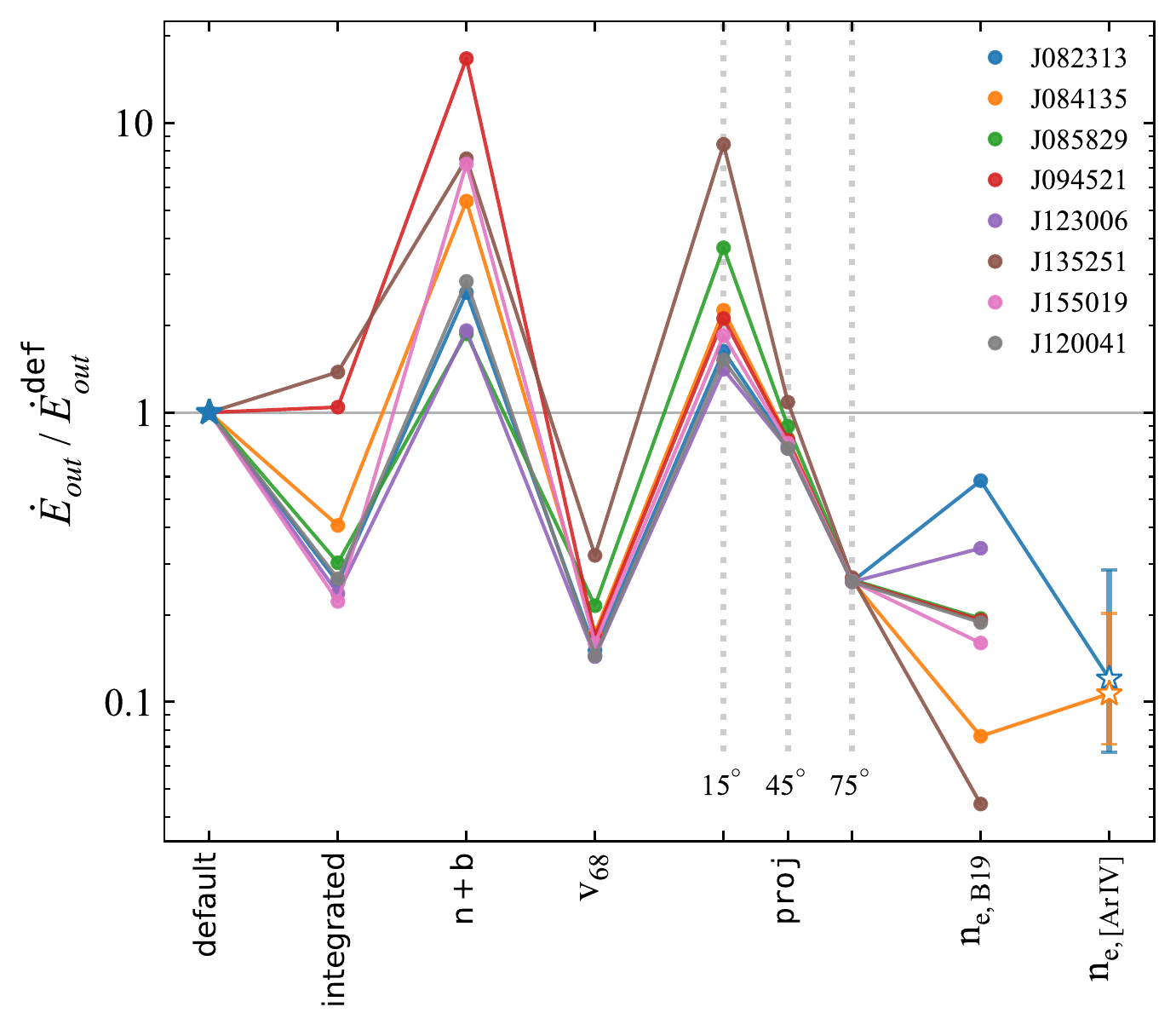}
    \caption{The effect of each test on the calculation of the outflow power ($\dot{E}_{out}$) in comparison to the {\default} assumptions (See Section\,\ref{sec:disc-methods} for details, and Table\,\ref{tab:met} for a summary).
    {\integrated}: method that assumes radially constant powers;
    {\nb}: uses the full extent of the {\ha} emitting region (\narrowbroad) to calculate M$_{\hii}$ (instead of only that of the {\broad} component);
    $v_{68}$: outflow velocity as $v_{68}$ instead of $v_{95}$;
    {\proj}: $v_{out}$ and $\delta r$ corrected for projection effects, for inclinations $i\,{=}$\,15,\,45 and 30$\,^{\circ}$;
    $n_{e,\rm{BN19}}$: density from \citet[BN19]{bar19} method (instead of {$n_{e,\sii}$});
    $n_{e,\ariv}$: density from the {\ariv} emission line ratio.
	}
    \label{fig:comp_method}
\end{figure}

Since different authors use different assumptions, and there is not a consensus about the best practice, in the following subsections we test how $\dot{M}_{out}$ and $\dot{E}_{out}$ are affected by different assumptions on: the ionized gas mass $M_{\hii}$, the outflow velocity $v_{out}$, and the electron density $n_e$. In Table\,\ref{tab:met} we provide a quick reference to the parameters of the {\default} method, and what was changed in each test. Fig\,\ref{fig:comp_method} visually shows how much the tests change the values of the {\default} calculation of $\dot{E}_{out}$ (the values are displayed in Tables\,\ref{tab:methods1} and \ref{tab:methods2}).  
A version of the Fig.\,\ref{fig:comp_method} for $\dot{M}_{out}$ would be equal, except for the  smaller relative variation due to the `$v_{68}$' and `\proj' tests (discussed in Section\,\ref{sec:disc-met-vout}), since $\dot{E}_{out} / \dot{M}_{out} \propto v_{out}^2$.

\subsubsection{The {\integrated} method (constant feedback rates)}\label{sec:disc-met-inte} 

We first tested the effect of using integrated and constant values within the outflow radius, which we have called the {\integrated} method (Table\,\ref{tab:methods1}). Here, we still applied Eqs.\,\ref{eq:mii}\,--\,\ref{eq:mdot} but for $\delta r = R_{\bb}$ annular width, with the electron density and the outflow velocity becoming the average value within the whole region covered by outflow radius ($\bar{n}_{e,\sii,\,\nb}$ and $\bar{v}_{95}$), while the ionized mass is summed over the same region ($\sum M_{\hii,\,\bb}$). We still use only the flux from the {\broad} component: $F_{\haa,\,\bb}$. Table\,\ref{tab:methods1} displays the values of these parameters, along with the resulting $\dot{M}_{out}^{\inte}$ and $\dot{E}_{out}^{\inte}$.

Fig\,\ref{fig:comp_method} doesn't show a clear trend, with the majority of the QSOs lowering $\dot{E}_{out}^{\inte}$ in comparison with its {\default} value, and only two objects showing a increase, although small, resulting in a $\dot{E}_{out}^{\inte}/\dot{E}_{out}^{\defa}$ range of [1/4.5,\,1.4]. Therefore, $\dot{M}_{out}$ and $\dot{E}_{out}$ values calculated with the {\integrated} method are different from the peak of the corresponding radial profiles by a factor of $\sim$\,3 on average, but are generally smaller.

\begin{table}
	\centering
	\caption{
	\textbf{Tests (part 1)}.  Results of the tests with the {\integrated} method (see also Table\,\ref{tab:met} and Section\,\ref{sec:disc-met-inte}).
	(2) Mean of electron density obtained with {\sii} (from the {\nb} component, in cm$^{-3}$);
    (3) Total ionized mass (from the {\bb} component, in M$_\odot$);
    (4) Mean of the $v_{95}$ outflow velocity (in km\,s$^{-1}$);
    (5) Mass outflow rate (in $\mathrm{M_{\odot}\,yr^{-1}}$);
	(6) Outflow power (in $\mathrm{10^{42}\,erg\,s^{-1}}$).
	The values of the parameters (2)--(4) were calculated pixel-by-pixel, inside the outflow radius $R_\bb$.	}
	\label{tab:methods1}
\begin{tabular}{lccccr}
\hline
        & \multicolumn{5}{c}{{\integrated} ($\delta r{=}R_b$)}                                                   
 \\[-1.5\medskipamount] \\  \cline{2-6}  \\[-1.3\medskipamount] 
Name    & $\bar{n}_{e,\sii{\,\nb}}$ & $\sum{M_{\hii,\mathrm{b}}}$ & $\bar{v}_{95}$ & $\dot{M}_{out}^{\inte}$ & $\dot{E}_{out}^{\inte}$
 \\[-1.3\medskipamount] \\
(1)     & (2)                & (3)                         & (4)            & (5)                        & (6)                        \\
\hline
J082313 & 290                & 2.3                         & 540            & 0.97                       & 0.091                      \\
J084135 & 120                & 1.4                         & 710            & 1.7                        & 0.27                       \\
J085829 & 340$^{\dagger}$    & 2.5                         & 1000           & 3.2                        & 1.0                        \\
J094521 & 720$^{\dagger}$    & 2.2                         & 2000           & 22                         & 29                         \\
J123006 & 370                & 1.2                         & 980            & 0.98                       & 0.41                       \\
J135251 & 260                & 1.6                         & 1900           & 3.1                        & 8.1                        \\
J155019 & 280                & 1.2                         & 430            & 0.98                       & 0.058                      \\
J120041 & 420                & 1.6                         & 980            & 3.1                        & 0.95                       \\ \hline
\end{tabular}
\\
 {\raggedright $\dagger$ Measured from the corresponding the SDSS spectrum\par}
\end{table}

\subsubsection{Ionized gas mass (\,{\broad} or {\narrowbroad})}\label{sec:disc-met-mass}
The total flux from {\ha} -- here characterized by the {\narrowbroad} components -- includes not only the outflowing gas, but also gas in the galaxy and its vicinity (e.g. gas with ordered motion, spread by merges, etc). Hence, in order to calculate the mass outflow rates and powers it is important to include only the flux coming from the outflows -- here traced by the {\broad} component. This way we avoid overestimating the ionized gas mass, and subsequently, the mass outflow rate and its power, since both depend linearly on $M_\hii$. In order to quantify the effect of including also the mass of the narrow components in the calculations, we recalculate the feedback properties, using the total ionized gas mass (from $F_{\haa,\nb}$) instead of only the outflowing one (from $F_{\haa,\bb}$). These values are displayed in Table\,\ref{tab:methods2}.

Fig\,\ref{fig:comp_method} shows that the results lead to an increase in $\dot{E}_{out}^{\nb}/\dot{E}_{out}^{\defa}$, with the values ranging between [1.9,\,17]. Therefore, without isolating the contribution from the outflows in the total flux observed, the powers can be overestimated up to one order of magnitude. 

Some authors, instead of decomposing the emission-line profiles into multiple Gaussians, estimate the outflowing gas mass from the percentage of the emission-line flux that has velocities above certain threshold. For example, Ruschel-Dutra et al. (submitted); \citet{sun17,kak20} used a threshold of $w_{80}>600\,\rm{km\,s^{-1}}$ as the minimum value to consider the gas to be in outflow, where $w_{80}$ is the width of an emission line profile above which 80\% of the total flux is emitted. This condition helps to avoid false outflow identifications, including only the the highest gas velocities that cannot be in mere orbital motion. On the other hand, a spectrum with multiple narrow components -- that could be related with mergers -- will also result in a larger $w_{80}$. For example, our J084135 spectra -- without considering the {\broad} component -- have three {\narrow} components (see Fig.\,\ref{fig:spec-comp}), that would increase the value of $w_{80}$ if this method would be applied here. Other issue is that this condition does not include the contribution of less powerful winds.

The discussion above highlights the importance of accessing how much does the outflowing component contributes to the measured flux. This can be best done with IFS \citep[e.g.][]{gal19}, or imaging\,+\,long-slit observations \cite[e.g.][]{fal20}. Without resolved spectroscopy, the calculation can be improved by using profile decomposition -- as done here -- to measure the contribution of the outflows in a single spectrum  \cite[e.g.][]{ros18,bar19}.

\subsubsection{Outflow velocity}\label{sec:disc-met-vout}

\begin{figure}
    \centering
	\includegraphics[width=0.95\columnwidth]{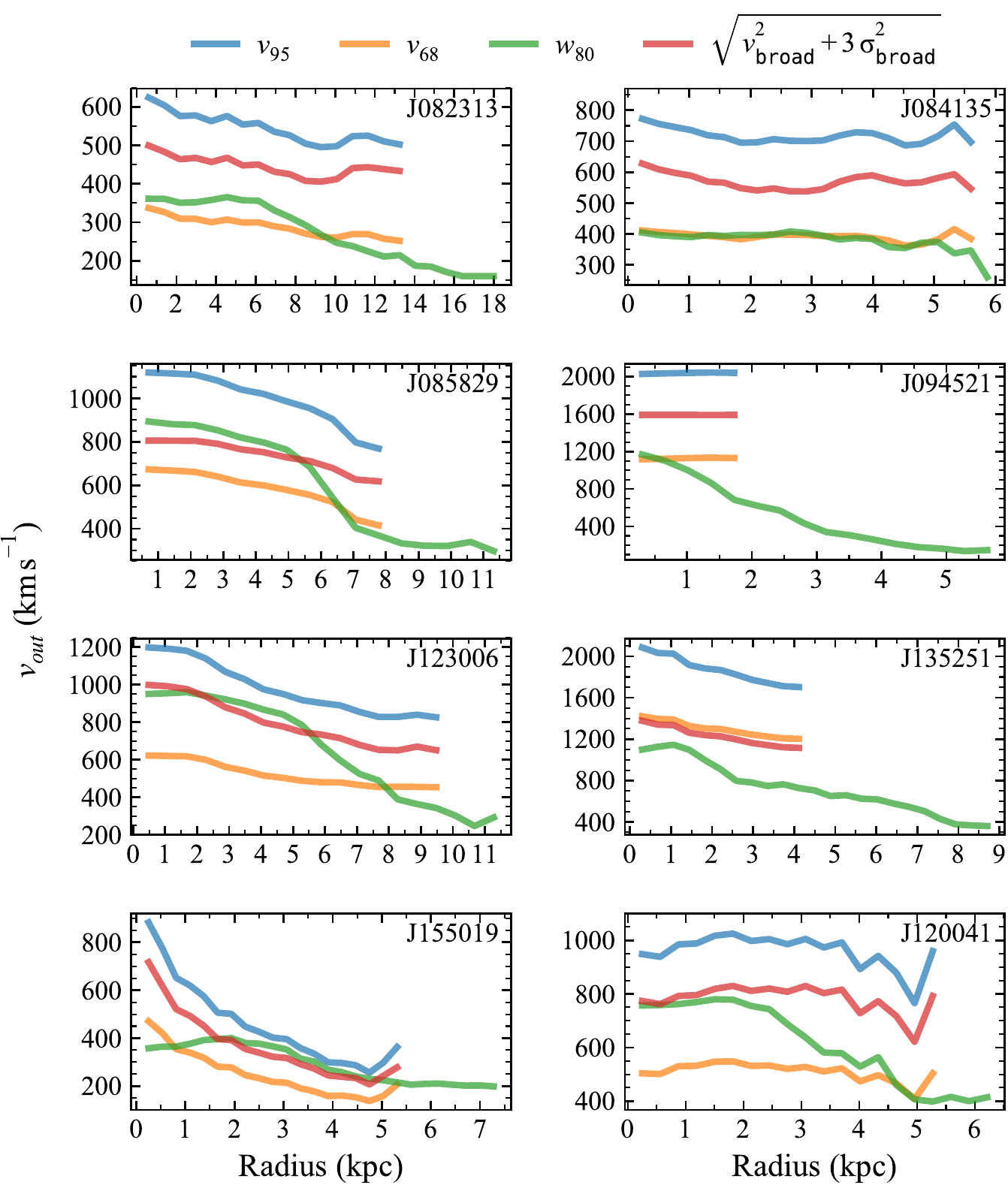}
    \caption{Radial profiles of different definitions for the outflow velocity $v_{out}$ for each QSO of the sample: $v_{95}$ in blue, $v_{68}$ in orange, $w_{80}$ in green and $\sqrt{v_{\broad}^2+3\,\sigma_{\broad}^2}$ in red (Section\,\ref{sec:disc-met-vout}).
    }
    \label{fig:vout}
\end{figure}

\begin{table*}
	\centering
	\caption{
	\textbf{Tests (part 2)}: Results of the tests about effect of the mass,  the outflow velocity and the electron density in the calculations of the feedback properties (see also Table\,\ref{tab:met} and Sections\,\ref{sec:disc-met-mass}--\ref{sec:disc-met-ne}). Mass outflow rates ($\dot{M}_{out}$) in units of $\mathrm{M_{\odot}\,yr^{-1}}$, and outflow powers ($\dot{E}_{out}$) in units of $\mathrm{10^{42}\,erg\,s^{-1}}$.
	(1) Galaxy short name;
	(2)--(3) Results of using the full profile (\nb) of {\ha}  to calculate $M_{\hii}$, instead of only the {\bb} component;
	(4)--(5) $v_{68}$ as outflow velocity, instead of $v_{95}$; 
	(6)--(7) outflow velocity corrected for inclinations $i\,{=}\,45\pm30\,^{\circ}$; 
	(8) average of the density calculated with the \citet[][BN19]{bar19} method (not used to calculate $\dot{E}_{out}^{\rm{BN19}}$ and $\dot{E}_{out}^{\rm{BN19}}$);
	(9)--(10) Results of using $n_{e,\rm{BN19}}$ with radial dependence (see Fig.\,\ref{fig:ne}) to obtain  $M_{\hii}$, $\dot{E}_{out}$ and $\dot{E}_{out}$; 
	(11) density calculated from the \ariv lines ratios;
	(12)--(13) Results of using $n_{e,\ariv}$.
	}
	
	\label{tab:methods2}
\begin{tabular}{lcccccccccccccr}
\hline
\multicolumn{1}{c}{}     & \multicolumn{2}{c}{Ionized Mass}                      &  & \multicolumn{4}{c}{Outflow velocity}                                                            &  & \multicolumn{6}{c}{Electron density}                                                                                                  
 \\[-1.5\medskipamount] \\  \cline{2-3} \cline{5-8} \cline{10-15}  \\[-1.3\medskipamount] 
\multicolumn{1}{c}{Name} & $\dot{M}_{out}^{\nb}$ & $\dot{E}_{out}^{\nb}$ &  & $\dot{M}_{out}^{v68}$ & $\dot{E}_{out}^{v68}$ & $\dot{M}_{out}^{\proj}$ & $\dot{E}_{out}^{\proj}$ &  & $\bar{n}_{e,\rm{BN19}}$ & $\dot{M}_{out}^{\rm{BN19}}$ & $\dot{E}_{out}^{\rm{BN19}}$ & $n_{e,\ariv}$       & $\dot{M}_{out}^{\ariv}$ & $\dot{E}_{out}^{\ariv}$
 \\[-1.3\medskipamount] \\
(1) & (2) & (3) & & (4) & (5) & (6) & (7) & & (8) & (9) & (10) & (11) & (12) & (13) \\
\hline
J082313 & 8.6 & 0.91 & & 1.8 & 0.053 & $2.4_{1.5}^{1.4}$ & $0.27_{0.18}^{0.3}$ & & 370 & 2.1 & 0.2 & $1450_{840}^{1150}$ & $0.38_{0.17}^{0.53}$ & $0.042_{0.019}^{0.058}$ \\[-0.8\medskipamount]\\
J084135 & 24 & 3.7 & & 2.6 & 0.12 & $3.4_{2.2}^{2.4}$ & $0.56_{0.38}^{0.99}$ & & 1300 & 0.65 & 0.052 & $870_{410}^{430}$ & $0.41_{0.13}^{0.36}$ & $0.073_{0.024}^{0.065}$ \\[-0.8\medskipamount]\\
J085829 & 16 & 6.3 & & 5.2 & 0.73 & $6.6_{4.3}^{6.4}$ & $3_{2.1}^{9.5}$ & & 1100 & 1.7 & 0.66 & - & - & - \\[-0.8\medskipamount]\\
J094521 & 360 & 460 & & 16 & 4.7 & $21_{13}^{15}$ & $22_{15}^{36}$ & & 3800 & 15 & 5.3 & - & - & - \\[-0.8\medskipamount]\\
J123006 & 7.6 & 3.3 & & 2.1 & 0.25 & $2.9_{1.8}^{1.5}$ & $1.3_{0.86}^{1.2}$ & & 740 & 0.49 & 0.041 & - & - & - \\[-0.8\medskipamount]\\
J135251 & 71 & 46 & & 12 & 2 & $14_{9.7}^{22}$ & $6.7_{5}^{45}$ & & 10000 & 3.2 & 0.67 & - & - & - \\[-0.8\medskipamount]\\
J155019 & 12 & 1.9 & & 1.5 & 0.041 & $2_{1.3}^{1.4}$ & $0.2_{0.14}^{0.28}$ & & 1700 & 0.49 & 0.041 & - & - & - \\[-0.8\medskipamount]\\
J120041 & 31 & 10 & & 6.5 & 0.51 & $8.9_{5.7}^{4.7}$ & $2.7_{1.7}^{2.7}$ & & 1100 & 3.2 & 0.67 & - & - & - \\
\hline
\end{tabular}
\end{table*}

Another issue concerning feedback properties of outflows is what velocity should be used to characterize the outflow. In this work, we used $v_{95}=|v_{\broad}|+2\,\sigma_{\broad}$, where $\sigma_{\broad}$ and $v_{\broad}$ are the velocity dispersion of the {\broad} component, and its velocity shift from the systemic velocity, respectively, measured from \oiii. This parametrization weights the contribution of the highest velocities in the outflow, using a value close to its maximum \citep[e.g.][]{rup13,fio17}. Therefore, since $\dot{M}_{out} {\propto}\,v_{out}$ and $\dot{E}_{out}\,{\propto}\,v_{out}^3$, using $v_{95}$ as the outflow velocity results in higher values of mass outflow rates and powers when compared with other $v_{out}$ definitions. To check this, we tested the effect of using another parametrization $v_{68}=|v_{\broad}|+\,\sigma_{\broad}$. This definition returns smaller values, as shown in Fig.\,\ref{fig:vout}, where we compare the radial velocity profiles of $v_{95}$ and $v_{68}$, along with other two $v_{out}$ parametrizations, that are discussed later in this Section.

As expected, Fig\,\ref{fig:comp_method} shows that this test decreases the the values of the powers, with $\dot{E}_{out}^{v68}/\dot{E}_{out}^{\defa}$ ranging between [1/7.1,\,1/3.1], an average of $\sim$\,5 times lower than the {\default} assumptions, while $\dot{M}_{out}^{v68}$ decreased by a factor of $\sim$\,2.  

Several other parametrizations for $v_{out}$ have been used in the literature, such as $\sqrt{v_{\broad}^2 + \sigma_{\broad}^2}$  \citep[e.g.][]{kar16,bar19} and $|v_{\broad}|+\rm{FWHM/2}$ \citep[e.g.][]{bis19b,gal19}.
Another approach is to use non-parametric definitions, that have the advantage of being non-dependent of a rigorous emission line fit, which is a source of error in our measurements. One of the most used  is $w_{80}$ \citep[e.g.][]{liu13b,sun17}. Comparing its radial profile with those of $v_{95}$ and $v_{68}$ in Fig.\,\ref{fig:vout}, the first noticeable feature is that the $w_{80}$ profiles are the most extended. This happens because $w_{80}$ -- differently from the other definitions -- is not limited by the S/N of the {\broad} component, but by S/N of the full {\oiii} profile (\narrowbroad). Differently from \citet{fio17}, that argued that $w_{80}$ is close to $v_{95}$, we found that this happens only in the outer parts of the J155019's emitting region, being lower in the other regions in this object, and also for the remaining QSOs. Besides, $w_{80}$ is close to or smaller than $v_{68}$ in four other QSOs, and stays between $v_{68}$ and $v_{95}$ in the remaining three: there isn't a clear pattern, other than $w_{80}<v_{95}$, in general. 

We note also that sometimes the velocity dispersion of the outflowing gas is included in the evaluation of outflow power:  
$\dot{E}_{out}=\dot{M}_{out}\,(v_{out}^2 + 3\,\sigma_{out}^2)/2$ \citep[e.g.][]{har18, ros18}.
If we use $v_{out}=v_{\broad}$ and $\sigma_{out}=\sigma_{\broad}$ in this formula, it will be equivalent to use $v_{out}=\sqrt{v_{\broad}^2+3\,\sigma_{\broad}^2}$ in our Eq.\,\ref{eq:edot}. In this case, Fig.\,\ref{fig:vout} shows that the results would be between our previous ones for $v_{95}$ and $v_{68}$, except for J135251, which closely follows $v_{68}$.

\subsubsection{Projection effects}

Projection effects due to the outflows inclinations ($i$)  -- angle between the outflow axis and the plane of the sky --  will affect the velocity and distance measurements.
We tested this effect by calculating deprojected velocities ($v_{0}$) and annular widths ($\Delta x_{0}$) as: $v_0 = v_{obs}/sin(i)$ and $\Delta x_0 = \Delta x_{obs}/cos(i)$, where $v_{obs}$ and $\Delta x_{obs}$ are the observed values.
Applying these corrections for $v_{out,0}=v_{95,0}=\left(|v_{\broad}|/\sin(i)+2\,\sigma_{\broad}\right)$ and $\delta r_0 = \delta r/cos(i)$, we recalculated $\dot{M}_{out}$ and $\dot{E}_{out}$, for $i=15,\,\,45$ and $75^{\circ}$.

Fig.\,\ref{fig:comp_method} displays the results for each of the inclinations above. On average, $\dot{E}_{out}^{\proj}/\dot{E}_{out}^{\defa}$ decreases by a factor of $\sim$\,4 for $i=75^{\circ}$ (range of [1/3.8,\,1/3.7]), and increases by a average factor of $\sim$\,3 for $i=15^{\circ}$ (range of [1.4,\,8.4]), with small changes for $i=45^{\circ}$ (range of [1/1.3,\,1.1]).

Ignoring the $\sigma_{\broad}$ term in $v_{out,0}$, this variation comes from the definitions used in Eqs.\,\ref{eq:mdot} and \ref{eq:edot}: $\dot{M}_{out}^{\proj} \,\propto\, v_{out}/\delta r \,\propto \left[ \cos(i)/\sin(i) \right]$ and $\dot{E}_{out}^{\proj} \,\propto\, v_{out}^3/\delta r \,\propto \left[ \cos(i)/\sin^3(i) \right]$.
Consequently, $\dot{M}_{out}^{\proj}$ decreases for $i>45^{\circ}$ and increases for lower inclinations, while for $\dot{E}_{out}$, the `turning point' occurs at $i\sim55^{\circ}$. Therefore, since we did not apply these corrections to the  {\default} method, the obtained values correspond to inclinations of $i\sim50^{\circ}$.

We conclude that the projection effects are a important source of uncertainty, inducing changes in $\dot{E}_{out}^{\proj}/\dot{E}_{out}^{\defa}$ between [1/3.8,\,8.4], for $i=75^{\circ}$ and $15^{\circ}$, and even higher for inclinations outside this range.

\subsubsection{Electron density}\label{sec:disc-met-ne}

Recent studies have been pointing out problems with the use of the {\sii}$\lambda\lambda$6718,31 ratio to obtain the density of the ionized gas in outflow \citep{har18,dav20}, suggesting that this density estimator results in smaller values of $n_e$ than those representative of these regions. \cite{ros18}, using the auroral  {\oii}$\lambda\lambda$7319,31 lines and the trans-auroral lines of {\sii}$\lambda\lambda$4068,76 \citep[method introduced by][]{hol11}, found densities between $10^{3.4}\,{-}\,10^{4.8}\,\mathrm{cm^{-3}}$, one to two orders of magnitude higher than the values obtained from \sii$\lambda\lambda6718,31$. 
\citet{bar19} presented another method, based on the ionization parameter $U$, obtaining even higher values for some objects, with $n_e$ ranging between ${\sim}\,10^{3}\,{-}\,10^{6}\,\mathrm{cm^{-3}}$.

A value of the electron density representative of the outflows is important for the calculation of the mass outflow rates and powers since the ionized gas mass is inversely proportional to $n_e$, and consequently, $\dot{M}_{out},\dot{E}_{out} \,\propto\, 1/n_e$. Therefore, we tested other density estimators, starting with the method introduced by \citet{bar19} to obtain $n_{e,BN19}$. Figure\,\ref{fig:ne} displays the radial profiles of $n_{e,BN19}$ as black dotted bars for annuli with a width corresponding to $\rm{FWHM_{PSF}}$. In this method the density is highly dependent on the outflow radius, with $n_{e,BN19}\propto 1/r^2$, as seen in the radial profiles decay. For reference, in Table\,\ref{tab:methods2} we present the average density values $\bar{n}_{e,BN19}$ along each radial profile, with an average among all QSOs of  2500\,cm$^{-3}$ and high standard deviation of 3500\,cm$^{-3}$. In comparison, the $n_{e,\sii\,\nb}$ average value is only 350\,$\pm$\,170\,cm$^{-3}$ (Table\,\ref{tab:methods1}). Individually, J082313 and J120041 present $n_{e,BN19}(r)$ radial profiles values similar to those calculated from \sii. \citet{dav20} obtained similar results by also testing different methods for calculating $n_e$: an average of 1900\,cm$^{-3}$ with the \citet{bar19} method,  4800\,cm$^{-3}$ from the trans-auroral/auroral lines, and only 350\,cm$^{-3}$ from the {\sii} ratio.

Using the radial profiles of $n_{e,BN19}$, we obtained $\dot{E}_{out}^{\rm{BN19}}/\dot{E}_{out}^{\defa}$ in the range [1/42,\,1/1.7], with values $\sim$\,5 times lower on average than the {\default} values, as shown in Fig.\,\ref{fig:comp_method}. This is expected because  $n_{e,BN19}>n_{e,\sii\nb}$, with the large variance in $\dot{E}_{out}^{\rm{BN19}}/\dot{E}_{out}^{\defa}$ reflecting the large range of $n_{e,BN19}$ values among the sample.  
In particular, for J120041, \citet{fal20} obtained the density from a photoionization model with a constant ionization parameter, resulting in lower peak values than $\dot{M}_{out}^{\rm{BN19}}$ and $\dot{E}_{out}^{\rm{BN19}}$ by factors of $\sim$\,1.5 and 7, respectively.
In part, the difference arises from the smaller $v_{out}=|v_{\bb}|$ definition used by these authors, which affects more the $\dot{E}_{out}$ values.

One of the issues with the traditional \sii method is that the {\sii}$\lambda\lambda$6718,31 ratio is sensitive to $n_e$ in the range of 
50\,--\,5000\,$\rm{cm^{-3}}$ \citep{har18}, with high uncertainties at the extreme values. Another problem is that the ionization potential of $\ion{S}{i}\rightarrow\ion{S}{ii}$ (10.36\,eV) is much lower than that of $\ion{O}{ii}\rightarrow\ion{O}{iii}$ (35.1\,eV) \citep{pro14}, whose emission lines are the usual tracers of the gas in outflow, as we have used here (the profile is dominated by more perturbed kinematics than most of the other emission lines). This difference in the ionization potential, plus the difference in critical densities, indicates that the {\sii} and {\oiii} emission may originate in different regions within the galaxies. A possible solution is to use {\ariv}$\lambda\lambda$4011,40 emission lines -- that have an ionization potential  $\ar{iii}\rightarrow\ar{iv}$ of 40.7\,eV \citep{pro14}. However, its usual small S/N ratio is an obstacle. Only two of our objects -- J082313 and J084135 -- have their integrated spectrum with S/N > 3 for these lines. The corresponding values of $n_{e,\ariv}$ are $1450_{840}^{1150}$ and $870_{410}^{430}$\,$\mathrm{cm^{-3}}$, respectively. Comparing with the average values obtained from the \sii ratio (see Fig.\,\ref{fig:ne}), we see that both are higher, with $n_{e,\ariv}/\bar{n}_{e,\sii\nb}\sim$\,$5\pm4$ and $7\pm4$, respectively.

Once more, the corresponding powers decrease (Fig.\,\ref{fig:comp_method}), with $\dot{E}_{out}^{\ariv}/\dot{E}_{out}^{\defa}$ values of $0.12_{0.22}^{0.17}$ for J082313 and $0.11_{0.26}^{0.19}$ for J084135, $\sim$\,9 times lower, on average. We highlighted the new values in Fig.\,\ref{fig:literature}, to show that better density calculations lead to values below the reference AGN coupling efficiencies (discussed in the next Section).

The {\ariv} lines are not commonly used in the literature to obtain gas densities for AGN-like objects. Two recent papers have  used them: \citet{may18} found $n_{e,\ariv}\sim10^{3.6}\,\mathrm{cm^{-3}}$ for Low-luminosity AGN; \citet{cer11} used them for three Coronal-Line Forest AGN (CLiF\,AGN), finding a range of $\sim10^{3.4}-10^{4.2}\,\mathrm{cm^{-3}}$, with $n_{e,\ariv} / n_{e,\sii}\sim5-17$ (in agreement with to our values).
The main reason for these lines not to be used frequently seems to be the weak intensity of {\ariv}: for the two QSOs for which we could measure them -- J082313 and J084135 -- the ratio $F_{\oiii\lambda5007} / F_{\ariv\lambda4011}$ are $\sim$\,450 and 380, respectively.

Another source of uncertainty in our analysis comes from the determination of the bolometric luminosities (\lbol), which were obtained from the total {\oiii}$\lambda$5007 luminosity \citep{tru15}. Dust extinction is an issue in this method \citep{hec14}, although we followed the \citet{lam09} prescription for correcting it for reddening.
Errors in {\lbol} affect the results displayed in Fig.\,\ref{fig:literature}, since, for example, lower values would bring our points closer to the $\varepsilon_f=0.5\%$ line. More importantly, it introduces uncertainties in the measurement of $\varepsilon_f$. 
We note that there are still further sources of uncertainties that were not taken into account, such as those due to reddening, density gradients, and geometry.

\subsubsection{AGN coupling efficiency}\label{sec:efficiency}

In studies about the impact of AGNs on their host galaxies, the AGN coupling efficiency -- the percentage of the {\lbol} that couples with the ISM --  is often calculated, since it has been suggested that its value may define if the feedback is powerful enough to affect the evolution of the host galaxy. Using simulations of galaxy mergers as triggers of AGNs, \citet{dimat05} found that a value of $5\,\%L_{bol}$ reproduces the $M_{BH}-\sigma_*$ relation. However, assuming that the feedback only needs to drive winds on the hot gas, that subsequently propagates to the cold part, \citet{hop10} found a threshold coupling efficiency of $0.5\,\%$, an order of magnitude lower. \citet{zub18} obtain similar values for higher redshidt ($z\sim3$) AGN, but higher at low redshifts, of $13\,\%$ at $z\sim0.5$.
In hydrodynamical simulations, for example, \citet{sch15} used $20\,\%L_{bol}$, while \citet{nel19} used $15\,\%$ for the radiative-mode feedback, both  injecting the energy thermally into the ISM. In this work, we provide the kinetic coupling efficiency $\varepsilon_f=\dot{E}_{out}/L_{bol}$ for the ionized gas, since $\dot{E}_{out}$ as we have calculated measures only this kinetic component.

Using the {\default} assumptions, the kinetic coupling efficiencies $\varepsilon_f^{\defa}$ vary from 0.0007\,\% for J082313, to 0.4\,\% for J094521. Only the latter shows a value of $\varepsilon_f^{\defa}$ close to the above threshold from \citet{hop10}, with the remaining QSOs showing values below $\varepsilon_f=0.05$\,\%. The average and median values are 0.6 and 0.1\,\%, respectively.
However, our tests show that these measurements are highly uncertain. Putting together all tests results, $\varepsilon_f$ ranges between 0.00005 and 7\,\%. In particular, the densities obtained with {\sii} in the {\default} method are not ideal. The possibly more adequate values, calculated from the {\ariv} emission line ratio, lead to an order of magnitude decrease in the efficiencies.

Our values of $\varepsilon_f$ and the uncertainties involved in the quantities inferred from the observations support the results from many recent studies \citep[e.g][]{kar16b,vil14,vil16,fis18,hus16,spe18,ros18}.
However, a direct comparison between our kinetic coupling efficiencies, and the ones obtained in models and simulations, cannot be made.
As pointed out previously by \citet{har18}, $\varepsilon_f$ as we have calculated  determines the coupling efficiency only of the kinetic component obtained from the ionized gas, while values provided by models and used in simulations usually refer to the total energy deposited in ISM, where only a fraction may actually induce outflows observed in ionized gas. For example, part of the energy is used to heat and ionize the gas \citep{zub18}.
Even more important is the fact that the we are measuring only a fraction of the  gas affected by the feedback  -- the ionized phase. \citet{fio17} found that molecular winds contribute more to the energy released in the AGN feedback than the ionized gas, although ionized gas outflows increase their impact for higher {\lbol}  sources. Other important phases that should be considered include the neutral (e.g. \ion{H}{i}\,21cm,  \naD, \ion{c}{ii}) and the highly ionized (X-ray absorption lines) \citep{cic18,har18,fio17}.
Hence, these effects may explain our lower coupling efficiencies in comparison to those predicted by models.

\section{Conclusions}\label{sec:conclusions}

We have analysed optical integral field spectroscopy data of a sample of 8 QSO\,2's at $0.1\le z \le 0.5$ in order to map the ionized gas kinematics and quantify mass outflow rates and powers from their AGN. The main conclusions of this paper are:

\begin{itemize}
\item Most of the emission-line profiles were best fitted by one or two {\narrow} components -- attributed to ambient gas, and sometimes associated with mergers -- 
plus a broader one (\broad) -- attributed to a nuclear outflow;

While the total ionized gas emission extents reach distances from the nucleus of $5 \lesssim R_{\rm{ELNR},0} \lesssim 26$\,kpc, the outflows show smaller extents in the range $1 \lesssim R_{out,0} \lesssim 8$\,kpc. Therefore, only part of the ionised gas is disturbed by the outflows, with an average ratio of $R_{out}/R_{\rm{ENLR}} \sim0.5\pm0.3$.

\item We have measured the mass outflow rates ($\dot{M}_{out}$) and powers  ($\dot{E}_{out}$) from the kinematics of the broader component as a function of distance from the nucleus. These measurements were first made using the {\default} assumptions: gas density determined from the \sii line ratio, outflow velocity as $v_{95}$, and the gas mass from the luminosity of the {\broad} component only. With these assumptions, we found peak values of $\dot{M}_{out}^{\defa}$ ranging between $\sim$\,3\,--\,30\,M$_{\odot}$\,yr$^{-1}$, and peak outflow powers $\dot{E}_{out}^{\defa}$ values between $\sim\,3\times10^{41}\,-\,3\times10^{43}\,\mathrm{erg\,s^{-1}}$.

\item We performed tests to check the influence in the resulting $\dot{M}_{out}$ and $\dot{E}_{out}$ of varying assumptions regarding: the ionized gas mass, the outflow velocity, the ionized gas density and orientation of the outflow. When compared with the {\default} method, these tests resulted in typical variations of one order of magnitude for each test, and being larger if considered together.

\item Including the total ionized gas mass in the calculations (instead of only the contribution from the {\broad} component) lead to overestimates of the outflow powers by factors of $\sim$\,2\,--\,17. 
Values based on integrated measurements --  as in unresolved observations -- differ from the peak of radial profile values by factors of up to $\sim$\,5.

\item The use of lower outflow velocities ($v_{out}$) than the {\default} $v_{95}$ lead to a decrease of $\sim$\,5 times in $\dot{E}_{out}$, and $\sim$\,2 times in $\dot{M}_{out}$, on average.  Corrections for projection effects also change $\dot{E}_{out}$, decreasing it -- on average -- by a factor of $\sim$\,4 for inclinations $i=75^{\circ}$, and increasing it by a factor of $\sim$\,2 for $i=15^{\circ}$.

\item The use of improved density tracers instead of the {\sii} ratio, resulted in higher density values, with an average increase of $n_{e,\rm{BN19}}/\bar{n}_{e,\sii}\sim$\,7 (with high variations) for the method based on the ionization parameter, and an increase of 
$\bar{n}_{e,\ariv}/\bar{n}_{e,\sii}\sim$\,$6\pm3$ for values obtained from the {\ariv} ratio. The resulting outflow powers decrease by the same factors. We recommend the use of {\ariv}, since its ionization potential is similar to the {\oiii} one;

\item From the peak values of $\dot{E}_{out}^{\defa}$, we calculated the corresponding kinetic coupling efficiencies $\varepsilon_f^{\defa}$, finding values in the range $7\,{\times}\,10^{-4}$\,--\,0.05\,\%, except for J094521, that has $\varepsilon_f^{\defa}\,{\sim}$\,0.4\%. 
Including the results from the tests, the range becomes $5\times10^{-5}$\,--\,7\,\%. In particular, using the densities calculated from the {\ariv} tracer, we observe almost one order of magnitude decrease in the coupling efficiencies.

\end{itemize}

We finally point out that the {\default} assumptions used in this study do not necessarily represent the best choice of parameters, but are useful as a reference baseline for the tests. We also remind the reader that, since this work only refers to the kinetic feedback from the ionized gas, it underestimates the actual fraction of the bolometric luminosity that couples with the ISM, as contributions of feedback from other gas phases, such as the molecular gas and hotter X-ray emitting gas should also be considered, as well as other forms of feedback.

\section*{Acknowledgements}
This study was financed in part by the Coordena{\c c}{\~a}o de Aperfei{\c c}oamento de Pessoal de N\'ivel Superior (CAPES-Brasil, 88887.478902/2020-00) and  Conselho Nacional de Desenvolvimento Cient\'ifico e Tecnol\'ogico (CNPq-Brasil, 130574/2018-0). 

\section*{Data availability}
Based on observations obtained at the international Gemini Observatory, a program of NSF’s NOIRLab, and available from Gemini Observatory Archive (program ID in Table\,\ref{tab:obs}), which is managed by the Association of Universities for Research in Astronomy (AURA) under a cooperative agreement with the National Science Foundation. on behalf of the Gemini Observatory partnership: the National Science Foundation (United States), National Research Council (Canada), Agencia Nacional de Investigaci\'{o}n y Desarrollo (Chile), Ministerio de Ciencia, Tecnolog\'{i}a e Innovaci\'{o}n (Argentina), Minist\'{e}rio da Ci\^{e}ncia, Tecnologia, Inova\c{c}\~{o}es e Comunica\c{c}\~{o}es (Brazil), and Korea Astronomy and Space Science Institute (Republic of Korea).



\bibliographystyle{mnras}
\bibliography{outflow2021} 

\appendix

\section{}

\begin{figure}
    \centering
	\includegraphics[width=1\columnwidth]{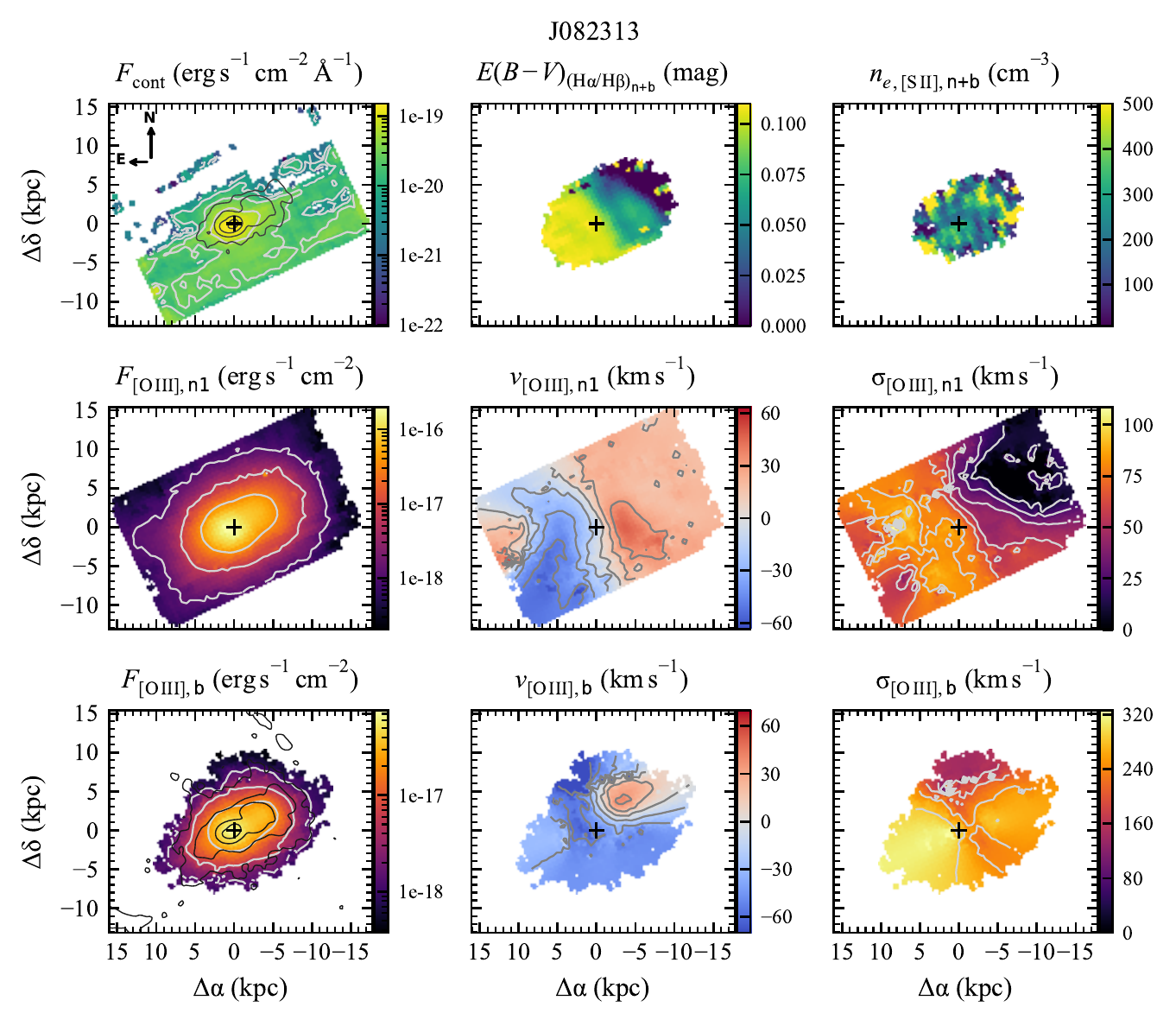}
\caption{Maps of J082313, similar to Fig.\,\ref{fig:mapJ135251}. The systemic velocity were calculated from $v_{\oiii,\n{1}}$\,(see Section\,\ref{sec:velocity}). Only pixels with S/N > 3 are shown. Note that the colorbars have different ranges.}
    \label{fig:mapJ082313}
\end{figure}

\begin{figure}
    \centering
	\includegraphics[width=1\columnwidth]{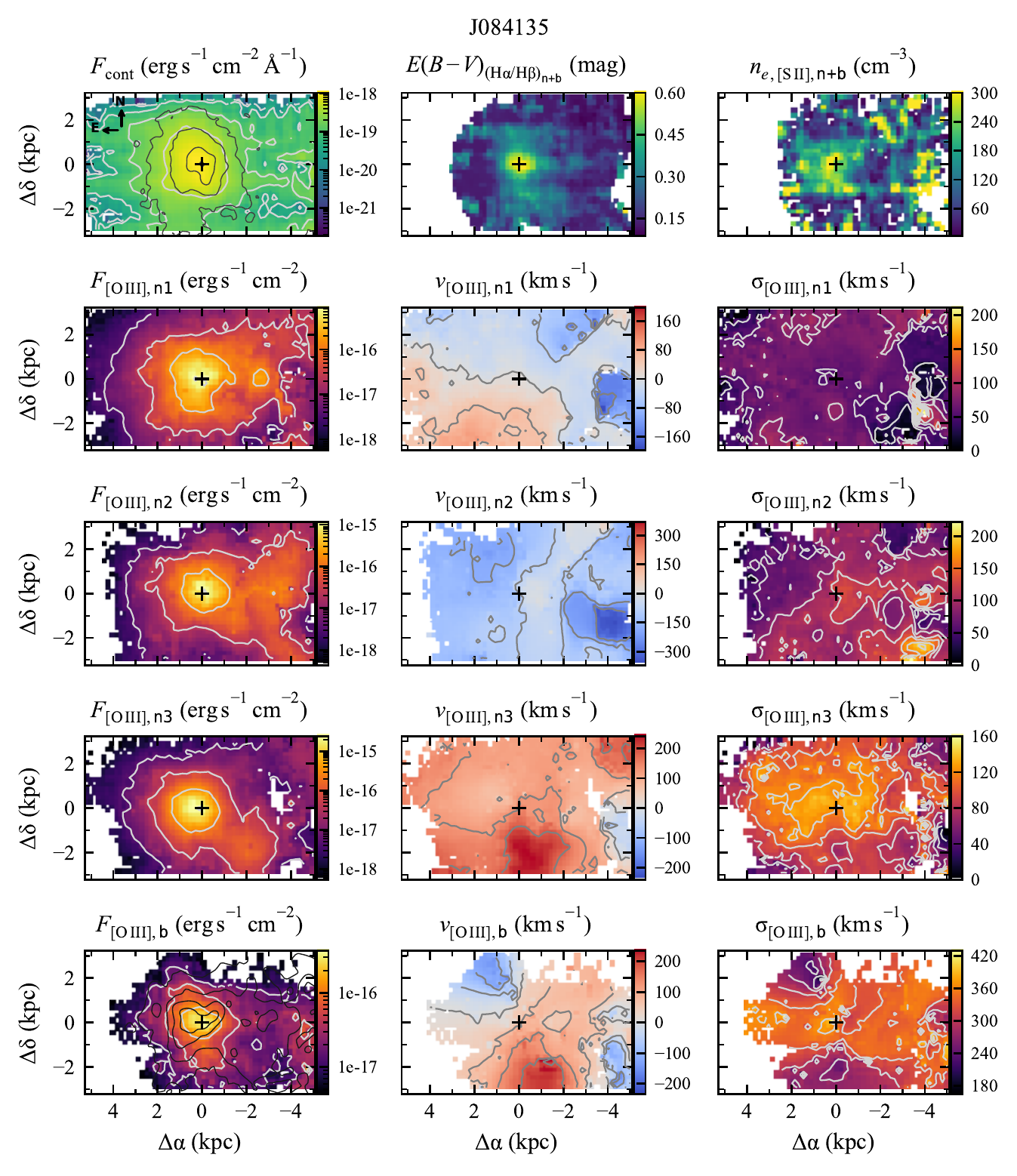}
    \caption{Same as Fig.\,\ref{fig:mapJ082313}, but for J084135. Systemic velocity calculated from $v_{\oiii,\n{1}}$.}
    \label{fig:mapJ084135}
\end{figure}

\begin{figure}
    \centering
	\includegraphics[width=1\columnwidth]{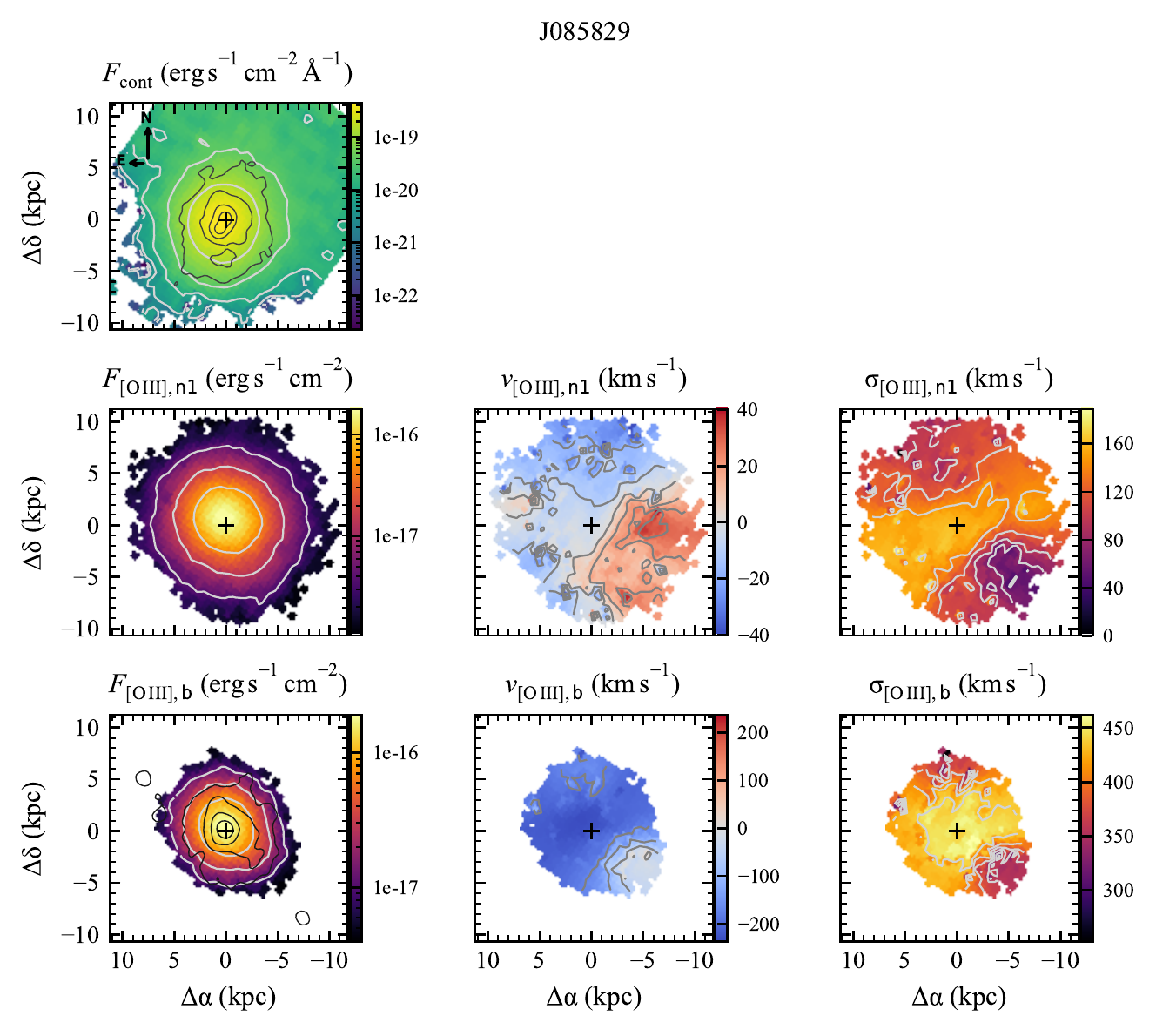}
    \caption{Same as Fig.\,\ref{fig:mapJ082313}, but for J085829. Systemic velocity calculated from $v_{\oiii,\n{1}}$.}
    \label{fig:mapJ085829}
\end{figure}

\begin{figure}
    \centering
	\includegraphics[width=1\columnwidth]{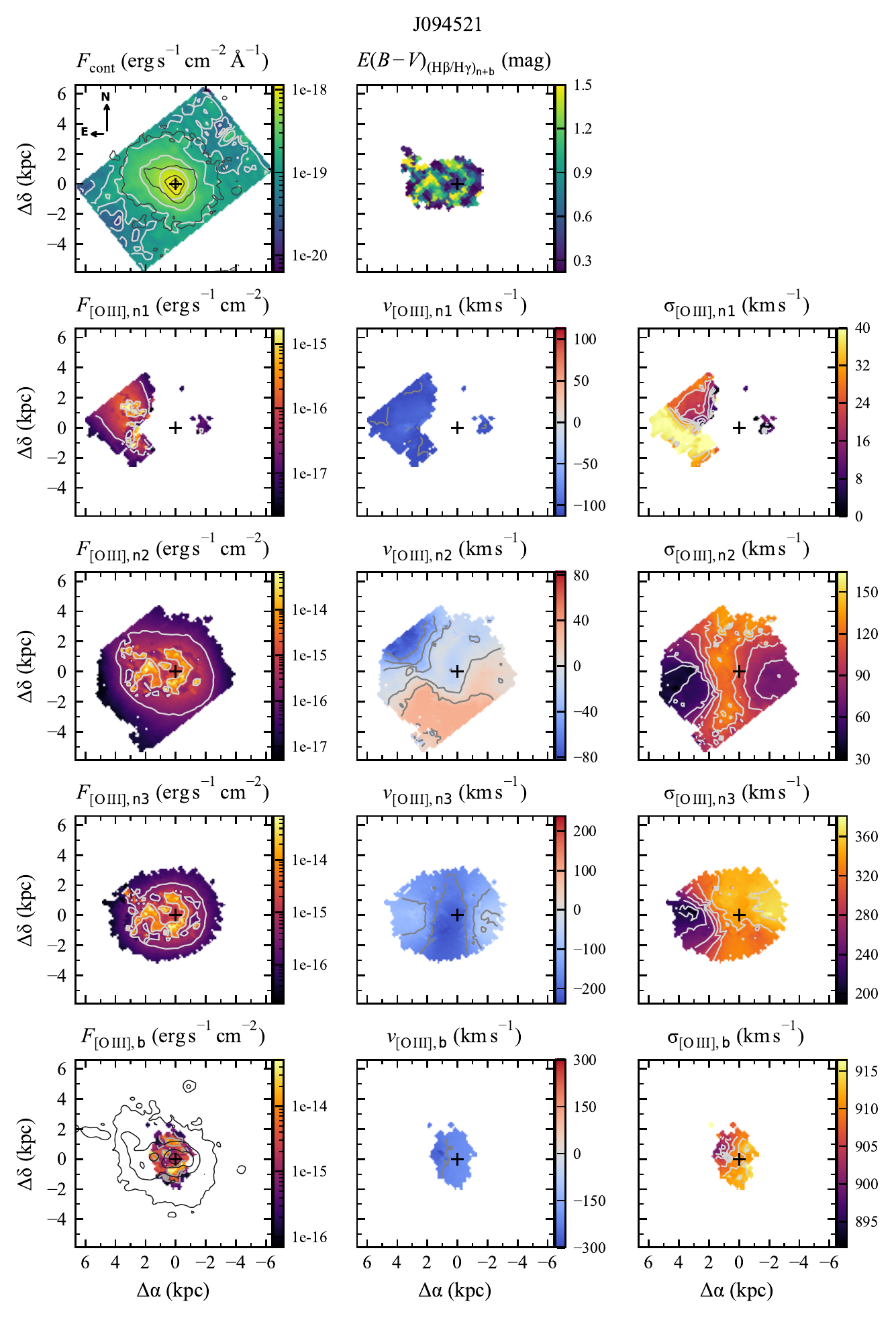}
    \caption{Same as Fig.\,\ref{fig:mapJ082313}, but for J094521. Systemic velocity calculated from $v_{\oiii,\n{2}}$.}
    \label{fig:mapJ094521}
\end{figure}

\begin{figure}
    \centering
	\includegraphics[width=1\columnwidth]{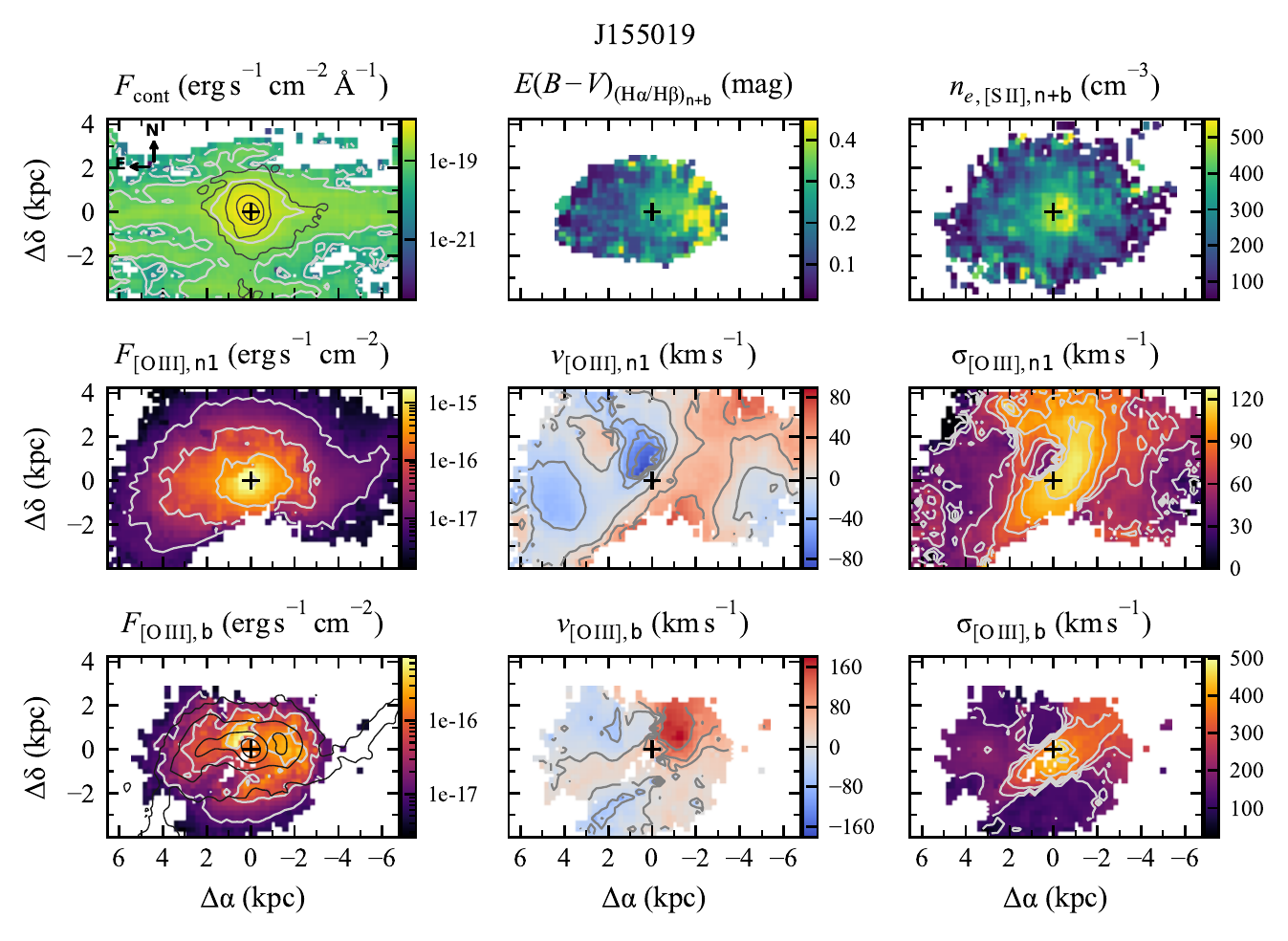}
    \caption{Same as Fig.\,\ref{fig:mapJ082313}, but for J155019. Systemic velocity calculated from $v_{\oiii,\n{1}}$.}
    \label{fig:mapJ155019}
\end{figure}

\begin{figure}
    \centering
	\includegraphics[width=1\columnwidth]{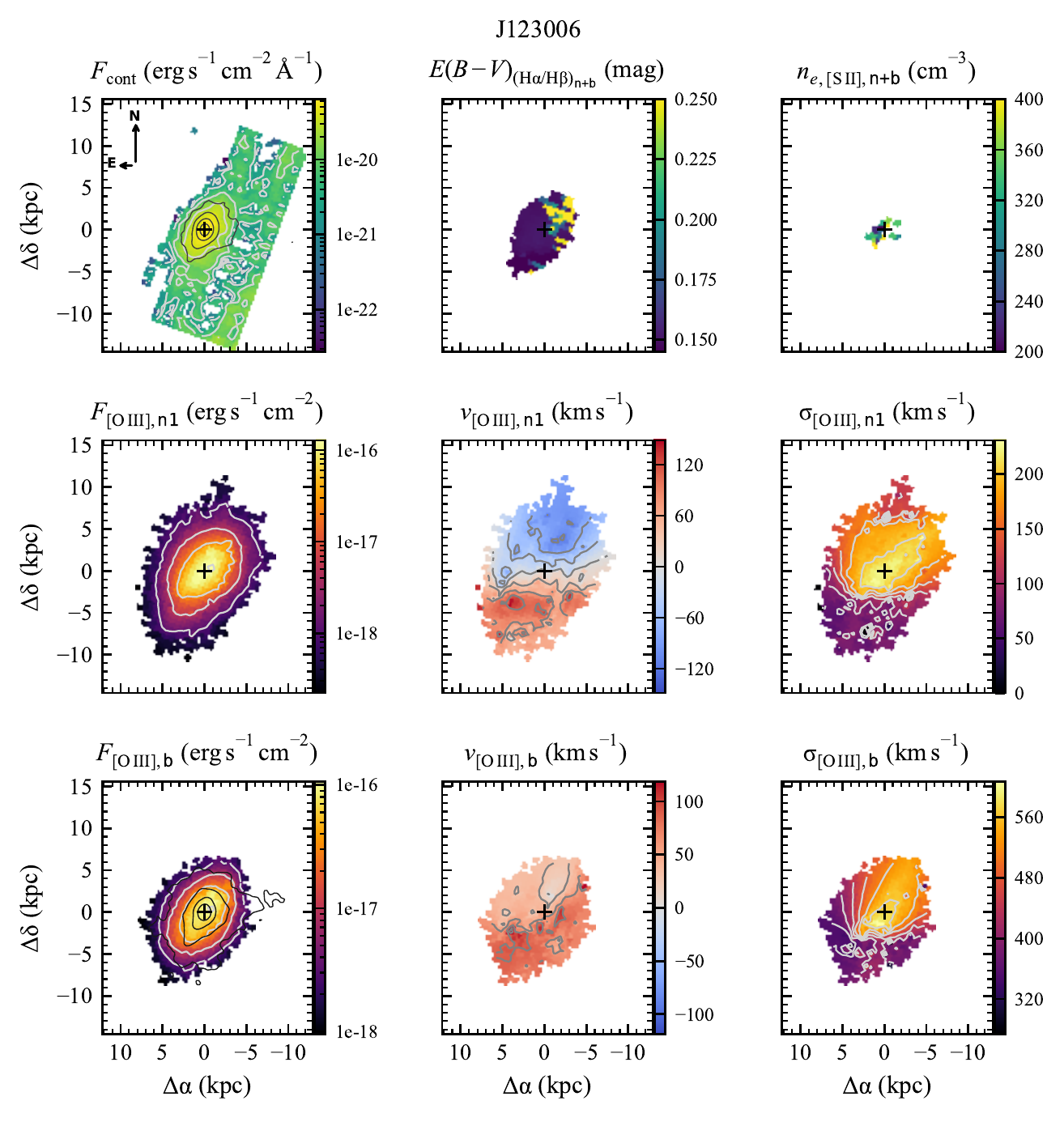}
    \caption{Same as Fig.\,\ref{fig:mapJ082313}, but for J123006. Systemic velocity calculated from $v_{\oiii,\n{1}}$.}
    \label{fig:mapJ123006}
\end{figure}

\begin{figure}
    \centering
	\includegraphics[width=1\columnwidth]{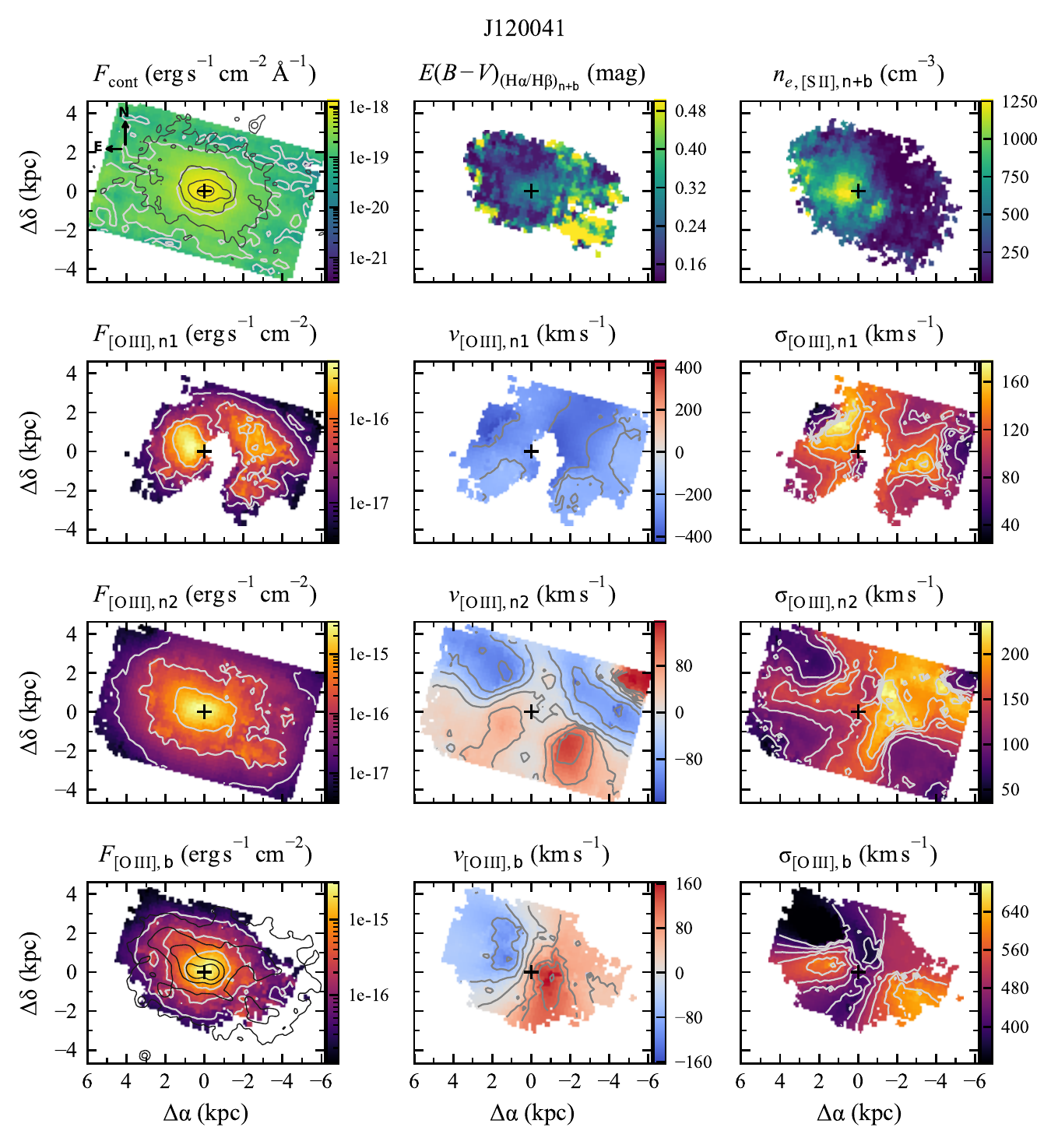}
    \caption{Same as Fig.\,\ref{fig:mapJ082313}, but for J120041. Systemic velocity calculated from $v_{\oiii,\n{2}}$.}
    \label{fig:mapJ120041}
\end{figure}


\bsp	
\label{lastpage}
\end{document}


\label{firstpage}
\pagerange{\pageref{firstpage}--\pageref{lastpage}}
\maketitle

\appendix

\addtocounter{section}{1}
\section{}

\subsection{Telluric absorption correction}\label{ap:tell}
\begin{figure}
	\includegraphics[width=.95\columnwidth]{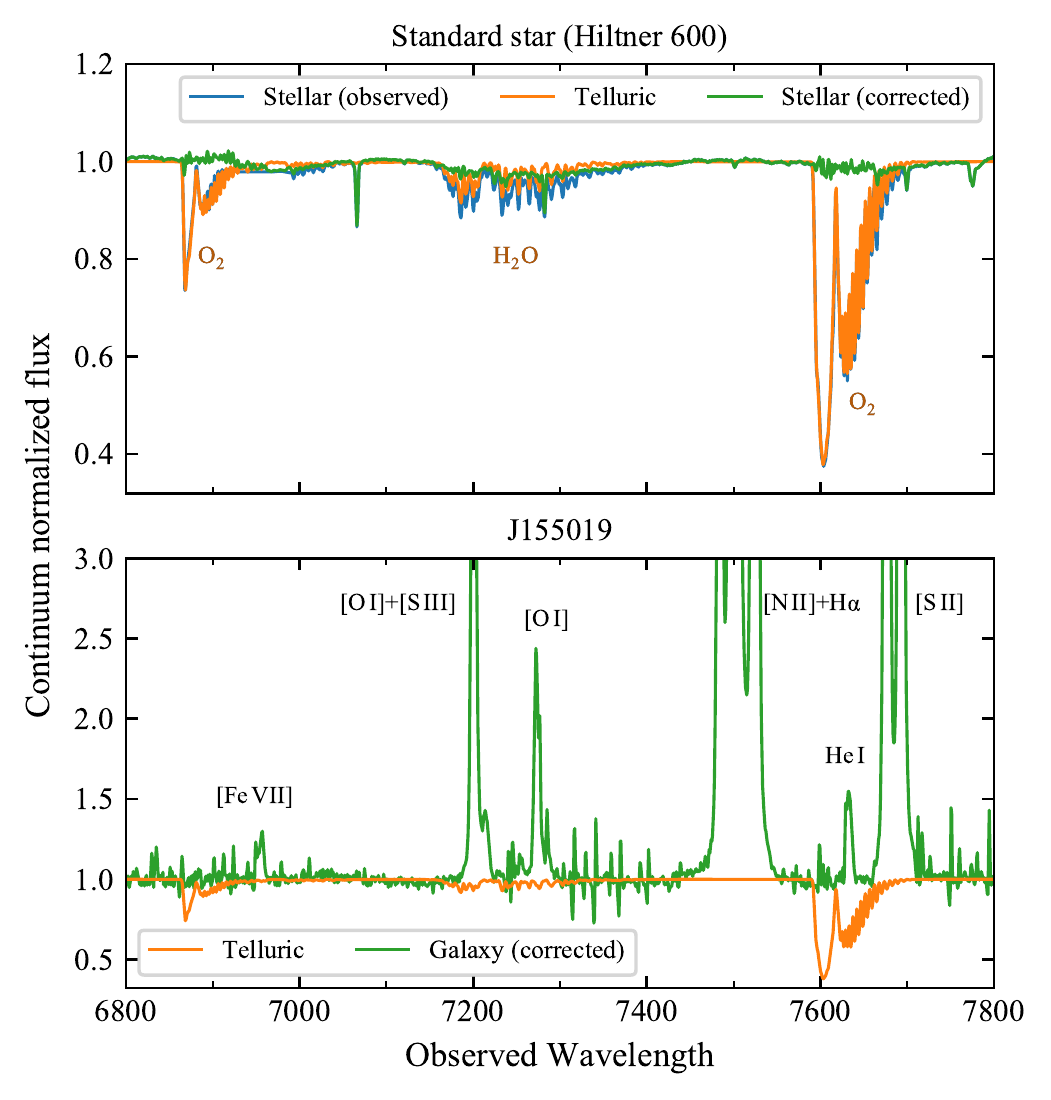}
    \caption{Telluric absorption correction for J155019. The top panel displays the telluric fit (orange) made to the observed standard star continuum-normalized flux (blue), and the result of the correction (green).
    The bottom axis shows a 2\,kpc aperture over the final galaxy data cube (green), after the individual observation were merged and the flux were normalize by the continuum. The telluric transmission (orange) shows how much were corrected it from the O$_2$ and H$_2$O aborptions bands.
    }
    \label{fig:tell}
\end{figure}

Before merging the individual data cubes of each object, we corrected the spectra for telluric absorption using the \texttt{MOLECFIT} software \citep{molec1,molec2}. Besides removing the absorption effect under some emission lines, we retrieved the spectral resolution 
($\sigma_{inst}$, presented in velocity units in Table\,2 in paper) from the model, used later to correct the Gaussian profiles for the instrumental spectral broadening. For each galaxy, we modeled the telluric contribution (O$_2$ and H$_2$O bands) in its corresponding standard star, and applied the solution to the science data cubes (see Fig\,\ref{fig:tell}). We performed the fit in the stellar spectrum because of its higher signal-to-noise. However, since the observations of the galaxy and the standard star can be months apart, the atmospheric conditions are not the equal. Therefore, before correcting it from the galaxy spectra, we scaled the telluric transmission by a multiplicative factor, minimizing the residuals near an absorption feature.

\subsection{Broad Line Region (BLR)}\label{ap:blr}
Even though our sample consists of QSOs classified as type II \citep{rey08}, one of them -- J082313 -- showed signs of a weak BLR contribution (only visible in the \ha profile). 

\begin{figure}
    \centering
	\includegraphics[width=0.95\columnwidth]{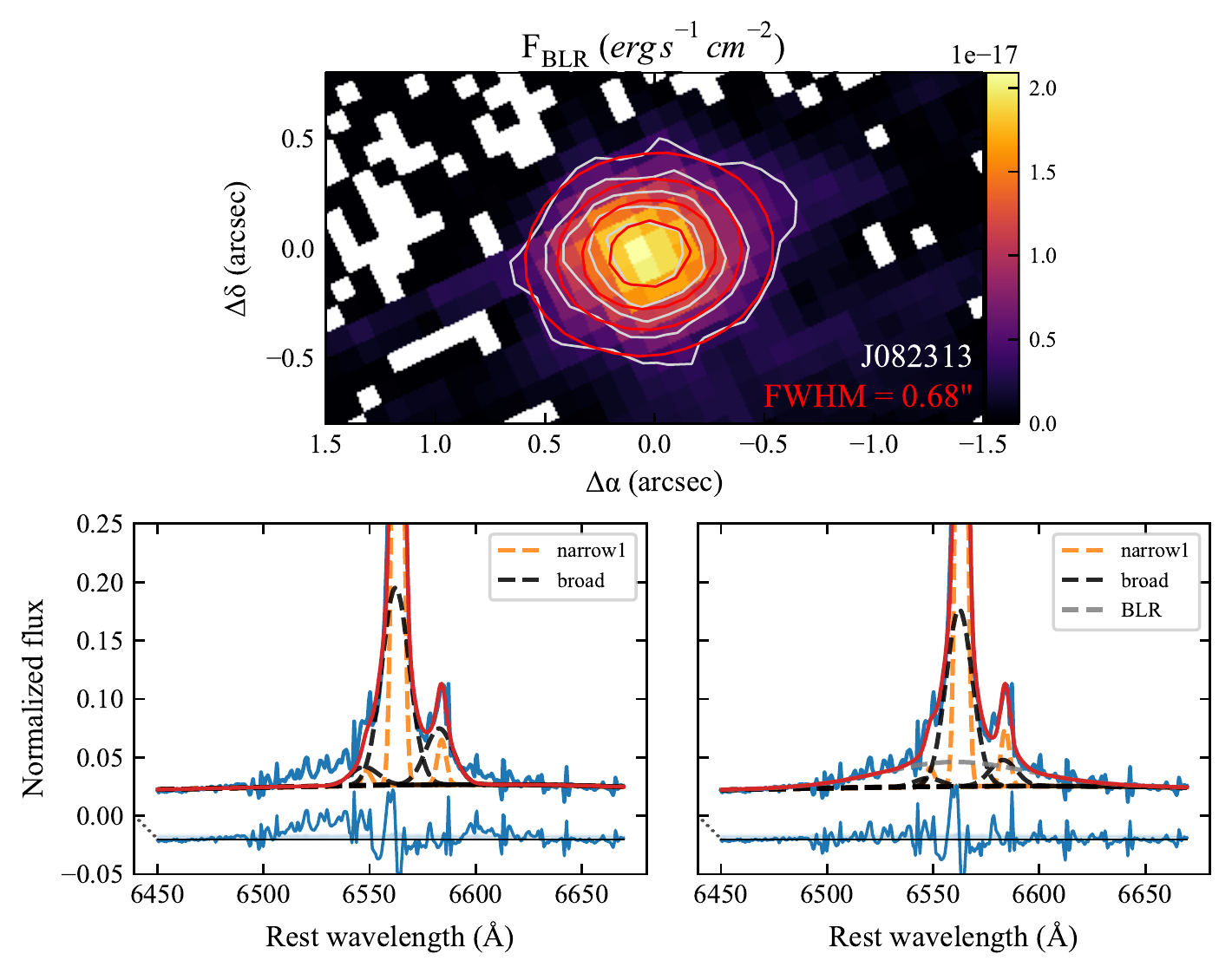}
    \caption{Flux distribution (top) of the J082313  Broad Line Region (BLR) from \ha, along with its the 2D Gaussian fit (white contours). In the bottom, we show the seeing-size aperture spectrum, fitted without (left) and (with) a BLR component (gray). Although the BLR intensity is weak, compared to the NLR components, its addition avoided a wrongly inclusion of another NLR component, which would change the results of our analysis.}
    \label{fig:BLR}
\end{figure}

To model it, we fitted the spectra -- obtained from a PSF-size aperture -- modeling it with one {\broad} and one {\narrow} components for the NLR gas, along with a broader Gaussian representing the BLR. The other emission emission were also fitted, which helped the NLR model, since we fixed the kinematics of the components between different lines. Fig.\,\ref{fig:BLR} (below) shows the resulting fit without (left) and with (right) the BLR component, where $v_{\BLR}=-140\mathrm{\,km\,s^{-1}}$ and $\sigma_{\BLR}=1560\mathrm{\,km\,s^{-1}}$.

With the the BLR component modelled, we repeated the fitting process, now, for all spectra of the data cube. However, in order to account for the smearing effect of the atmospheric seeing, we fixed $v_{\BLR}$ and $\sigma_{\BLR}$ at the above values, with the amplitude being the only free parameter of the BLR component. Fig.\,\ref{fig:BLR} (top) shows the resulting BLR flux distribution, along with white contours from the 2D-Gaussian best fit model. The resulting FWHM of $\sim$\,0.68\,arcsec is in agreement with the seeing obtained from the field in the acquisition image 
(Table\,2 in the paper).

\subsection{Source of ionization}\label{ap:bpt}
In order to identify the main source of ionization in our objects, we generated a BPT diagram \citep{bal81} using the total flux ({\nb}) of each emission line (Fig.\,\ref{ap:bpt}). In this diagram, emission lines with ratios in the region to the of left of the dashed and dotted lines \citep{sta06,kew01} can be produced by the radiation originated in star-forming regions in the galaxy. At the right, the solid line \citep{kew06} separates emission line ratios originated from high-ionization sources (AGNs, above) and low-ionization nuclear emission-line regions (LINERs, below). The parametrizations of the lines were taken from \citet{cid10}. Since the GMOS-IFU spectra of J085829 and J094521 doesn't cover \nii{+}\ha, we measured $\nii\lambda6584/\haa$ in the SDSS spectrum. The diagram confirms that the observed emission lines of our sample are mostly ionized by AGN sources. 

\begin{figure}
    \centering
	\includegraphics[width=0.95\columnwidth]{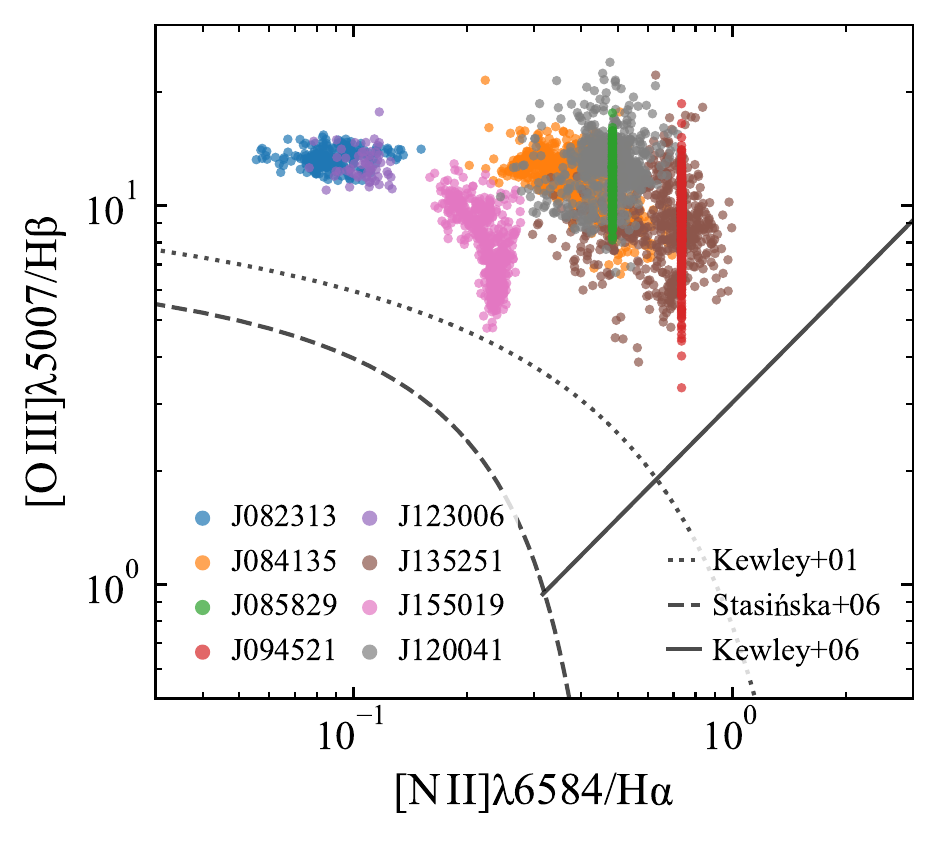}
    \caption{BPT diagram for the galaxies of our sample. The dashed, dotted and solid lines are from \citet{sta06}, \citet{kew01} and \citep{kew06}.}
    \label{fig:bpt}
\end{figure}

\subsection{[Ar\,{\scriptsize IV}] fit}\label{sec:ariv}
\begin{figure}
    \centering
	\includegraphics[width=1\columnwidth]{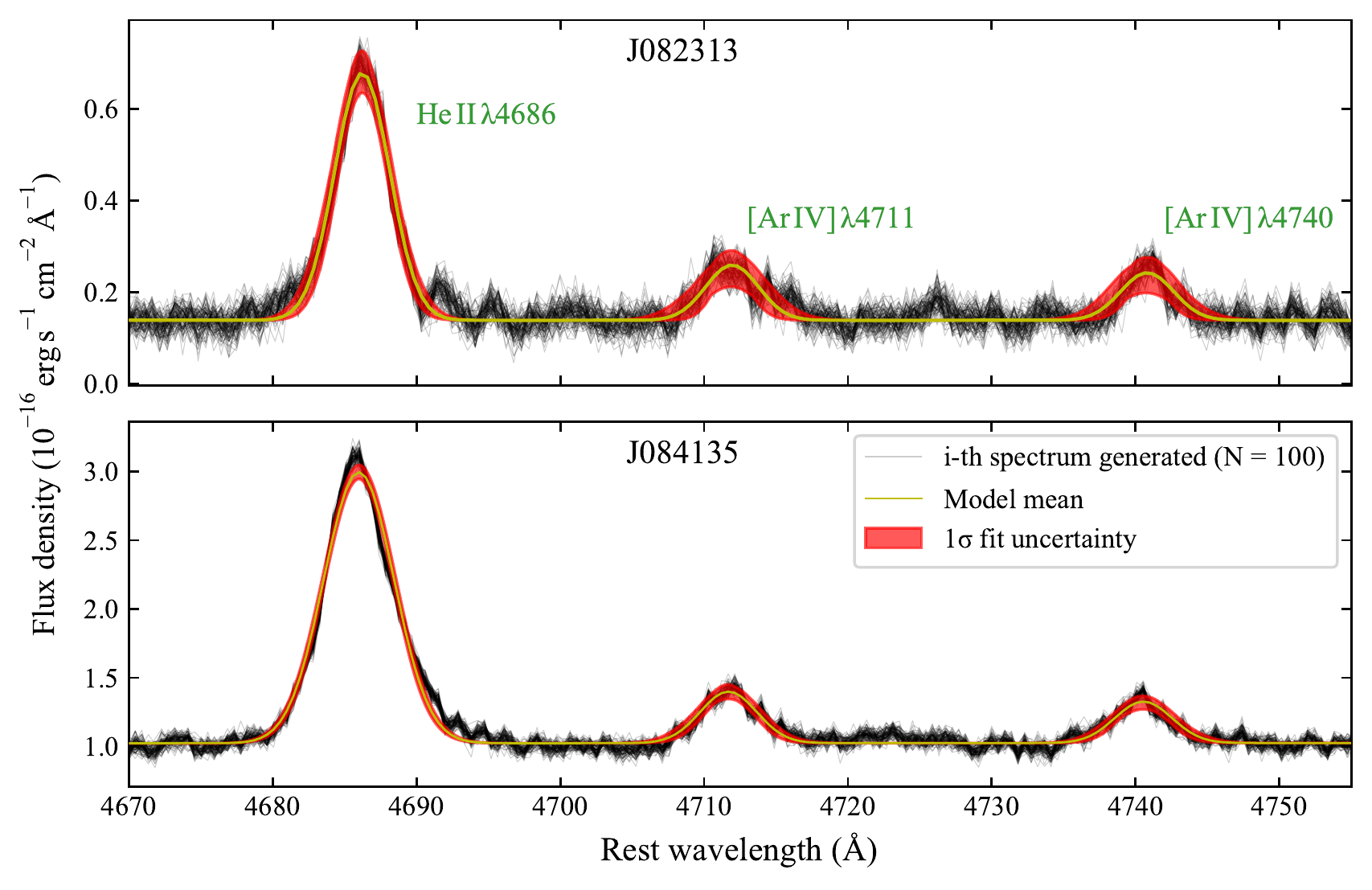}
    \caption{Result of the Montecarlo method, used to fit the {\ariv}$\lambda4711,40$ and \ion{He}{ii}$\,\lambda4686$ emission lines of J082313 and J08413. We fitted the spectra generated in each run (black lines), with model corresponding to the mean of the parameters shown in yellow. In the red filled region, we display the limits of 100 models, generated using the mean and standard deviation values of the parameters previously fitted.}
    \label{fig:ariv}
\end{figure}

We applied a Montecarlo method -- fitting single Gaussians for each {\ariv}$\lambda4711,40$ emission line (and the near line of \ion{He}{ii}$\,\lambda4686$) -- for the seeing-size apertures spectra of J082313 and J084135 (Fig.\,\ref{fig:ariv}). We ran 100 iterations, generating new spectra with gaussian noise (standard deviation from the original spectra), and fitting the results, where the {\ariv} kinematics were kept equal. From the fits, we obtained the mean and standard deviation values of the parameters of the lines, which uncertainties were propagated to the flux ratio of the {\ariv} lines, used to obtain $n_{e,\ariv}$.

\subsection{Emission line fitting results}\label{ap:specs}
\begin{figure*}
	\includegraphics[width=.73\linewidth]{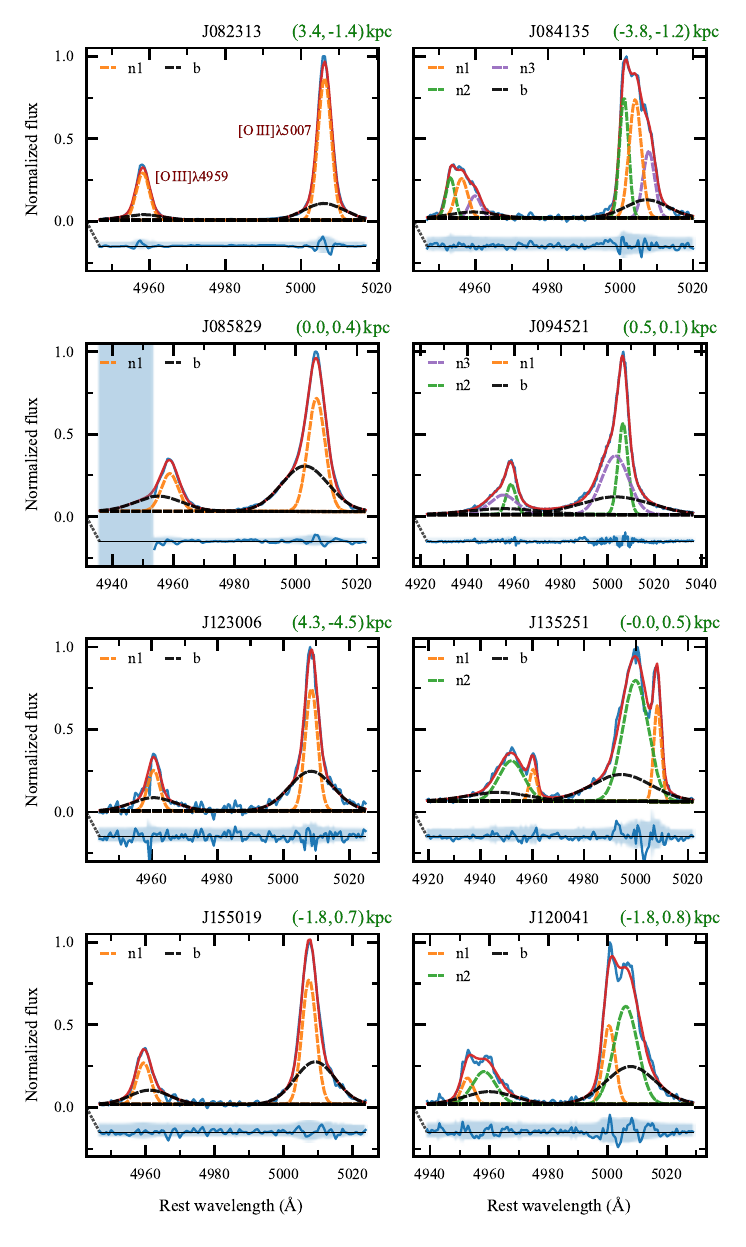}
    \caption{Examples of fitting results for the {\oiii} line profiles for each galaxy of the sample, highlighting the decomposition of the spectral flux density in {\narrow} (\n{1}, \n{2}, \n{3}) and {\broad} (\bb) components. The pixel location (projected distance from the nucleus, in kpc) is shown in top right of each panel. Four objects have more than one {\narrow} component, as it can clearly in both {\oiii} profiles.  Here, J094521 does not show its {\n{1}} component, that appears in another region of the galaxy, away from the nucleus (see Fig.\,\ref{fig:specJ094521} and Fig.\,\ref{fig:mapJ094521})}
    \label{fig:spec-comp}
\end{figure*}

An alternative and larger version of Fig.\,3 from the paper is shown in Fig.\ref{fig:spec-comp}.

\begin{figure*}
    \centering
	\includegraphics[width=1\linewidth]{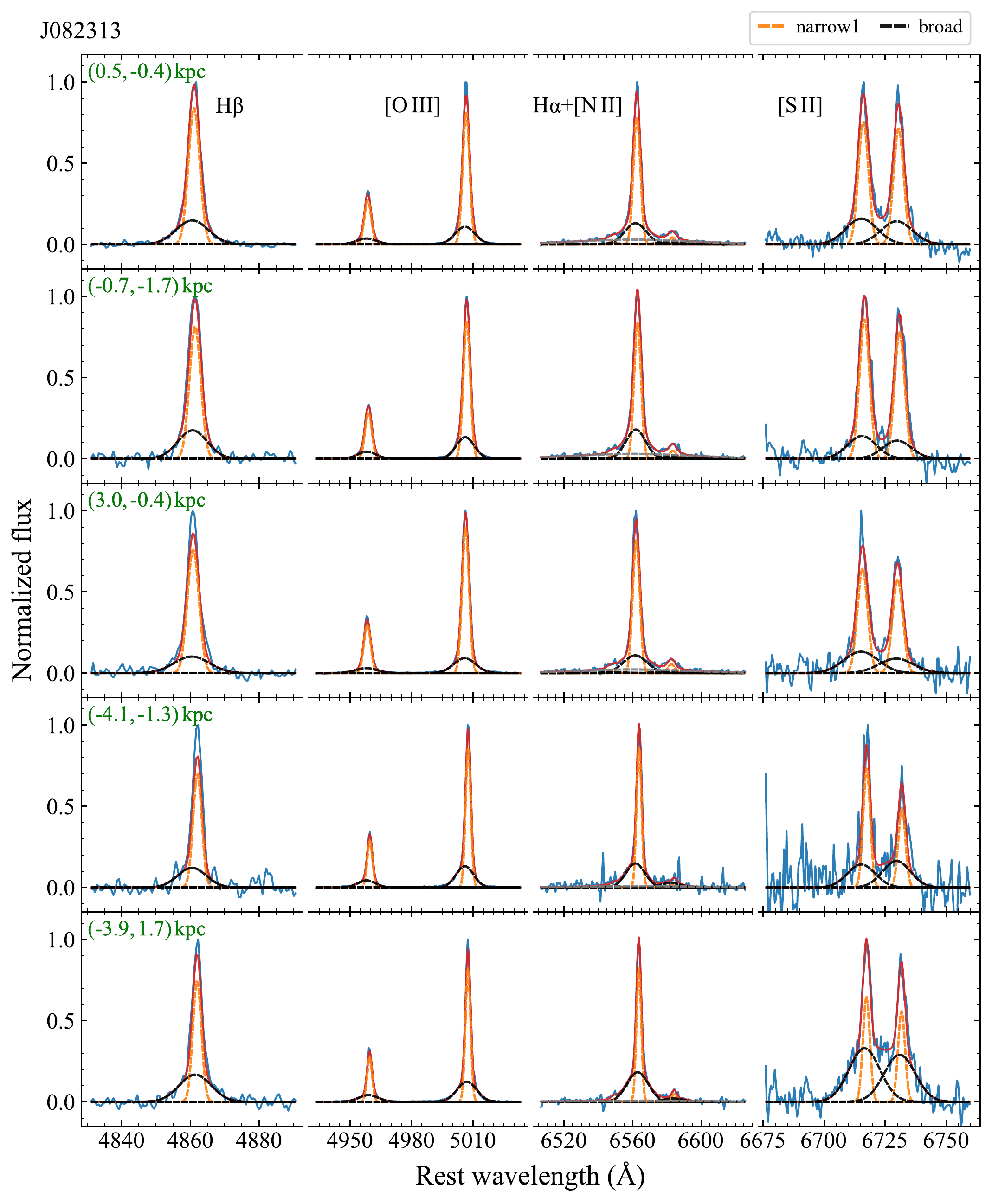}
    \caption{Fitting results for a few spectra of J082313. Each row of the figure corresponds to a different position in the galaxy, identified by $(\Delta \alpha, \Delta \delta)$, the distance from the nucleus (in green at the upper left corner of each row). The columns are zoom-in in the emission lines: \hb, \oiii, (\ha+\nii) and \sii.}
    \label{fig:specJ082313}
\end{figure*}

\begin{figure*}
    \centering
	\includegraphics[width=1\linewidth]{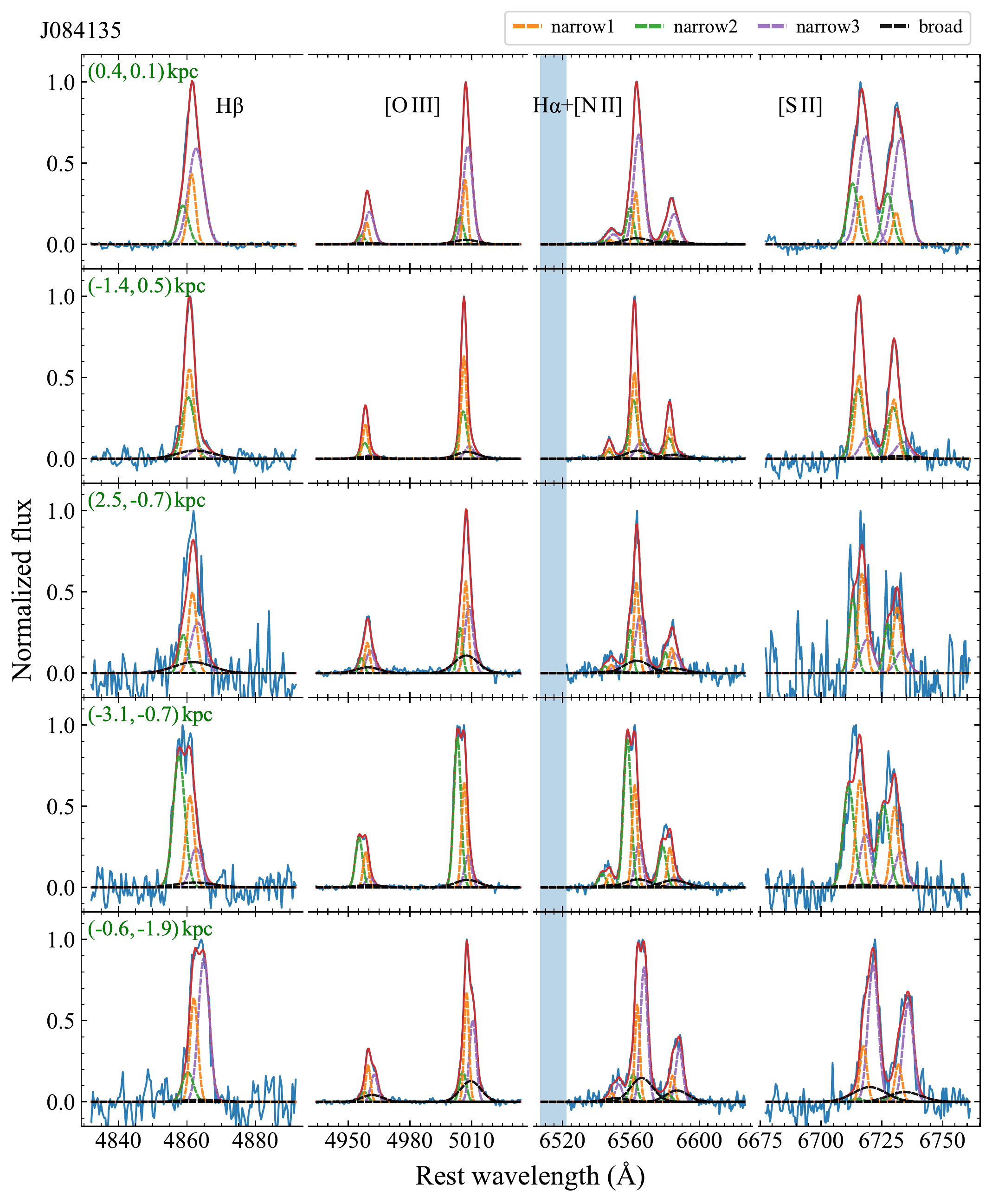}
    \caption{Same as Fig.\,\ref{fig:specJ082313} but for J084135.}
    \label{fig:specJ084135}
\end{figure*}

\begin{figure}
    \centering
	\includegraphics[width=1\columnwidth]{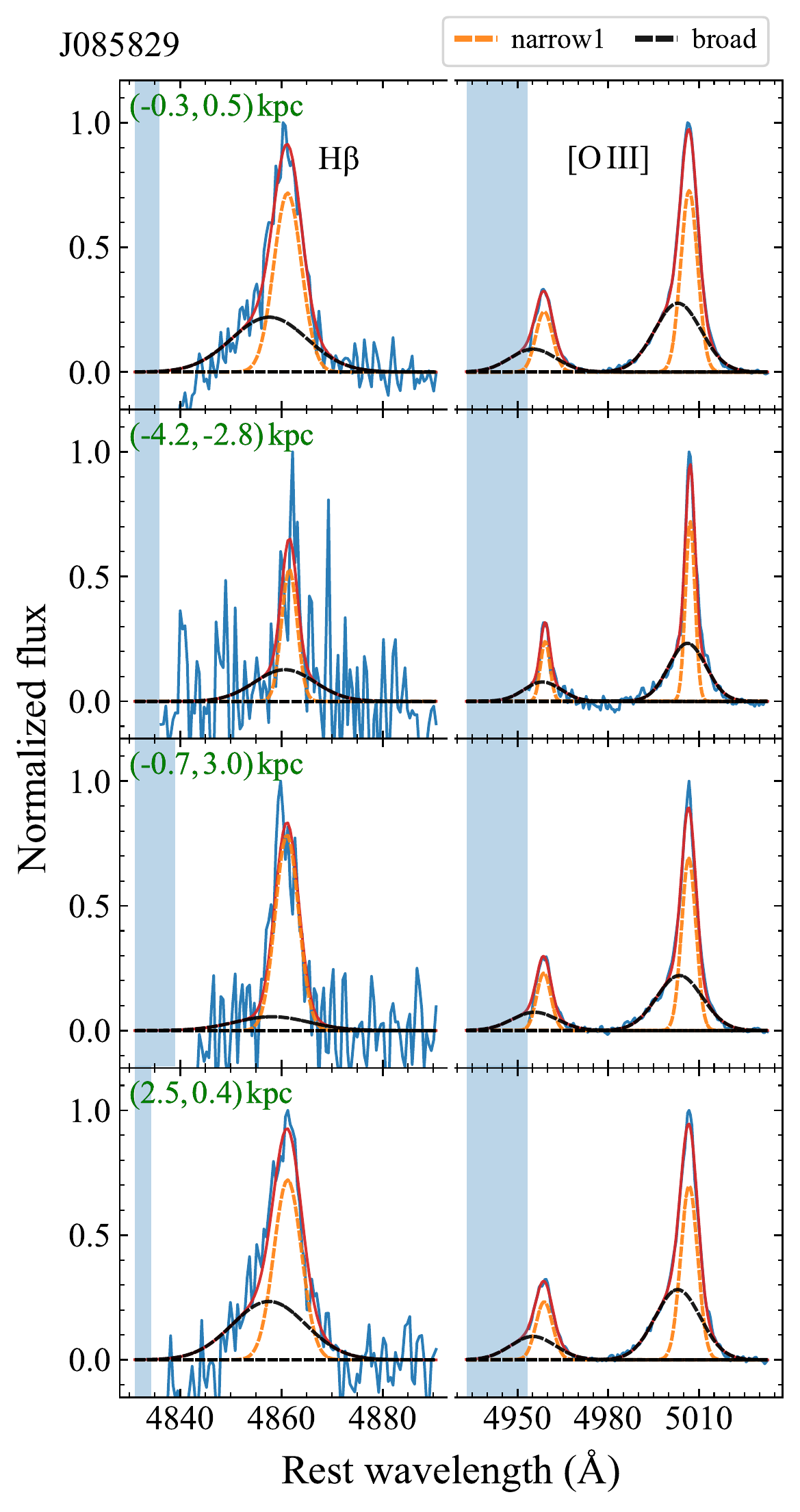}
    \caption{Same as Fig.\,\ref{fig:specJ082313} but for J085829. There were no data in the blue region (due to CCD gaps or bad pixels).}
    \label{fig:specJ085829}
\end{figure}

\begin{figure}
    \centering
	\includegraphics[width=1\columnwidth]{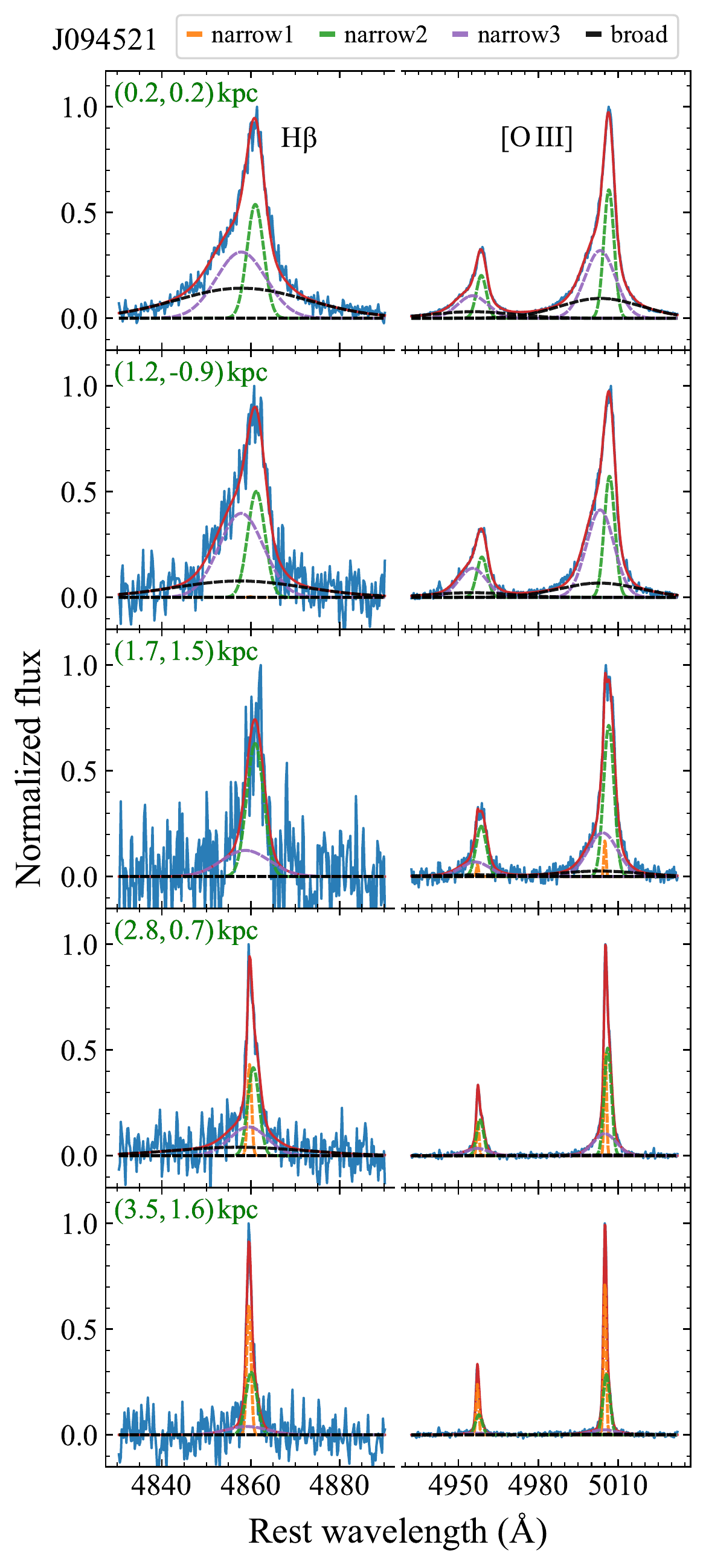}
    \caption{Same as Fig.\,\ref{fig:specJ082313} but for J094521.}
    \label{fig:specJ094521}
\end{figure}

\begin{figure*}
    \centering
	\includegraphics[width=1\linewidth]{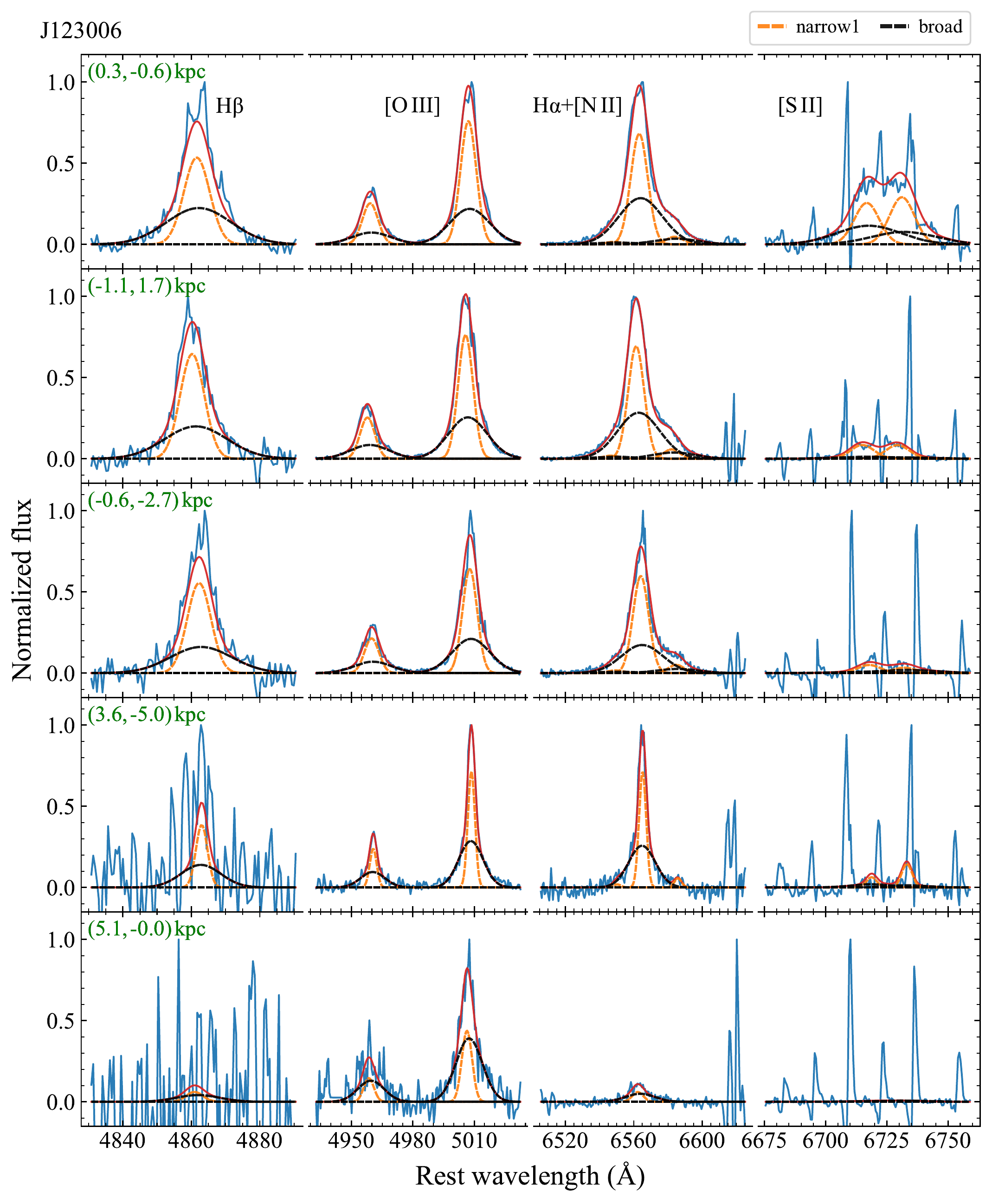}
    \caption{Same as Fig.\,\ref{fig:specJ082313} but for J123006. 
    The telluric emission features were not completely removed from the red part of the spectra, affecting the {\sii} measurements. As a consequence, few spaxels had S/N > 3 in $F_{\sii}$, resulting in a low number of $n_{e,\nb}$ data points (see Fig.\,6 of the paper).
    }
    \label{fig:specJ123006}
\end{figure*}

\begin{figure*}
    \centering
	\includegraphics[width=1\linewidth]{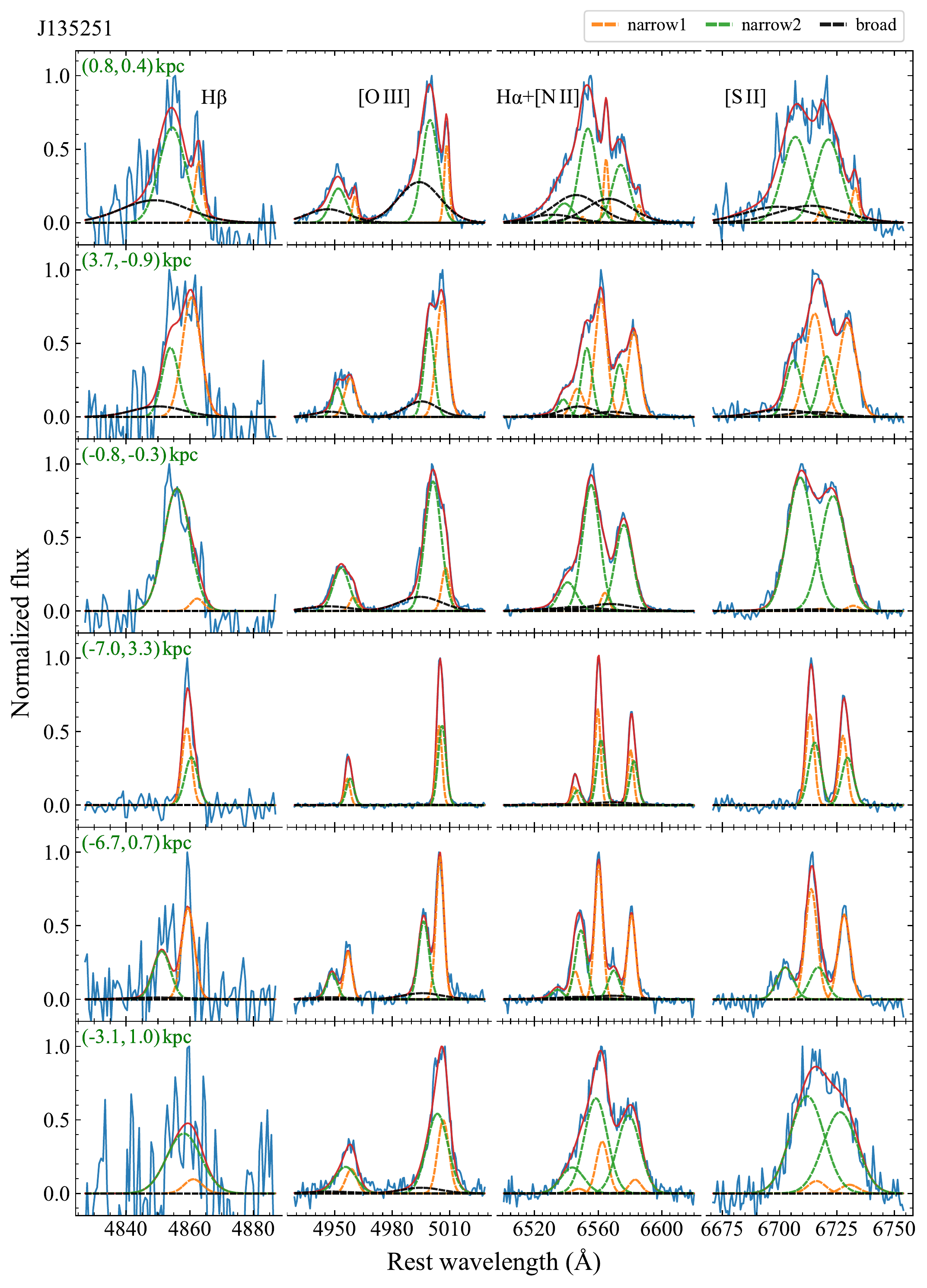}
    \caption{Same as Fig.\,\ref{fig:specJ082313} but for J135251.}
    \label{fig:specJ135251}
\end{figure*}

\begin{figure*}
    \centering
	\includegraphics[width=1\linewidth]{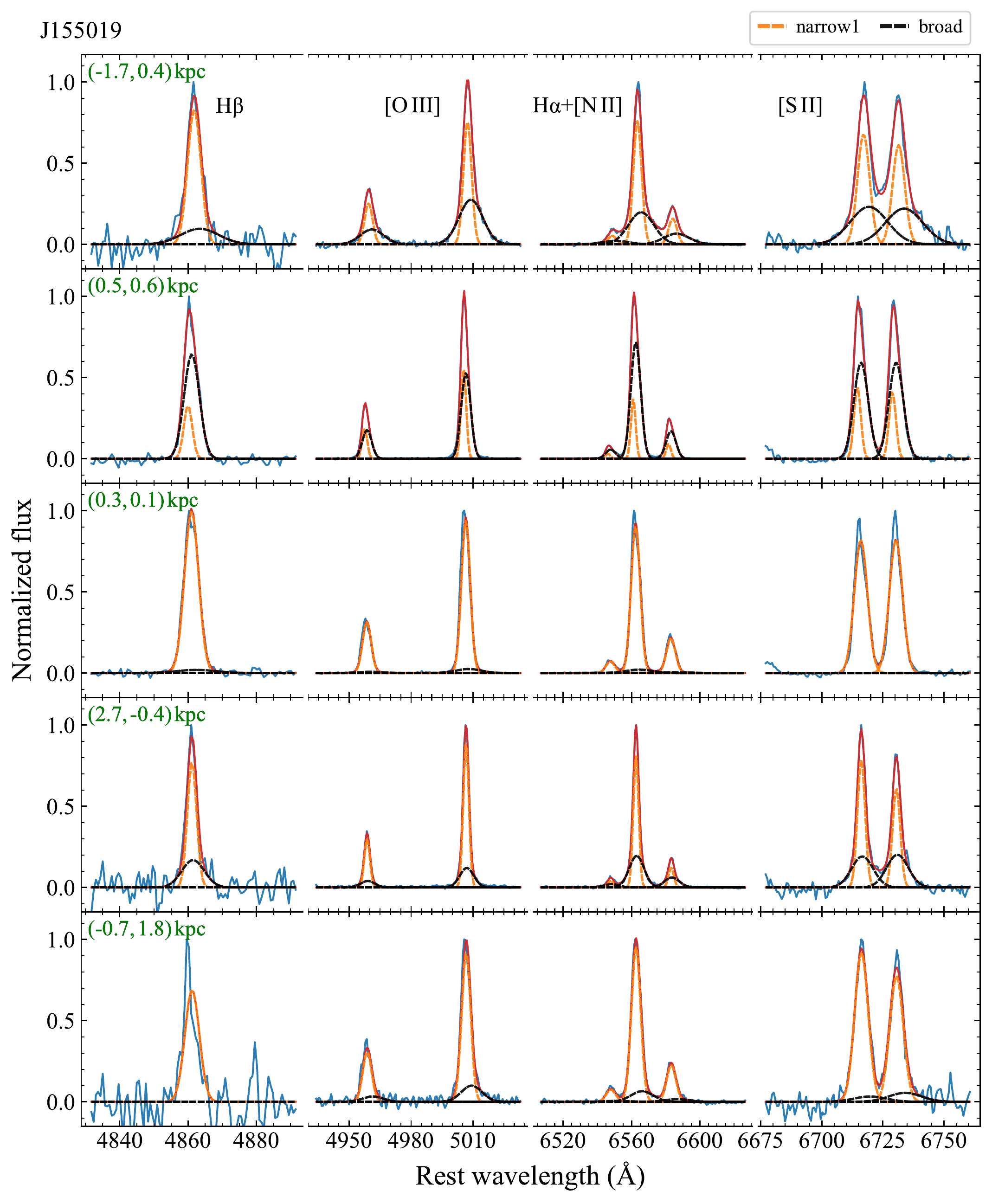}
    \caption{Same as Fig.\,\ref{fig:specJ082313} but for J155019.}
    \label{fig:specJ155019}
\end{figure*}

\begin{figure*}
    \centering
	\includegraphics[width=1\linewidth]{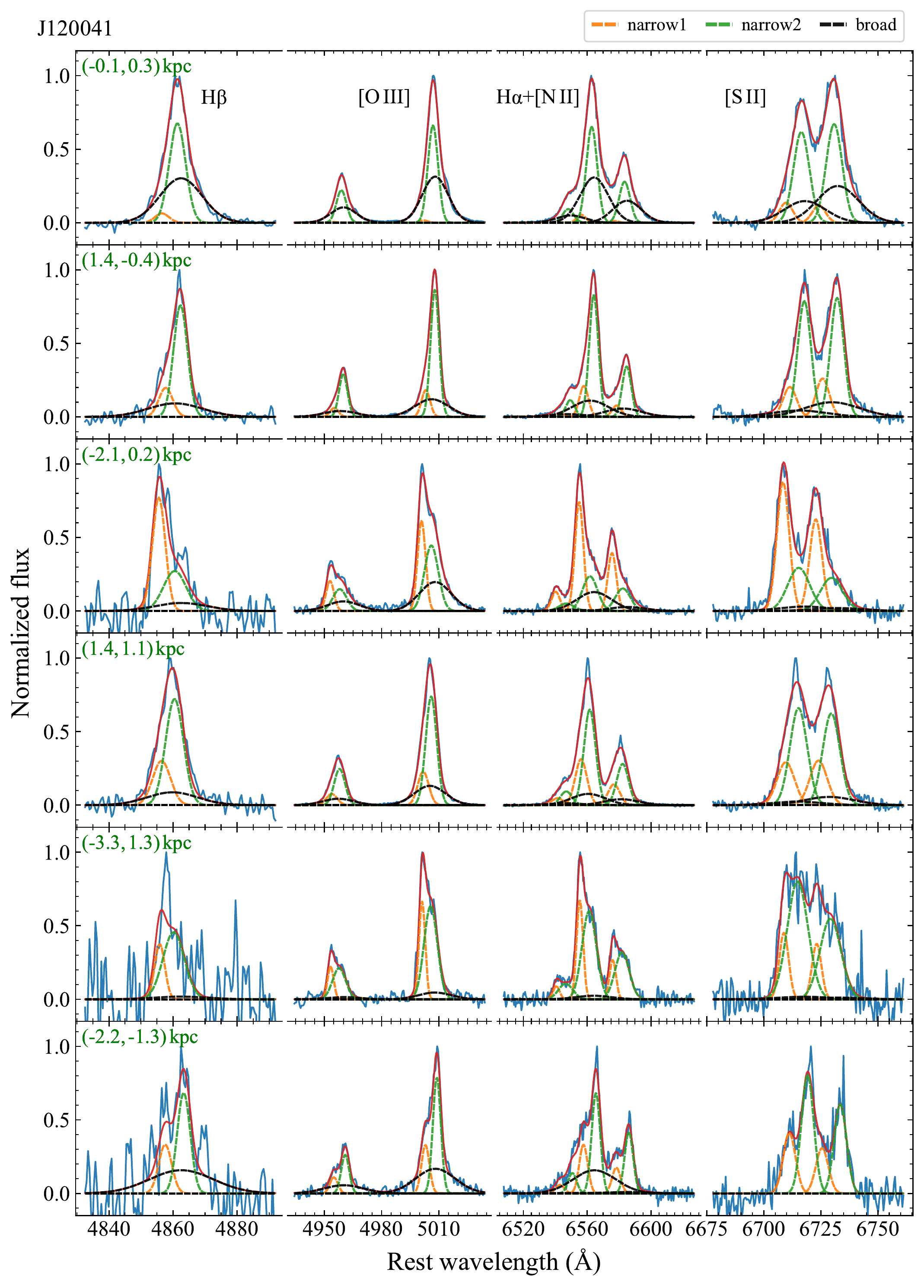}
    \caption{Same as Fig.\,\ref{fig:specJ082313} but for J120041.}
    \label{fig:specJ120041}
\end{figure*}

In Figs.\,\ref{fig:specJ082313}--\ref{fig:specJ120041}, we present results of the fitting process for each QSO of our sample. Different rows correspond to line profiles from different spaxels over the FoV, with the columns displaying a zoom-in on the emission line profiles of \hb, \oiii, \nii, \ha and \sii.

\subsection{Maps}\label{sec:maps}
\begin{figure*}
    \centering
	\includegraphics[width=1\linewidth]{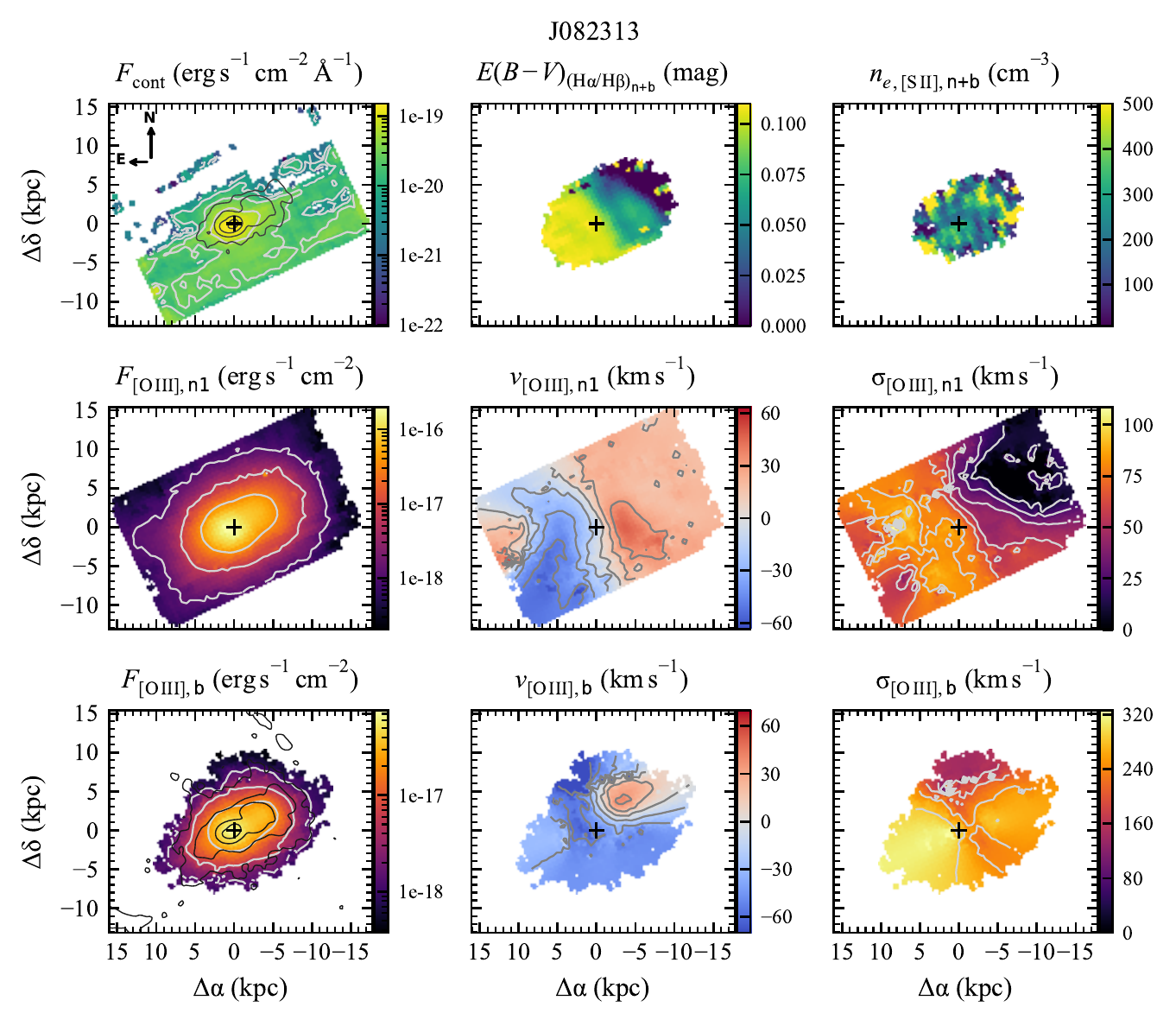}
\caption{Maps of the measured properties of J08231.
In the first row, from left to right, we show: the continuum flux ($F_{\rm{cont}}$), the gas reddening ($E(B\,{-}\,V)$), and the electron density ($n_e$) maps.
The remaining rows display maps of the parameters of each {\oiii}$\lambda$5007 component: flux ($F_{\oiii}$, left); radial velocity ($v_{\oiii}$, centre); and the velocity dispersion ($\sigma_{\oiii}$, right column).
 The last row refers to the {\broad} component, while the remaining middle rows refer to the {\narrow} components (\n{1} nad \n{2}, in this case).
We overplotted the \textit{HST} contours (in black, starting in 3\,$\rm{\sigma_{sky}^{HST}}$) of the continuum (top left) and \oiii (bottom left) images.
North is up and East is left, with the right ascension and declination distances given relative to the continuum peak (black cross).
The systemic velocity were calculated from $v_{\oiii,\n{1}}$. 
Only pixels with S/N > 3 are shown. Note that the colorbars have different ranges.
}
    \label{fig:mapJ082313}
\end{figure*}

\begin{figure*}
    \centering
	\includegraphics[width=1\linewidth]{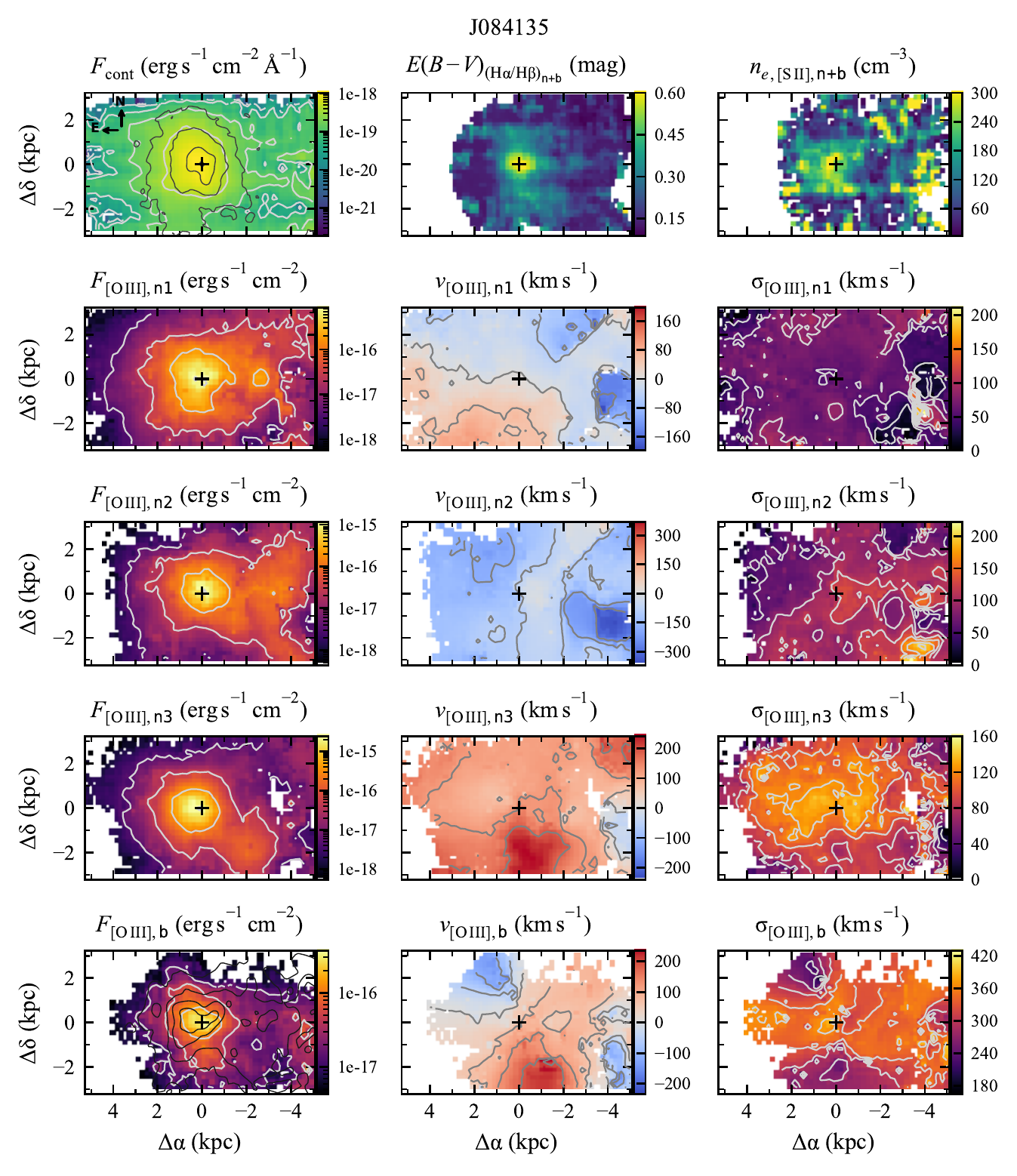}
    \caption{Same as Fig.\,\ref{fig:mapJ082313}, but for J084135. Systemic velocity calculated from $v_{\oiii,\n{1}}$.}
    \label{fig:mapJ084135}
\end{figure*}

\begin{figure*}
    \centering
	\includegraphics[width=1\linewidth]{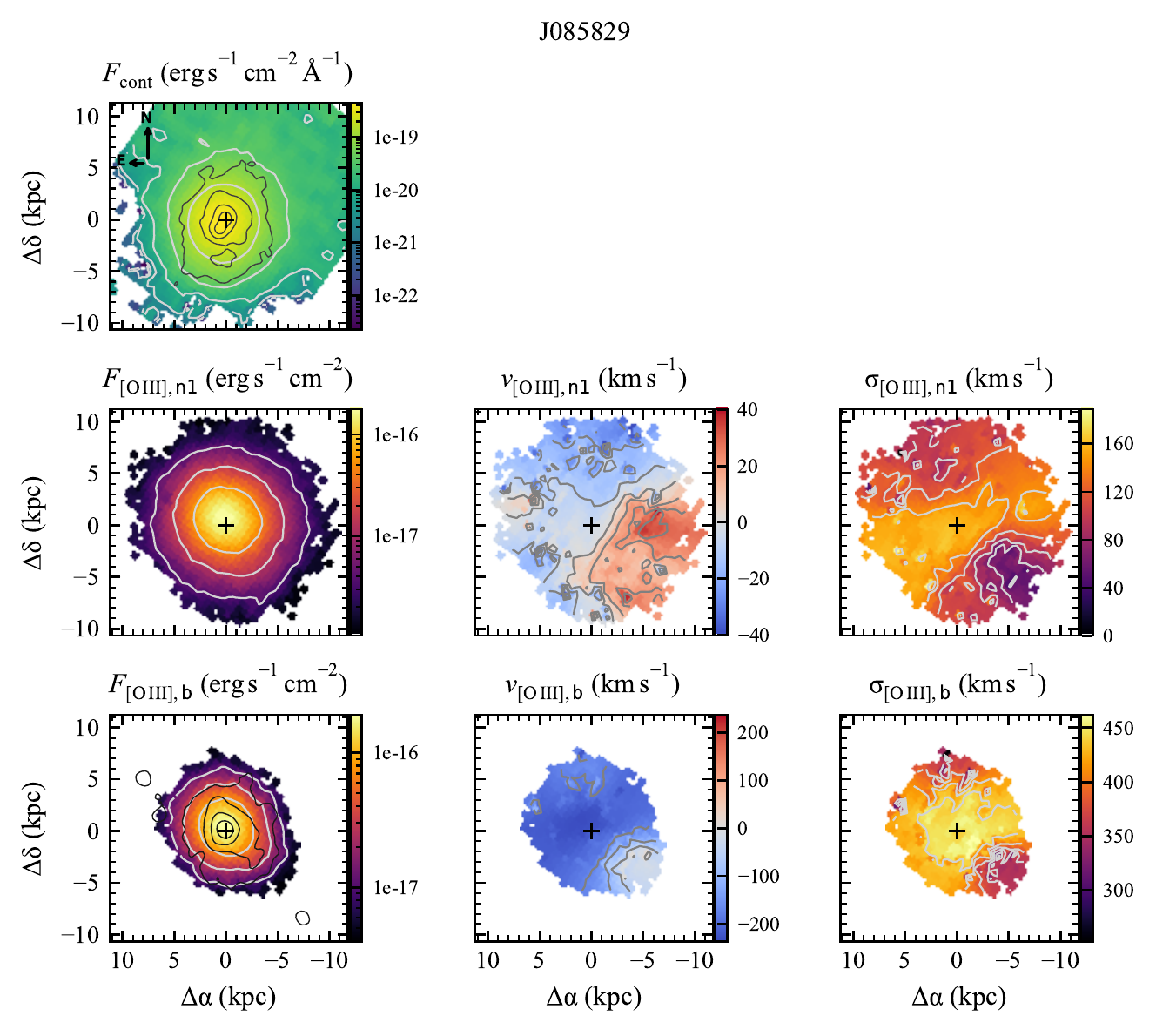}
    \caption{Same as Fig.\,\ref{fig:mapJ082313}, but for J085829. Systemic velocity calculated from $v_{\oiii,\n{1}}$.}
    \label{fig:mapJ085829}
\end{figure*}

\begin{figure*}
    \centering
	\includegraphics[width=.9\linewidth]{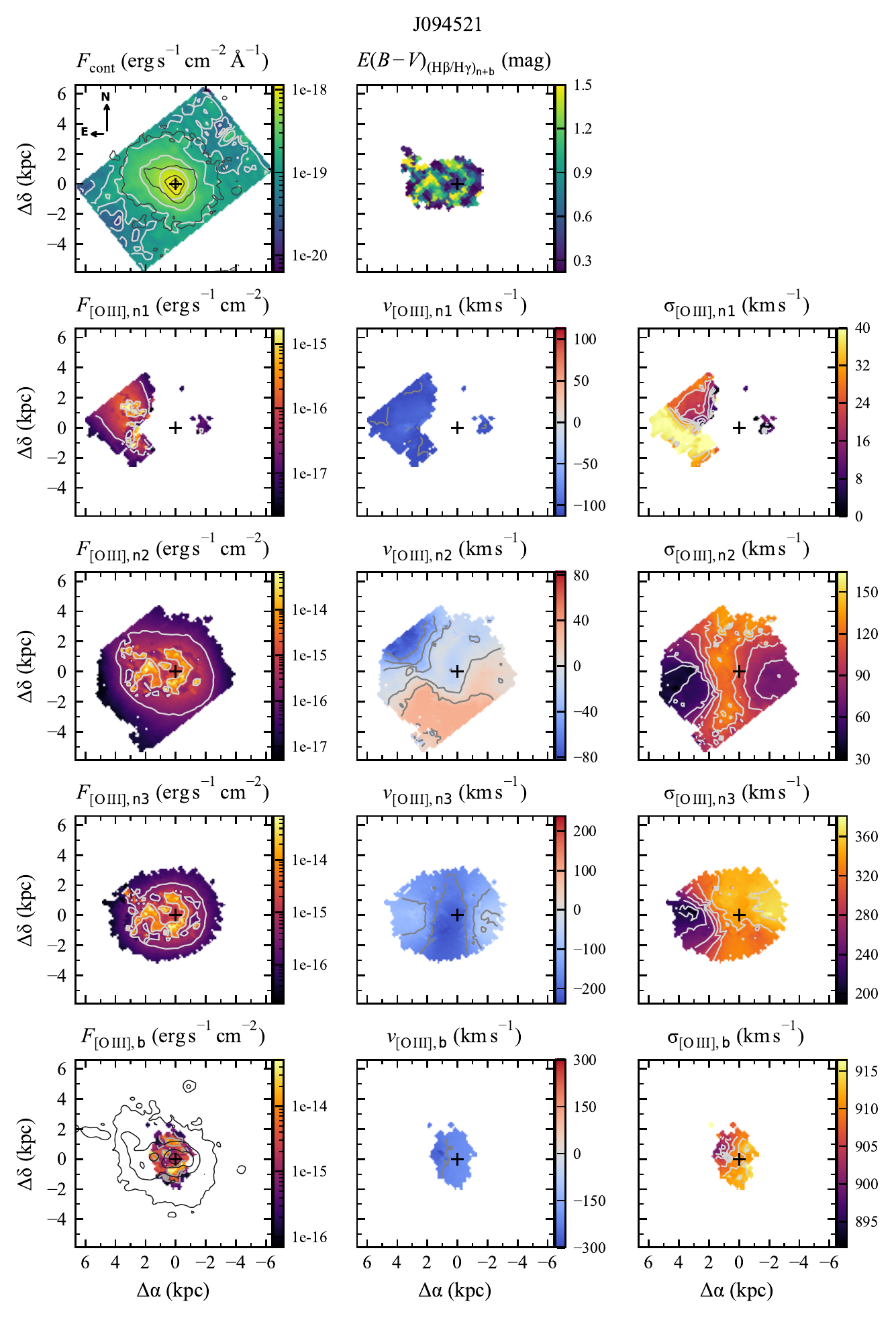}
    \caption{Same as Fig.\,\ref{fig:mapJ082313}, but for J094521. Systemic velocity calculated from $v_{\oiii,\n{2}}$.}
    \label{fig:mapJ094521}
\end{figure*}

\begin{figure*}
    \centering
	\includegraphics[width=1\linewidth]{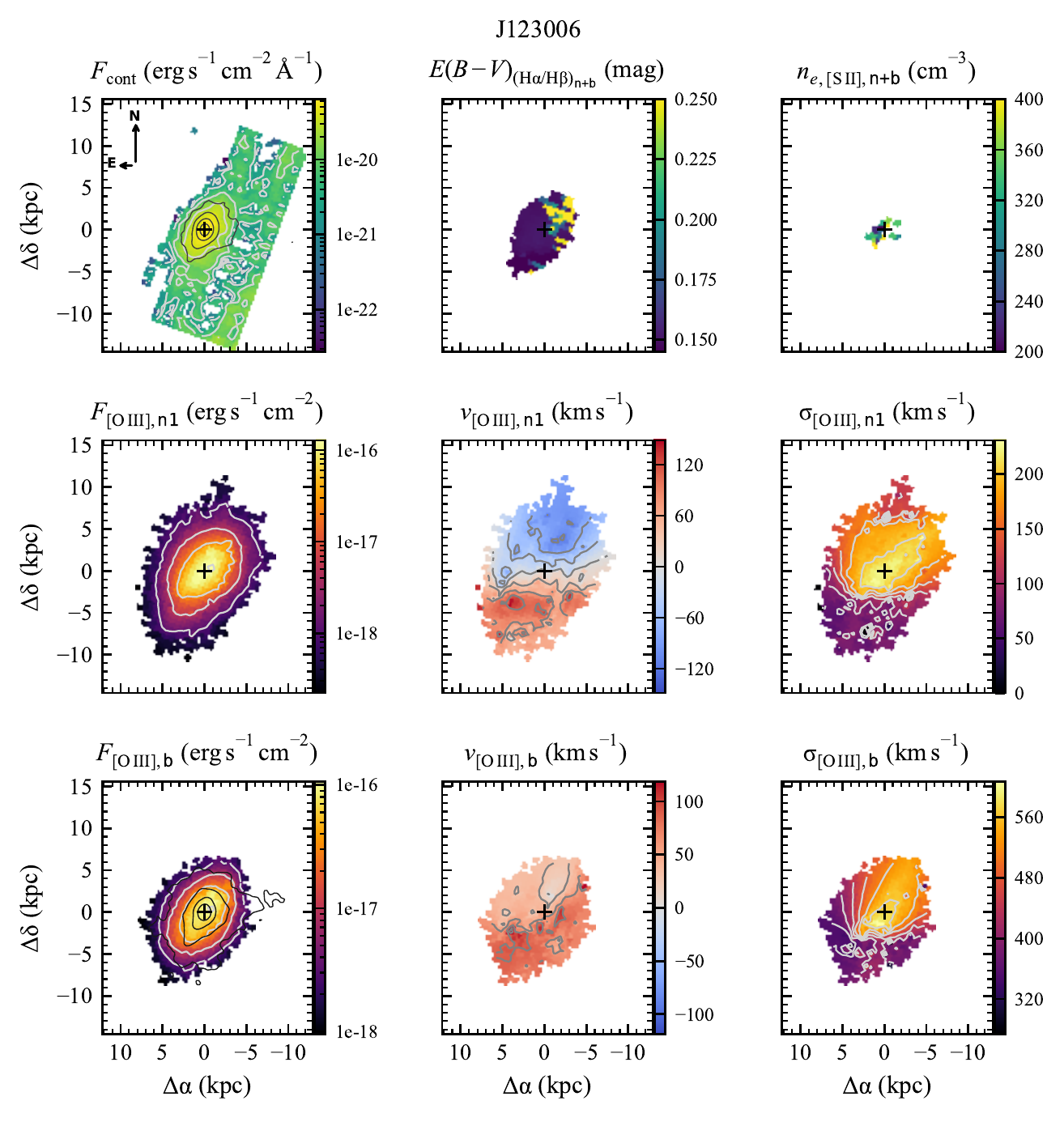}
    \caption{Same as Fig.\,\ref{fig:mapJ082313}, but for J123006. Systemic velocity calculated from $v_{\oiii,\n{1}}$.}
    \label{fig:mapJ123006}
\end{figure*}

\begin{figure*}
	\includegraphics[width=1\linewidth]{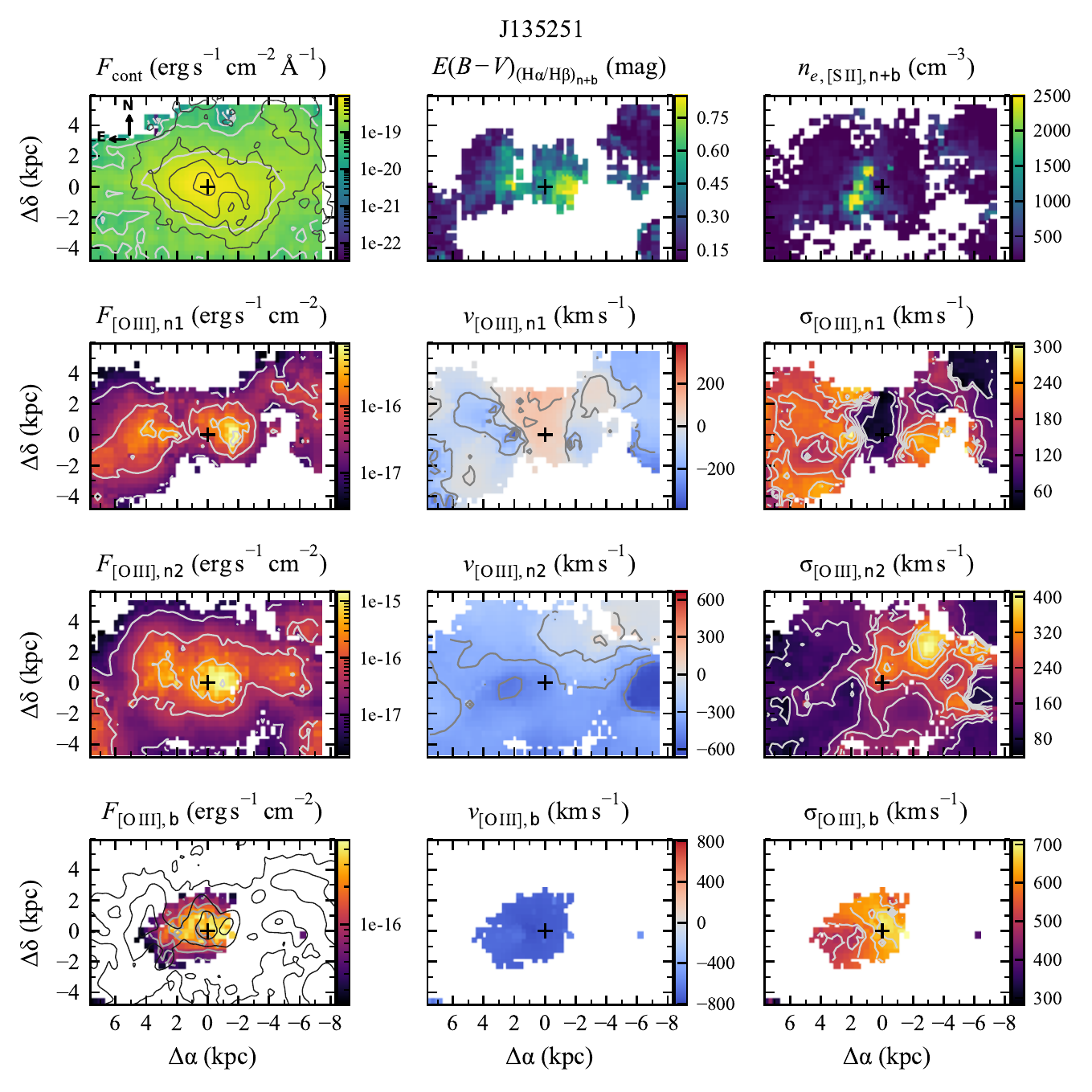}
\caption{Same as Fig.\,\ref{fig:mapJ082313}, but for J135251.
The systemic velocity were calculated from $v_{\oiii,\n{1}}$.} 
    \label{fig:mapJ135251}
\end{figure*}

\begin{figure*}
    \centering
	\includegraphics[width=1\linewidth]{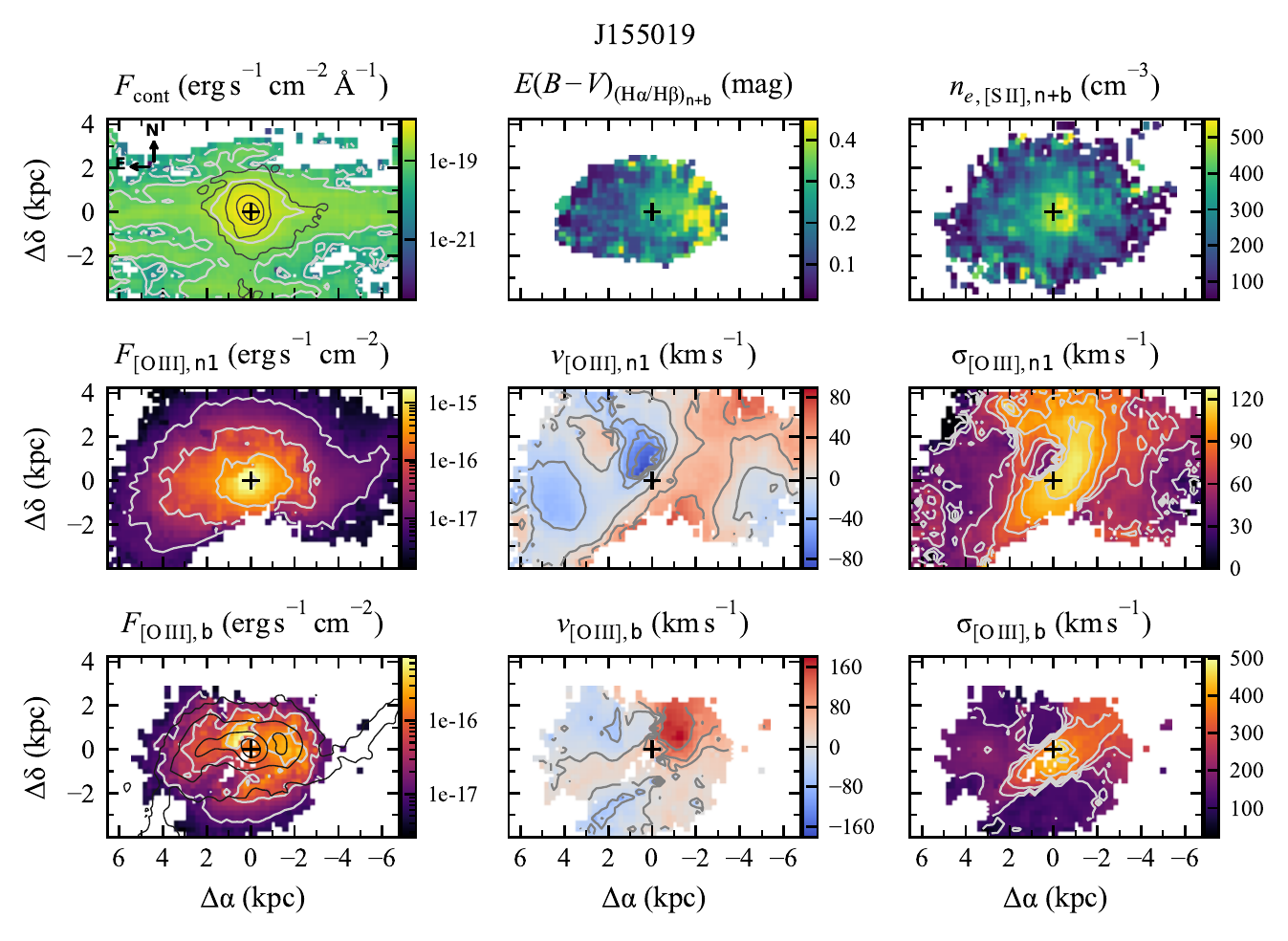}
    \caption{Same as Fig.\,\ref{fig:mapJ082313}, but for J155019. Systemic velocity calculated from $v_{\oiii,\n{1}}$.}
    \label{fig:mapJ155019}
\end{figure*}

\begin{figure*}
    \centering
	\includegraphics[width=1\linewidth]{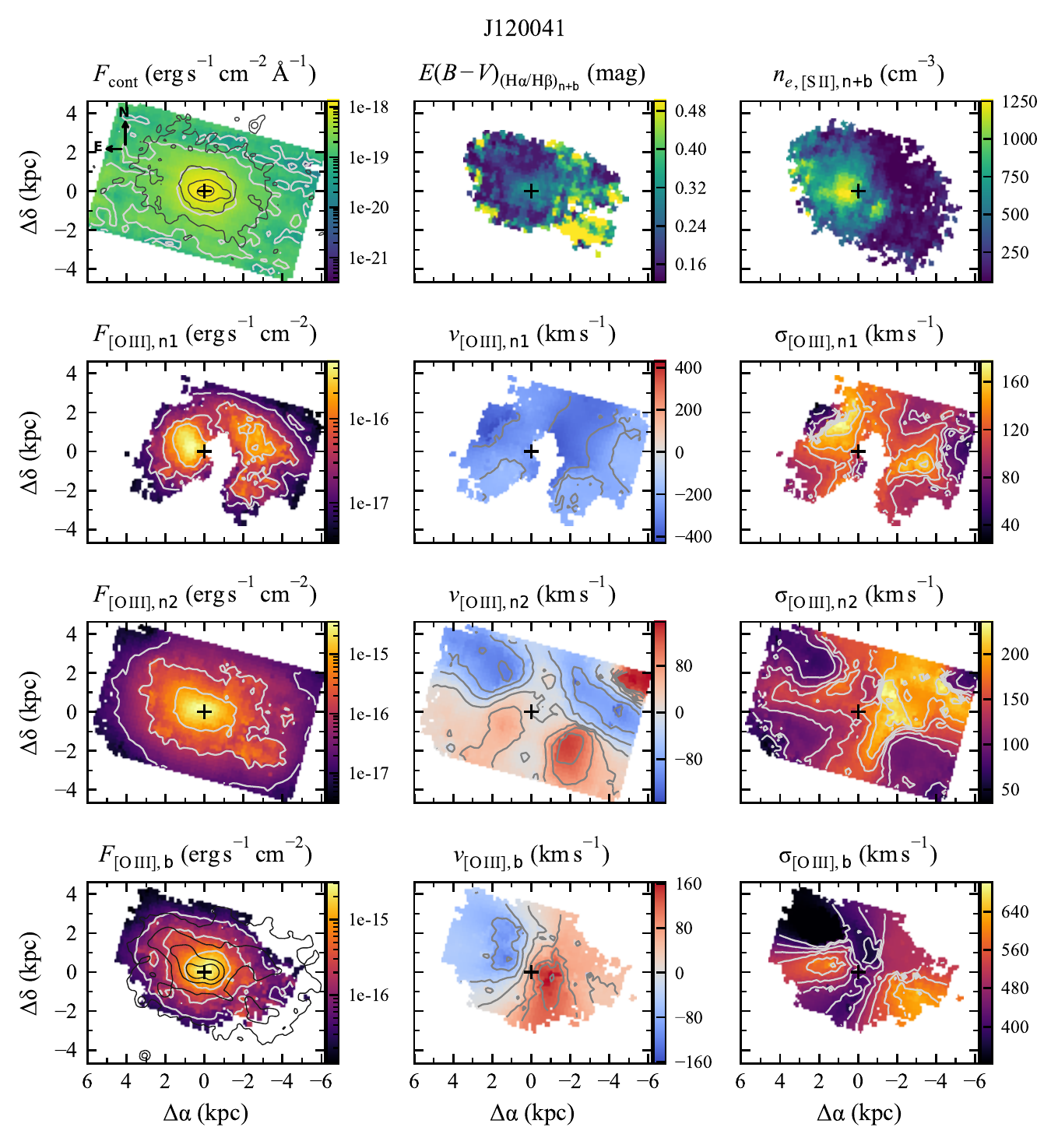}
    \caption{Same as Fig.\,\ref{fig:mapJ082313}, but for J120041. Systemic velocity calculated from $v_{\oiii,\n{2}}$.}
    \label{fig:mapJ120041}
\end{figure*}

In Figs.\,\ref{fig:mapJ082313}--\ref{fig:mapJ120041}, we present larger versions for the maps of the fitted parameters.

\bibliographystyle{mnras}
\bibliography{outflow2021} 

\bsp	
\label{lastpage}